


\hsize=5.0truein
\vsize=7.8 truein
\hoffset=1.9cm
\voffset=1.5cm
\baselineskip=13pt plus 1pt minus 1pt
\tolerance=1000
\parskip=0pt
\parindent=15pt

\font\tenbf=cmbx10
\font\tenrm=cmr10

\font\bfit=cmbxti10
\font\ninebf=cmbx9
\font\ninerm=cmr9
\font\nineit=cmti9

\font\eightrm=cmr8
\font\eightit=cmti8
\font\eightmit=cmmi8
\font\eightcal=cmsy8
\font\sevenrm=cmr7
\font\sevenit=cmti7
\font\sevenmit=cmmi7
\font\fivemit=cmmi5
\font\fiverm=cmr5

\font\tenss=cmss10

\def\smallGamma{{\char'0}}

\def\smallPi{{\char'5}}

\def\sevenmu{\hbox{\sevenmit\char'26}}
\def\sevennu{\hbox{\sevenmit\char'27}}
\def\smallL{\hbox{\eightcal\char'114}}
\def\nineneq{ {\ninerm = \kern-1.em\hbox{/}} }

\nopagenumbers

\def\firstfootline{\eightrm\hss\folio\hss}
\def\otherfootline{\eightrm\hss\folio\hss}
\footline={\ifnum\pageno=0\firstfootline\else\otherfootline\fi}

\def\firstheadline{\hfil}
\def\otherheadline{\eightit
   \ifodd\pageno \hss Neutral and Charged Anyon Fluids \hss
     \else \hss Yutaka Hosotani \hss \fi}
\headline={\ifnum\pageno=1 \firstheadline \else\otherheadline\fi}

\def\qed{\hbox{${\vcenter{\vbox{
    \hrule height 0.4pt\hbox{\vrule width 0.4pt height 6pt
    \kern5pt\vrule width 0.4pt}\hrule height 0.4pt}}}$}}


\def\normal{\baselineskip=13pt plus 1pt minus 1pt}
\def\figure{\baselineskip=10pt \eightrm }
\def\sectionskip{\vskip 12pt}
\def\myref#1{$^{#1}$}

\def\next{~~,~~}
\def\big{\displaystyle \strut }

\def\x{ {\bf x} }
\def\y{ {\bf y} }
\def\z{ {\bf z} }
\def\p{ {\bf p} }
\def\q{ {\bf q} }
\def\u{ {\bf u} }
\def\L{ {\cal L} }
\def\H{ {\cal H} }
\def\K{ {\cal K} }
\def\E{ {\cal E} }
\def\D{ {\cal D} }
\def\S{ {\cal S} }
\def\F{ {\cal F} }
\def\D{ {\cal D} }
\def\N{ {\hat N} }

\def\Ang{ {\,\hbox{\AA}} }
\def\A{ {\rm \AA} }

\def\ep{\epsilon}
\def\eps{\varepsilon^{\mu\nu\rho}}

\def\smallneq{ {\eightrm = \kern-1.em\hbox{/}} }

\def\d{\partial}
\def\la{\raise.16ex\hbox{$\langle$} }
\def\ra{\raise.16ex\hbox{$\rangle$} }
\def\st{\, \raise.16ex\hbox{$|$} \, }
\def\go{\rightarrow}

\def\sing{\Omega_{\rm sing}}
\def\fPhi{\Phi^{\rm f}}

\def\abar{ \bar a }
\def\ahat{ \hat a }

\def\Dbar{ D \kern-.7em\raise.65em\hbox{$-$} }
\def\Dsquare{ D_k \kern-1.2em\raise.63em\hbox{$-${\sevenrm 2}} }
\def\Dbarky{ D_k \kern-1.2em\raise.63em\hbox{$-${\sevenrm y}} }
\def\Dkstarbar{ D_k \kern-1.2em\raise.63em\hbox{$-*$} }
\def\Pibar{ \Pi \kern-.78em\raise.6em\hbox{$-$} }
\def\squarePibar{ \Pi_1 \kern-1.18em\raise.6em\hbox{$-${\sevenrm 2}}\, }
\def\Gbar{ G \kern-.7em\raise.63em\hbox{$-$} }

\def\rhobar{ \rho \kern-.6em\raise.5em\hbox{$-$} }
\def\psibar{ \psi \kern-.65em\raise.6em\hbox{$-$} }
\def\Lbar{ {\cal L} \kern-.65em\raise.6em\hbox{$-$} }

\def\jzbar{ j^0 \kern-.9em\raise.7em\hbox{--} ~}
\def\jkbar{ j^k \kern-.9em\raise.7em\hbox{--} ~}
\def\jmubar{ j^\mu \kern-.9em\raise.7em\hbox{--} ~}
\def\Jzbar{ J^0_{\rm ind} \kern-1.6em\raise.7em\hbox{$-$} ~~~}
\def\Jkbar{ J^k_{\rm ind} \kern-1.6em\raise.7em\hbox{$-$} ~~~}
\def\Jmubar{ J^\mu_{\rm ind} \kern-1.6em\raise.7em\hbox{$-$} ~~~}

\def\indJ{ J_{\rm ind} }

\def\psibar{ \psi \kern-.65em\raise.6em\hbox{$-$} }
\def\Delbar{ \Delta \kern-.8em\raise.65em\hbox{$-$} }
\def\delbar{ \delta \kern-.4em\raise.65em\hbox{--} }
\def\Sbar{ S \kern-.6em\raise.65em\hbox{$-$} }
\def\zz{ \hbox{\tenss Z} \kern-.4em \hbox{\tenss Z} }

\def\tr{{\rm Tr}~}

\def\lam{ \lambda_0 }
\def\blam{ \bar \lambda }

\def\lambar{ \lambda \kern-.6em\raise.65em\hbox{$-$} }
\def\sqlamb{ \lambda \kern-.6em\raise.65em\hbox{$-${\sevenrm 2}} }

\def\Qc{ {Q_c} }
\def\Qn{ {Q_n} }

\def\Dn{ {D_{\rm n}} }

\def\om{ \omega \kern-.7em\raise.45em\hbox{$-$} }
\def\Om{ \Omega \kern-.72em\raise.67em\hbox{$-$} }
\def\delbar{ \delta \kern-.4em\raise.65em\hbox{--} }

\def\film{ {\rm film} }
\def\twoD{ {\rm 2D} }
\def\ext{ {\rm ext} }
\def\ind{ {\rm ind} }
\def\tot{ {\rm tot} }
\def\eff{ {\rm eff} }
\def\HF{ {\rm HF} }
\def\SCF{ {\rm SCF} }
\def\CS{ {\rm CS} }
\def\EM{ {\rm EM} }
\def\vol{ {\it \, vol~} }
\def\inte{ {\rm int} }


\global\newcount\refno
\global\refno=1 \newwrite\reffile
\newwrite\refmac
\newlinechar=`\^^J
\def \ref#1#2{\the\refno\nref#1{#2}}
\def\nref#1#2{\xdef#1{\the\refno}%
	\ifnum\refno=1\immediate\openout\reffile=refs.tmp\fi%
	\immediate\write\reffile{\noexpand\item{\noexpand#1.}#2\noexpand\nobreak}
	\immediate\write\refmac{\def\noexpand#1{\the\refno}}%
	\global\advance\refno by1}


\def\semi{ ^^J}

\def\refn#1#2{\nref#1{#2}}

%

\def\ap#1#2#3{{\nineit Ann.\ Phys.\ (N.Y.)} {\ninebf {#1}}, #3 (19{#2})}
\def\cmp#1#2#3{{\nineit Comm.\ Math.\ Phys.} {\ninebf {#1}}, #3 (19{#2})}
\def\ijmpA#1#2#3{{\nineit Int.\ J.\ Mod.\ Phys.} {\ninebf {A#1}}, #3 (19{#2})}
\def\ijmpB#1#2#3{{\nineit Int.\ J.\ Mod.\ Phys.} {\ninebf {B#1}}, #3 (19{#2})}
\def\jmp#1#2#3{{\nineit  J.\ Math.\ Phys.} {\ninebf {#1}}, #3 (19{#2})}
\def\mplA#1#2#3{{\nineit Mod.\ Phys.\ Lett.} {\ninebf A{#1}}, #3 (19{#2})}
\def\mplB#1#2#3{{\nineit Mod.\ Phys.\ Lett.} {\ninebf B{#1}}, #3 (19{#2})}
\def\plB#1#2#3{{\nineit Phys.\ Lett.} {\ninebf {#1}B}, #3 (19{#2})}
\def\plA#1#2#3{{\nineit Phys.\ Lett.} {\ninebf {#1}A}, #3 (19{#2})}
\def\nc#1#2#3{{\nineit Nuovo Cimento} {\ninebf {#1}B}, #3 (19{#2})}
\def\np#1#2#3{{\nineit Nucl.\ Phys.} {\ninebf B{#1}}, #3 (19{#2})}
\def\prl#1#2#3{{\nineit Phys.\ Rev.\ Lett.} {\ninebf #1}, #3 (19{#2})}
\def\prB#1#2#3{{\nineit Phys.\ Rev.} {\ninebf B{#1}}, #3 (19{#2})}
\def\prD#1#2#3{{\nineit Phys.\ Rev.} {\ninebf D{#1}}, #3 (19{#2})}
\def\prp#1#2#3{{\nineit Phys.\ Report} {\ninebf {#1}C}, #3 (19{#2})}
\def\rmp#1#2#3{{\nineit Rev.\ Mod.\ Phys.} {\ninebf {#1}}, #3 (19{#2})}

\def\ibid#1#2#3{{\nineit ibid.} {\ninebf {#1}}, #3 (19{#2})}

\def\cline{\hfil\noexpand\break ^^J}  

\immediate\openout\refmac=refno.list


\refn\ReviewOne{
For other reviews on the subject, see
F. Wilczek, {\nineit Fractional Statistics and Anyon Superconductivity},
(World Scientific 1990), and refs. 2 -- 10 below.}

\refn\ReviewTwo{
E. Fradkin, {\nineit Field Theories of Condensed Matter Systems},
(Addison-Wesley 1991).}

\refn\ReviewThree{
A.L. Fetter, C.B. Hanna, and R.B. Laughlin, \ijmpB {5} {91} {2751}.}

\refn\ReviewFour{
A. Zee, {\nineit From Semionics to Topological Fluids},  ITP preprint
NSF-ITP-91-129, (1991).}

\refn\ReviewFive{
A.P. Balachandran, E. Ercolessi, G. Morandi, and A.M. Srivastava,
{\nineit The Hubbard Model and Anyon Superconductivity},   (World Scientific
1990).}

\refn\ReviewSix{
Proceedings of the TCSUH Workshop on {\nineit Physics and Mathematics
of Anyons}, \ijmpB {5} {91} {1487}. }

\refn\ReviewSeven{
D. Lykken, J. Sonnenschein,  and N. Weiss, \ijmpA {6} {91} {5155}.}

\refn\ReviewEight{
S. Forte, \rmp {64} {92} {193}.}

\refn\ReviewNine{
R. Iengo and K. Lechner,  \prp {213} {92} {179}.}

\refn\ReviewLast{
D. Boyanovsky,   {\nineit Gauge invariance and broken symmetries in anyon
superfluids}, Pittsburgh preprint PITT-92-01.}

\refn\LaughlinOne{
V. Kalmeyer and R. B. Laughlin, \prl{59}{87}{2095};
\semi R. B. Laughlin, {\nineit Science} {\ninebf 242}, 525 (1988);
\semi R. B. Laughlin, \prl{60}{88}{2677}.   }

\refn\Anyon{J.M. Leinaas and J. Myrheim, \nc {37}{77} {1};
\semi G.A. Goldin, R. Menikoff, and D.H. Sharp, \jmp {22} {81} {1664};
\semi F. Wilczek, \prl{48} {82} {1144}; \ibid {49} {82} {957};
\semi Y.S. Wu, \prl {53} {84} {111}. }

\refn\LaughlinFQHE{
R.B. Laughlin, \prl {50} {83} {1395};
\semi R.B. Laughlin, in  {\nineit The Quantum Hall Effect},
 (ref. 17 below), p. 233. }

\refn\hierarchy{
F.D.M. Haldane, \prl {51} {83} {605};
\semi B.I. Halperin, \prl {52} {84} {1583}, \ibid {52} {84} {2390}(E).}

\refn\ASW{
D. Arovas, J.R. Schrieffer, and F. Wilczek, \prl {53} {84} {722};
\semi R. Tao and Y.S. Wu, \prB {31} {85} {6859};
\semi S. He, X-C. Xie, and F-C. Zhang, \prl {68} {92} {3460}.}

\refn\Tsui{
J.A. Simmons, H.P. Wei, L.W. Engel, D.C. Tsui, and M. Shayegan,
\prl {63} {89} {1731}.}

\refn\FQHEbook{
R.E. Prange and S.M. Girvin (eds.), {\nineit The Quantum Hall Effect},
(Springer-Verlag, 1987);
\semi T. Chakraborty and P. Pietil\"ainen, {\nineit The Fractional Quantum
Hall Effect},  (Springer-Verlag,1988). }


\refn\CSinFQHE{
S.C. Zhang, T. Hansson and S. Kivelson, \prl {62} {89} {82};
\semi N. Read, \prl {62} {89} {86}.}

\refn\Jain{
J.K. Jain, \prl {63} {89} {199}; \prB {40} {89} {8079};
\semi J.K. Jain, S.A. Kivelson, and N. Trevedi, \prl {64} {90} {1297};
\ibid  {64} {90} {1993} (E);
\semi G. Dev and J.K. Jain, \prB {45} {92} {1223}.}

\refn\Niu{
 X.G. Wen and Q. Niu,  \prB {41} {90} {9377}.}

\refn\FQHEmoreOne{
B. Blok and X.G. Wen, \prB {42} {90} {8133}; \ibid {42} {90} {8145};
\semi X.G. Wen and A. Zee, \prB {44} {91} {274};
\semi J. Fr\"ohlich and T. Kerler,\np {354} {91} {369};
\semi A. Lopez and E. Fradkin, \prB {44} {91} {5246};
\semi J. Fr\"ohlich and A. Zee, \np {364} {91} {517};
\semi D.H. Lee and S.C. Zhang,  IBM preprint, {\nineit Collective
excitations in the Ginzburg-Landau theory of the fractional quantum Hall
effect}.}

\refn\Ishikawa{
N. Imai, K. Ishikawa,  T. Matsuyama, and I. Tanaka, \prB {42} {90} {10610}.}

\refn\Ezawa{
Z.F. Ezawa and A. Iwazaki, \prB {43} {91} {2637}; {\nineit Chern-Simons gauge
theory for even-denominator fractional quantum hall states}, Tohoku preprint
TU-390 (1991); {\nineit Wigner crystal of quasiparticles and fractional
quantum Hall states}, TU-398 (1992); {\nineit Chern-Simons gauge theory
for doublelayer electron system}, TU-402;
\semi Z.F. Ezawa, M. Hotta, and A. Iwazaki, \prD {44} {91} {452};
{\nineit Field theory of anyons and fractional quantum Hall effect},
TU-376 (1991).}

\refn\Shizuya{
K. Shizuya, \prB {45} {92} {11143}.}

\refn\FQHEmoreTwo{
D.H. Lee, S.A. Kivelson, and S.C. Zhang, \prl {67} {91} {3302};
  \ibid {68} {92} {2386};
\semi S.C. Zhang, \ijmpB {6} {92} {25};
\semi S. Kivelson, D.H. Lee, and S.C. Zhang, \prB {46} {92} {2223}.}

\refn\FQHEmoreThree{
 M. Greiter, X.G. Wen, and F. Wilczek,  \prl {66} {91} {3205};
\np {374} {92} {567};
{\nineit Paired Hall states in double layer electron
systems}, IASSNS-HEP-92/1;
\semi F. Wilczek, {\nineit Disassembling anyons}, IASSNS-HEP-91/70;
\semi X.G. Wen and A. Zee, {\nineit A classification of Abelian quantum Hall
states and matrix formulation of topological fluids}, NSF-ITP-92-10;
{\nineit Neutral superfluid
modes and magnetic  monopoles in multi-layered quantum Hall systems},
NSF-ITP-92-22; {\nineit Tunnelling in double-layered quantum
Hall systems}, ITP preprint (May 1992);
\semi A. Balatsky, {\nineit Spin singlet quantum Hall effect and
non-Abelian Landau-Ginzburg theory},  Los Alamos preprint, LA-UR-91-3717;
\semi B.  Rejaei and C.W.J. Beenakker, {\nineit Vector-mean-field theory of the
fractional quantum Hall effect}, Leiden preprint (Aug. 1992).}

\refn\FHL{
A. L. Fetter, C. B. Hanna, and R. B. Laughlin, \prB{39} {89} {9679}.}

\refn\HLF{
C. B. Hanna, R. B. Laughlin, and A. L. Fetter, \prB{40}{89}{8745};
\ibid {43} {91} {309}.}

\refn\CWWH{
Y.-H. Chen, F. Wilczek, E. Witten and B. Halperin, \ijmpB {3} {89} {1001}.  }

\refn\WenZee{
X. G. Wen and A. Zee, \prB{41} {90} {240}.  }

\refn\CanGirv{
G. S. Canright, S. M. Girvin, and A. Brass, \prl {63} {89} {2291, 2295}. }

\refn\HosoChak{
Y. Hosotani and S. Chakravarty, IAS report, IASSNS-HEP-89/31,
 May 1989 ; \prB{42}{90}{342}; \prD{44}{91}{441}.}

\refn\Fradkin{
E. Fradkin, \prl{63} {89} {322}; \prB {42} {90} {570}. }

\refn\FisherLee{
M.P.A. Fisher and D.H. Lee, \prl{63} {89} {903}. }

\refn\Mori{
H. Mori, \prB {42} {90} {184}.}

\refn\Banks{
T. Banks and J. Lykken, \np {336} {90} {500}. }

\refn\BalatskyKal{
A. Balatsky and V. Kalmeyer, {\nineit Singlet pair superconductivity in the
two-component anyon gas},   Illinois preprint P/90/5/63.}

\refn\RDSS{
S. Randjbar-Daemi, A. Salam, and J. Strathdee, \np{340} {90} {403}. }

\refn\KitaMura{
Y. Kitazawa and H. Murayama, \np{338} {90} {777}; \prB{41}{90}{11101}.}

\refn\Sakita{
P. K. Panigrahi, R. Ray, and B. Sakita, \prB {42} {90} {4036}. }

\refn\Lykken{
J.D. Lykken, J. Sonnenschein and N. Weiss, \prD {42} {90} {2161};
\ijmpA{6}{91}{1335} }

\refn\Gerbert{
P. de Sousa Gerbert, \prD {42} {90} {543}.}

\refn\Ma{
 L. Zhang, M. Ma, and F.C. Zhang, \prB{42} {90} {7894};
\semi F.C. Zhang and M.R. Norman, \prB {43} {91} {6143}.}

\refn\HHL{
J. E. Hetrick, Y. Hosotani, and B.-H. Lee, \ap{209} {91} {151}. }

\refn\Aronov{
A.G. Aronov and A.D. Mirlin, \plA{152}{91}{371}.}

\refn\Khveshchenko{
B. Rejaei and C.W.J. Beenakker, \prB {43} {91} {11392};
\semi D.V. Khveshchenko, ETH reports, {\nineit Transport properties of anyons},
ETH-TH/91-8 and ETH-TH/91-12.}

\refn\Georgelin{
Y. Georgelin, M. Knecht, Y. Leblanc, and J.C. Wallet, \mplB {5} {91} {211};
\semi Y. Leblanc and J.C. Wallet, {\nineit More on finite temperature anyon
superconductivity}, UAHEP-929 (May 1992).}

\refn\Leggett{
A.J. Leggett, Minnesota report, {\nineit The mean field anyon gas;
superconductor
or superdiamagnetic insulator?},  TPI-MINN-90/36-T, 1990.}

\refn\Fetter{
A.L. Fetter and C. Hanna, \prB {45} {92} {2335}.}

\refn\radiative{
D.V. Khveshchenko and I.I. Kogan, \ijmpB {5} {91} {2355}.}

\refn\Wang{
X.C. Xie, S. He, and S. Das Sarma, \prl {65} {90} {649};
\semi B.H. Liu, X.Q. Wang and L. McLerran, \prB{43}{91}{13736}.}

\refn\Rao{
D.M. Gaitonde and S. Rao, \prB {44} {91} {929}.}

\refn\Joe{
J. Kapusta, M.E. Carrington, B. Bayman, D. Seibert, and C.S. Song,
 \prB {44} {91} {7519}. }

\refn\Eliashvili{
M. Eliashvili and G. Tsitsishvili, {\nineit On the anyon superconductivity
in thermo fields dynamics}, Tbilisi preprint (1991).}

\refn\Dai{
Q. Dai, J.L. Levy, A.L. Fetter, C.B. Hanna, and R.B. Laughlin, Stanford
preprint, {\nineit Quantum Mechanics of the Fractional Statistics Gas;
Random Phase Approximation}.  }

\refn\LaughZouTwo{
A.M. Tikofsky, R.B. Laughlin, and Z. Zou, {\nineit Optical conductivity of the
t-J model as a fractional statistics superconductor},  Stanford preprint,
 (Nov. 1991).}

\refn\YiCanright{
J. Yi and G.S. Canright, {\nineit Spontaneous magnetization
of anyons with long-range repulsion}, Tennessee preprint, (July 1992).}

\refn\Sumantra{
S. Chakravarty, {\nineit Linear response of charged and neutral semions},
Minnesota preprint UMN-TH-1015/91.}

\refn\Aharonov{
Y. Aharonov and D. Bohm, {\nineit Proc. Phys. Soc. (London)} {\ninebf
B62}, 8 (1949);
\semi A. Tonomura {\nineit et al.}, \prl  {48} {82} {1443};
    \ibid {51} {83} {331}. }

\refn\Arovas{
D. Arovas, J. R. Schrieffer, F. Wilczek, and A. Zee, \np {251} {85} {117}.}

\refn\Goldhaber{
A. Goldhaber, R. MacKenzie, and F. Wilczek, \mplA {4} {89} {21};
\semi X.G. Wen and A. Zee, {\nineit J. Phys. France} {\ninebf 50} 1623,
(1989).}

\refn\ABHosotani{
Y. Hosotani, \plB {126} {83} {309}; \ap {190} {89} {233};
\semi D. Toms \plB {126} {83} {445};
\semi E. Witten, \np {258} {85} {75}. }

\refn\CStheory{
J. Schonfeld, \np {185} {81} {157};
\semi R. Jackiw and S. Templeton, \prD{23}{81}{2291};
\semi S. Deser, R. Jackiw, and S. Templeton, \prl{48}{82}{975};  \ap{140}
{82}{372}; \ibid {185} {88} {406} (E).}

\refn\Redlich{
A.N. Redlich, \prl {52} {84} {18}; \prD {29} {84} {2366};
\semi R. Jackiw, \prD {29} {84} {2375}.}

\refn\Hagen{
C.R. Hagen, \ap{157}{84}{342}; \prD{31}{85}{2135}.  }

\refn\CSFracStat{
A.M. Polyakov, \mplA{3}{88}{325}.}

\refn\CsFracStatTwo{
 G. W. Semenoff, \prl{61} {88} {517};
\semi X. G. Wen and A. Zee, \prl {61} {88} {1025}, \prl{62} {89} {1937},
     \prl{63} {89} {461};
\semi J. Fr\"ohlich and P. Marchetti, Comm. Math. Phys. {\bf116}, 127 (1988);
  \ibid {121} {89} {117};
\semi J. Fr\"ohlich and T. Kerler, \np {354} {91} {369}. }

\refn\CSfieldTheory{
G.V. Dunne, R. Jackiw, and C.A. Trugenberger, \ap {194} {89} {197};
     \prD {41} {90} {661};
\semi R. Jackiw, \ap {201} {90} {83};
\semi R. Jackiw and V.P. Nair, \prD {43} {91} {1933}. }

\refn\Boyan{
D. Boyanovsky,  \prD {42} {90} {1179}; \ijmpA {7} {92} {4619};
 \semi D. Boyanovsky and D. Jasnow, {\nineit Physica} {\ninebf A177}
(1991) 537;
 {\nineit Gauge invariance, statistics and order in anyon
fluid}, PITT-91-07;
\semi D. Boyanovsky, E. Newman, and C. Rovelli, \prD {45} {92} {1210};
\semi C. Aragao de Carvalho and D. Boyanovsky, {\nineit Quasi-long range order
in a flavored anyon model}, PITT-92-03.}

\refn\JackiwPi{
R. Jackiw and S.Y. Pi, \prD {42} {90} {3500}. }

\refn\CStorus{
Y. Hosotani, \prl {62} {89} {2785};  \prl {64} {90} {1691}.}

\refn\WenTorus{
X.G. Wen, \prB {40} {89} {7387}; \ijmpB {4} {90} {239}.}

\refn\FradkinTwo{
E. Fradkin, \prl {63} {89} {322}. }

\refn\Lee{
K. Lee, Boston Univ. report,
               {\nineit Anyons on spheres and tori},  BU/HEP-89-28;
\semi A.P. Polychronakos, \ap{203}{90}{231};
Univ. of Florida preprint,
{\nineit Abelian Chern-Simons theories and conformal blocks} UFIFT-HEP-89-9;
       \plB {241} {90} {37}.  }

\refn\Conformal{
S. Elitzur, G. Moore, A. Schwimmer and N. Seiberg, \np {326} {89} {108};
\semi M. Bos and V.P. Nair, \plB {223} {89} {61}; \ijmpA {5} {90} {959};
\semi J.M.F. Labastida and A.V. Ramallo, \plB {227} {89} {92}; \plB {228}
{89} {214};
\semi H. Murayama, {\nineit Z.\ Phys.} {\bf C48}, 79 (1990). }

\refn\CStorusW{
E. Witten, \cmp {121} {89} {351}.}

\refn\Einarsson{
T. Einarsson, \prl {64} {90} {1995};  \ijmpB {5} {91} {675};
\semi A.P. Balachandran, T. Einarsson, T.R. Govindarajan, and R. Ramachandran,
 \mplA {6} {91} {2801}. }

\refn\Segre{
G.C. Segre,  Pennsylvania preprint, {\nineit A model of superconductivity}. }

\refn\DaemiTwo{
S. Randjbar-Daemi, A. Salam, and J. Strathdee, \plB {240} {90} {121}. }

\refn\Iengo{
R. Iengo and K. Lechner, \np {346} {90} {551}; \np {364} {91} {551};
\semi K. Lechner, \plB {273} {91} {463};
\semi R. Iengo, K. Lechner, and D. Li, \plB {269} {91} {109}. }

\refn\Musto{
G. Cristofano, G. Maiella, R. Musto and F. Nicodemi, \plB {262} {91} {88};
\mplA {6} {91} {1779}. }

\refn\lattice{
X.G. Wen, E. Dagotto, and E. Fradkin, \prB {42} {90} {6110}.}


\refn\Lechner{
K. Lechner, Trieste ISAS preprint, {\nineit Anyon physics on the torus},
thesis,
(Apr. 91).}

\refn\WuCylin{
Y.S. Wu, Y. Hatsugai, and M. Kohmoto, \prl {66} {91} {951};
\semi Y. Hatsugai, M. Kohmoto, and Y.S. Wu, \prB {43} {91} {2661}.}

\refn\Ho{
C.-L. Ho and Y. Hosotani,  {\nineit Anyon equation on a torus},
 {\nineit Int.\ J.\ Mod.\ Phys.} {\ninebf A} (in press). }

\refn\Burns{
See, for instance, G. Burns, {\nineit High-Temperature Superconductivity --
An Introduction},  (Academic Press, 1992).}

\refn\holon{
P. W. Anderson, {\nineit Science} {\ninebf 235}, 1196 (1987).}

\refn\holonTwo{
P. W. Anderson and Z. Zou, \prl {60} {88} {137};  \ijmpB {4}{90}{181};
\semi Z. Zou and P. W. Anderson, \prB {37} {88} {627};
\semi G. Baskaran and P. W. Anderson, \prB {37} {88} {580};
\semi P. W. Anderson, G. Baskaran, Z. Zou, and T. Hsu, \prl {58} {87} {2790} ;
\semi V. Kalmeyer and R. B. Laughlin, \prB {39} {89} {11879}.}

\refn\holonThree{
 E. Dagotto, \ijmpB {5} {91} {907};
\semi Y. Hasegawa, P. Lederer, T.M. Rice, and P.B. Wiegmann, \prl {63}
  {89} {907};
\semi P. Lederer, D. Poilblanc, and T.M. Rice, \prl {63} {89} {1519};
\semi J.P. Rodriguez and B. Doucot, \prB {42} {90} {8724};
   \prB {43} {91} {6209} (E).}

\refn\LaughlinLattice{
R. B. Laughlin, \ap {191} {89} {163}.}

\refn\LaughlinZou{
R. B. Laughlin and Z. Zou, \prB{41}{90}{664}.}

\refn\spinModel{
E.J. Mele, G. Segre, and Ch. Bruder, \prB {43} {91} {5576}. }

\refn\EffecTh{I. Dzyaloshinskii, A. Polyakov and P. Wiegmann, \plA
{127} {88} {112};
\semi P. Wiegmann, {\nineit Physica} {\ninebf C153} (1988) 103;
\prl{60}{88}{821} ;
\semi J. B. Marston, \prl {61} {88} {1914};
\semi X. G. Wen, F. Wilczek, and A. Zee, \prB {39} {89} {11413};
\semi X. G. Wen, \prB {39} {89} {7223};
\semi F. D. M. Haldane, \prl {61} {88} {1029}. }

\refn\Wiegmann{
P. Wiegmann, {\nineit Topological Superconductivity}, Chicago preprint
(Dec. 1991)}

\refn\PathIntegral{
See, for instance, S. Coleman, {\nineit Aspects of Symmetry}, Chapter 5,
(Cambridge University Press, 1985).}

\refn\FTFT{R. Kubo and M. Toda (ed), {\nineit Statistical Physics},
(Iwanami, 1972), (in Japanese);
\semi R. Kubo, M. Toda, and N. Hashitsume, {\nineit Statistical
Physics II}, (Springer-Verlag,  1985);
\semi A. L. Fetter and J. D. Walecka, {\nineit Quantum Theory of Many Particle
Systems}, (McGraw-Hill, 1971). }

\refn\Abrikosov{A.A. Abrikosov, L.P. Gorkov, and I.E. Dzyaloshinski,
{\nineit Methods of Quantum Field Theory in Statistical Physics},
(Dover Publications,  1963).}

\refn\SCReview{For  BCS  superconductivity, see,  
M. Tinkham, {\nineit Introduction to Superconductivity}, (McGraw-Hill, 1975);
\semi P.G. De Gennes, {\nineit Superconductivity in Metals and Alloys},
(Benjamin Inc., 1966).}

\refn\Alphen{
See, for instance, A.A. Abrikosov, {\nineit Fundametals of the Theory
of Metals},  Chapter 10, (North-Holland, 1988).}

\refn\Uemura{
Y.J. Uemura et al., \prl {62} {89} {2317};
\semi Y.J. Uemura et al., \prl {66} {91} {2665}.}

\refn\TDLee{
R. Friedberg, T.D. Lee, and H.C. Ren, {\nineit Applications of the s-channel
theory to the $\mu$SR and Hall number experiments},  Columbia preprint,
CU-YP-483 (Oct. 1990). }

\refn\Josephson{
B.D. Josephson,  {\nineit Advances in Physics} {\ninebf 14}, 419 (1965);
\semi P.W. Anderson, {\nineit Prog. in Low Temp. Phys. vol.}{\ninebf V}, 1
(1967).}

\refn\interlayer{
A.G. Rojo and G.S. Canright, \prl {66} {91} {949};
\semi A.G. Rojo and A.J. Leggett, \prl {67} {91} {3614}.}

\refn\WenZeeTwo{
X. G. Wen and A. Zee, \prl {62} {89}  {2873}.  }

\refn\HRW{
B. I. Halperin, J. March-Russell, and F. Wilczek, \prB{40} {89} {8726}. }

\refn\PTVio{
J. March-Russell and F. Wilczek, \prl {61} {88} {2066};
\semi X.G. Wen, F. Wilczek, and A. Zee, \prB{39}{89}{11413};
\semi Y. Kitazawa, \prl {65} {90} {1275};
\semi P.B. Wiegmann, \prl{65} {90} {2070};
\semi G.S. Canright and M.D. Johnson, \prB{42} {90} {7931}. }

\refn\CanrightThree{
G.S. Canright and A.G. Rojo, \ijmpB {5} {91} {1553};
\semi T. McMullen, P. Jena, and S.N. Khanna, \ijmpB {5} {91} {1579}.}

\refn\Halperin{
B. I. Halperin, Harvard report, {\nineit The hunt for anyon superconductivity},
1991;
\semi I.E. Dzyaloshinskii, \plA {155} {91} {62};
\semi G.S. Canright and A.G. Rojo,   \prl {68} {92} {1601}.}

\refn\ExpPolarization{
K. B. Lyons et al. \prl{64} {90} {2949};
\semi S. Spielman et al. \prl{65} {90} {123}; \prB {45} {92} {3149};
  \prl {68} {92} {3472};
\semi H.J. Weber et al. {\nineit Solid State Com.} {\ninebf 76}, 511 (1990);
\semi K.B. Lyons and J.F. Dillon, Jr.,  \ijmpB {5} {91} {1523};
\semi T.W. Lawrence, A. Sz\"oke, and R.B. Laughlin, \prl {69} {92} {1439}.}

\refn\Hetrick{
J. Hetrick and Y. Hosotani, \prB {45} {92} {2981}.}

\refn\ExpHall{
M.A.M. Gijs et al. \prB{42} {90} {10789}.}

\refn\Goldman{
A. Goldman, T.-F. Wang, and A. Mack,  (private communication).}

\refn\ExpMuon{
R.E. Kiefl et al. \prl{64} {90} {2082}; {\nineit Hyperfine Interactions}
 {\ninebf 63}, 139 (1990);
\semi N. Nishida and H. Miyatake, {\nineit Hyperfine Interactions}
 {\ninebf 63}, 183 (1990).}


\global\newcount\secno \global\secno=0
\global\newcount\appno \global\appno=0
\global\newcount\meqno \global\meqno=1
\global\newcount\figno \global\figno=1
\newwrite\eqmac
\def\eqn#1{
        \ifnum\secno>0
            \eqno(\the\secno.\the\meqno)\xdef#1{\the\secno.\the\meqno}%
            \immediate\write\eqmac{\def\noexpand#1{\the\secno.\the\meqno}}%
        \else\ifnum\appno>0
                  \eqno(\romappno.\the\meqno)\xdef#1{\romappno.\the\meqno}%

\immediate\write\eqmac{\def\noexpand#1{\romappno.\the\meqno}}%
               \else
                  \eqno(\the\meqno)\xdef#1{\the\meqno}%
                    \fi
        \fi
        \global\advance\meqno by1
          }
\newwrite\figmac
\def\fig#1{\ifnum\secno>0
            \the\figno\xdef#1{\the\figno}%
        \fi
        \global\advance\figno by1
          }



\baselineskip=10pt
\line{\eightrm Preprint from the University of Minnesota\hfil}
\line{\eightrm UMN-TH-1106/92 \hfil}
\line{\eightrm September 15, 1992\hfil}
\vglue 5pc

\baselineskip=13pt
\centerline{\tenbf NEUTRAL AND CHARGED ANYON FLUIDS}

\vglue 24pt
\centerline{\eightrm YUTAKA HOSOTANI}

\baselineskip=12pt
\centerline{\eightit School of Physics and Astronomy, University of Minnesota}
\baselineskip=10pt
\centerline{\eightit Minneapolis, Minnesota 55455, U.S.A.}

\vglue 20pt
\centerline{\eightrm Type-set by plain \TeX}

\parindent=15pt
\vglue 16pt
\eightrm\baselineskip=10pt
{\narrower \noindent
Properties of neutral and charged anyon fluids are examined, with
the main focus  on the question whether or not a charged anyon fluid
exhibits a superconductivity at zero and finite temperature.
Quantum mechanics of anyon fluids is precisely described by Chern-Simons gauge
theory. The random phase approximation (RPA), the linearized
self-consistent field method (SCF), and the hydrodynamic approach employed
in the  early analysis of anyon fluids are all equivalent.  Relations  and
differences between neutral and charged anyon fluids are discussed.
It is necessary to go beyond RPA and the linearized SCF, and
possively beyond the Hartree-Fock approximation, to correctly describe
various  phenomena such as the flux quantization, vortex formation, and phase
transition.
\par}

\baselineskip=10pt\noindent
{
\parindent=13pt \eightrm

\halign{ \hskip 1.cm # \hskip .1cm & # \cr

\vtop{  \hsize=5cm

\item{1.}  Introduction

\item{2.}  Anyons

\item{3.}  Aharonov-Bohm effect

\item{4.}  Chern-Simons gauge theory

\item{5.}  Charged anyon fluid

\item{6.}  Mean field ground state

\item{7.}  Hartree-Fock ground state

\item{8.}  RPA and SCF

\item{9.}  Path integral representation

\item{10.}  RPA = linearized SCF

\item{11.}  Response functions

\item{12.}  Evaluation of the kernel

}

& \vtop{ \hsize=5.5cm

\item{13.}  Phonons and plasmons

\item{14.}  Hydrodynamic description

\item{15.}  Effective theory

\item{16.}  Meissner effect at {\eightit T}=0

\item{17.}  {\eightit T}{\smallneq}0 -- homogeneous fields

\item{18.}  de Haas -- van Alphen effect in SCF

\item{19.}  {\eightit T}{\smallneq}0 -- inhomogeneous fields

\item{20.}  Thermodynamic potential in inhomogeneous fields

\item{21.}  Partial Meissner effect in SCF

\item{22.}  {\eightit T}$_c$

\item{23.}  Other important issues

   } \cr}
}

\normal
\parindent=15pt


\tenrm

\vskip 13pt

\line{\tenbf 1. Introduction \hfil}
\vglue 5pt
Is a charged anyon fluid a superconductor?  Is there any
difference in its behaviour from traditional superconductors described by
Ginzburg-Landau-BCS theory?  Can newly discovered high $T_c$ superconductors
be anyon superconductors?   These are the main questions addressed in this
article.\myref{\ReviewOne -\ReviewLast}  Since Laughlin suggested that a high
$T_c$ superconductor may be viewed as an anyon fluid,\myref{\LaughlinOne} an
extensive study has been conducted by many authors by various methods.

After three years of investigation, we can now have a coherent assessment of
the
current understanding. A fair statement, at the moment, is that a charged anyon
fluid seems to  behave like  a superconductor, but its properties
have not been understood very well.  We are not even sure if we have found
 a good approximate ground state of
a charged anyon fluid, which  should serve as an alternative to the BCS ground
state of ordinary superconductors.  The approximate ground state employed in
the
early investigation, characterized by a state of completely filled
Landau levels, might miss many of important phenomena such as the flux
quantization, Josephson effect, and vortex formation.

Anyons\myref{\Anyon} exist in Nature.  Excited states (quasi particles) in
fractional quantum Hall effect (FQHE) obey fractional
statistics.\myref{\LaughlinFQHE - \Tsui}  Laughlin's
theory of  FQHE implies that for a filling factor $\nu=p/q$, where $p$ and $q$
are coprime numbers, quasi particles have a statistics phase $\pm \pi/q$.  The
theory predicts a hierarchy structure in the Hall conductivity $\sigma_{xy}$
as  the filling factor or magnetic field varies, which has been confirmed
experimentally at multiple levels.\myref{\FQHEbook -\FQHEmoreThree}

Where else are anyons?  Three years ago Laughlin made a bold hypothesis that
anyons are in newly discovered high $T_c$ superconductors.\myref{\LaughlinOne}

There are three issues involved here. (1)  First one has to show that anyons
really emerge as excitations, starting from some spin, or electron, systems.
There are arguments that the ground state of the Hubbard or $t$-$J$ model in
two dimensions near the half-filling may be in the flux phase or more
specifically in chiral spin state, in which excitations are anyons.   So far
only consistency arguments have been provided.  (2) Secondly, assuming
that there exist anyon fluids with a finite density, one needs to know if
such  fluids exhibit a superfluidity or superconductivity.
One has to understand physics of anyon fluids.
  (3)  Thirdly, one has to deduce physical properties of anyon fluids
which can be subject to experimental tests.  Specifically, one has to
know whether or not (a part of) high $T_c$ superconductors are anyon
superconductors.

The first issue is most difficult, and is outside the topics covered in this
article.  The third one is controversial, in both experimental and
theoretical viewpoints.    This article exclusively deals with the second
issue.

First Fetter, Hanna, and Laughlin showed,\myref{\FHL,\HLF} computating response
functions  in the random phase approximation  (RPA), that a neutral anyon
fluid has a massless excitation, and that with electromagnetic interactions
coupled the system becomes superconducting.  Several months later, various
authors  confirmed the superconductivity at $T=0$ by many different methods.
Chen, Wilczek, Witten, and Halperin generalized the RPA analysis, deriving
many physical consequences.\myref{\CWWH}  Wen
and Zee, taking the hydrodynamic approach, gave a physical meaning of a
massless
excitation as a breathing mode of density wave.\myref{\WenZee}  Canright,
Girvin,
and Brass's numerical analysis yielded a desired flux dependence in
superconductors.\myref{\CanGirv} The self-consistent field  analysis of a
charged
anyon fluid  by Hosotani and Chakravarty led to a new equation replacing the
London equation in the BCS theory.\myref{\HosoChak}  Fradkin gave an analysis
of
the model on a lattice.\myref{\Fradkin}

At first all these analyses looked quite different from each other, though
reaching to the same conclusion, namely a superconductivity at $T=0$.
In the mean time  the investigation has been extended in various directions,
which includes analyses of finite temperature, vortices, conductivity, $P$ and
$T$ violation,  higher order radiative corrections, and so
force.\myref{\FisherLee-\Sumantra}

It has been recently shown\myref{\HHL} that the RPA, self-consistent field
method
(SCF), and hydrodynamic approach yield the same results for many physical
quantities such as the excitation spectrum and response function.   In this
paper
we shall strengthen the statement.  We show that RPA is exactly the same as
the linearlized version of SCF,
and that the hydrodynamic approach describes the same physics in terms
of the density and velocity field as SCF does in terms of the gauge fields.
Hence all these three are equivalent.

Quantum mechanics of anyon systems is precisely described in
terms of Chern-Simons gauge theory.  The essence of anyon dynamics is
contained in the Aharanov-Bohm effect with respect to Chern-Simons gauge
fields.  We shall establish the equivalence between
the two descriptions in the following three sections.  It will be seen  that
the
language of Chern-Simons gauge theory facilitates and simplifies all the
discussions of anyon fluids.

\sectionskip

\secno=2  \meqno=1

\line{\tenbf 2.  Anyons \hfil}
\vglue 5pt
Under the interchange of two identical particles, the Schr\"odinger wave
function in quantum mechanics acquires a factor of either $+1$ or $-1$,
depending on whether the particles are bosons or fermions.  In two spacial
dimensions there can be other possibility.\myref{\Anyon}
The interchange of  two particles, say $a$ and $b$, defines two paths $C_1$
and $C_2$ along which the particles $a$ and $b$ are transported to the
original locations of $b$ and $a$, respectively.  $C_1$ and $C_2$ together
form an oriented closed loop.  Pick the paths such that none of the other
particles are inside the closed loop. Depending on whether the loop is
oriented in a clockwise or counter-clockwise direction,
the operation defines $P_{-}(a,b)$  or $P_{+}(a,b)$.  (Fig.\ 1)

In three dimensional space there can be no distinction between
$P_{+}(a,b)$ and $P_{-}(a,b)$ ($\equiv P(a,b)$), since $P_+$ can be
continuously deformed to $P_-$.   Hence
$P(a,b)^2=P_{+}(a,b) P_{-}(a,b)=1$, and  $P(a,b)= \pm 1$.
In two spatial dimensions, however, one can have
$$P_{\pm}(a,b) ~ \Phi(1, \cdots,q) = - e^{\pm i\theta_s} ~ \Phi(1, \cdots,q)
   \eqn\phase  $$
where $\Phi(1, \cdots,q)$ is the Schr\"odinger wave function for a $q$-particle
system.  The minus sign in (\phase),  retained for later convenience, may be
absorbed in the definition of the statistical phase $\theta_s$.
$\theta_s=0$ or $\pi$ ($mod ~2\pi$) corresponds to
fermions or bosons, respectively.  Otherwise the statistics of the particles is
in  between.  It is said that particles obey fractional statistics.  Such
particles are generically called anyons.  It is easy to see that (\phase)
satisfies, for instance,  an operational identity (Fig.\ 2)
$$P_{-}(a,b) P_{+}(a,c) P_{+}(b,c)=  P_{+}(a,c) ~~.
   \eqn\topological $$

If the wave function is to be well-defined in the limit  two of the
coordinates $\x_a$ and $\x_b$ coincide, the identity (\phase) leads to
$$\Phi(1, \cdots,q) \Big|_{\x_a=\x_b} =0 \quad {\rm for~}
   -e^{i\theta_s} \not= 1 ~.  \eqn\restrict $$
In other words, unless particles are bosons, the wave function must
vanish when two coordinates coincide.  Pauli's exclusion principle applies to
anyons.

\vglue 4.5cm
{\figure
$$\hbox{ Fig. 1} ~~
 \vtop{ \hsize=9cm  \noindent
  Interchange of two identical particles.  The closed contour formed by
{\eightit C}$_{\fiverm 1}$ and {\eightit C}$_{\fiverm 2}$
should not encircle any other particles.} $$
}

\vglue 4cm
{\figure
$$\hbox{\eightrm Fig. 2} ~~
\hbox{\eightrm The identity (2.2).} $$
}

A system of ``free'' anyons is described by the equation
$$\eqalign{
&i {\d\over \d t} \,\Phi = H \Phi \cr
&H = \sum_{a=1}^q  - {\hbar^2\over 2m} \nabla_a^2 \cr}  \eqn\SchEq  $$
where $\nabla_a = \d/\d \x_a$.  The equation (\SchEq) must be
solved with the boundary condition (\phase).   This system defines a
{\it neutral anyon fluid}.  We shall see below that except for the cases of
bosons
and fermions a neutral anyon fluid is not ``free''.  The energy of
a many-anyon system is not the sum of single particle energies.  An
interaction is hidden in the nontrivial boundary condition (\phase).

It is most instructive to go over to a new gauge.  We define
$$\eqalign{
&\fPhi = \sing \Phi \next \sing = e^{i\omega} \cr
&\omega(\x_1,\cdots,\x_q) = \sum_{a<b} {\theta_s\over \pi} ~
    \tan^{-1} {x_{a2} -x_{b2} \over x_{a1} - x_{b1}} ~~. \cr}
    \eqn\newWave  $$
In terms of the new wave function, the equation and boundary condition become
$$\eqalign{
&i {\d\over \d t} \, \fPhi = \sum_{a=1}^q  - {\hbar^2\over 2m}
 \Big( \nabla_a - i {\bf B}^{(a)}(\{ \x_b \}) \Big)^2  ~ \fPhi \cr
&\hskip 1.cm B^{(a)j}(\{\x_b\}) = \nabla^a_j \omega
= - {\theta_s\over \pi} \sum_{b\not= a}
       { \ep^{jk} (x_a - x_b)_k \over (\x_a -\x_b)^2} \cr
&\hskip 3.9cm =    - {\theta_s\over \pi}
   \sum_{b\not= a} \ep^{jk} \d^a_k ~\ln |\x_a -\x_b| \cr
&P_\pm (a,b) \fPhi(1, \cdots, q) = - \fPhi(1,\cdots,q) \cr}
    \eqn\EqNeutral  $$
In the new representation the particles behave as fermions, but with a
specific long range interaction described by ${\bf B}$.  $\fPhi$ is
the Schr\"odinger wave function in the fermion representation.

The ${\bf B}^{(a)}(\{ \x_b\})$ term in (\EqNeutral)
gives rise to two- and three-body interactions.  It involves a velocity
dependent potential. The interaction, which account for the
anyon nature of the particles, can be interpreted as an Aharonov-Bohm
effect or as a Chern-Simons gauge interaction, as we shall show in the
subsequent sections.   The gauge transformation
potential $\sing$ which connects the original $\Phi$ and new $\fPhi$ is
singular at $\x_a=\x_b$.  One advantage of working in the new gauge is
that  the wave function $\fPhi$ is a regular, single-valued function of the
coordinates $\{ \x_a \}$.

\sectionskip

\secno=3  \meqno=1

\line{\tenbf 3.  Aharonov-Bohm Effect \hfil}
\vglue 5pt
Suppose that there is a solenoid parallel to the $x_3$-axis at
$x_1=x_2=0$ with a total magnetic flux $\mu = \int dx_1dx_2 ~B_3$.
Outside the solenoid there results a vector potential
$$A_\phi = {\mu \over 2\pi r} \next A_r=0 \next A^3=0
    \eqn\solenoidAphi  $$
in cylindrical coordinates ($x_1$=$r\cos \phi$, $x_2$=$r\sin\phi$) or
$$A^k = - \ep^{kl} {\mu\over 2\pi} {x_l\over r^2}
  = {\mu \over 2\pi} {\d\over \d x_k} ~\phi \quad (k=1,2) .
       \eqn\solenoidAk  $$
As far as ${\bf E}={\bf B}=0$, the motion of electrons outside
the solenoid is not affected by the presence of the flux $\mu\not= 0$
in classical theory.

In quantum theory a non-vanishing gauge potential $A_\mu$, which
locally generates  vanishing field strengths, but is not
globally a pure gauge, affects the motion of electrons.  Let us focus on
the two-dimensional motion of electrons in the $x_1$-$x_2$ plane,
supposing that a momentum in the $x_3$-direction is zero.
The Schr\"odinger equation outside the solenoid ($r\ge R$) is
$$-{\hbar^2\over 2m} \bigg\{ {1\over r}{\d\over \d r} r {\d\over \d r}
 + {1\over r^2} \Big( {\d\over \d\phi}- i{e\mu\over 2\pi \hbar c} \Big)^2
 \bigg\} ~\Phi_0 = E ~\Phi_0 ~~.  \eqn\solenoidEq $$

The general solution is
$$\Phi_0= \sum_{l=-\infty}^\infty e^{il\phi} ~\big\{
  a_l J_{l-\alpha}(kr) + b_l J_{\alpha-l}(kr) \big\}  \eqn\solenoidSol$$
where $\alpha= e\mu/2\pi \hbar c$,  $k^2 = 2mE/\hbar^2$, and $J_\nu$ is
a Bessel function of fractional order $\nu$.  The wave function
is supposed to vanish at the boundary of the solenoid.
The wave function and energy eigenvalue depend on $\alpha$,
or on the  magnetic flux $\mu$ in a periodic fashion.  It is called the
Aharonov-Bohm effect.\myref{\Aharonov}

A basic assumption in deriving (\solenoidSol) is that the wave function
$\Phi_0$ is a singlevalued function of $r$ and $\phi$.  Let us define a
new wave function by $\Phi_0(r,\phi)= e^{i\alpha\phi} \Phi_1(r,\phi)$.
Then
$$\eqalign{
-&{\hbar^2\over 2m} \bigg\{ {1\over r}{\d\over \d r} r {\d\over \d r}
 + {1\over r^2} {\d^2 \over \d\phi^2 } \bigg\} ~\Phi_1 = E \, \Phi_1  \cr
&\Phi_1(r,\phi+2\pi)  = e^{-i(e\mu/\hbar c)} \Phi_1(r,\phi)  ~~. \cr}
   \eqn\newSolenoid $$
In this new gauge the electron wave function satisfies a free
equation, but obey a non-trivial boundary condition upon making a trip around
the solenoid.  In other words, the Aharonov-Bohm effect is traded for
the multi-valuedness of the wave function, the property anyons share.

Indeed, the analogy is exact.  Now we imagine that  particles (in two
spatial dimensions) have both charge $e$ and flux $\mu$,  and that there
exsit only charge-flux interactions, but no charge-charge or flux-flux
interactions.  Each particle, say $a$, creates a vector potential
$$A^{k}(\x) = - \ep^{kj} {\mu\over 2\pi}
     {x_j- x_{aj}\over|\x -\x_a|^2}   $$
which is felt by other particles by the minimal coupling.  Hence the
Schr\"odinger equation is given by
$$\eqalign{
&\sum_{a=1}^q  - {\hbar^2\over 2m} \Big( \nabla_a
   - i{e \over \hbar c}{\bf A}_a  \Big)^2  ~ \Phi  = E ~\Phi  \cr
&{e \over \hbar c} A_a^j =- {e\mu\over 2\pi \hbar c}
  \sum_{b\not= a} \ep^{jk} {(x_a-x_b)_k \over |\x_a -\x_b|^2}   ~~. \cr}
     \eqn\ChargeFluxEq  $$
This eqation is exactly the same as (\EqNeutral) upon identifying
$$\theta_s = {e\mu\over 2\hbar c} ~~~. \eqn\thetaChargeFlux $$

It is recognized that the anyon interaction is nothing but an Aharonov-Bohm
effect.\myref{\Arovas-\Goldhaber}  Since only the product of $e$ and $\mu$ is
relevant, one can phrase
$$ {\rm anyon} = \cases{ {\rm charge} &:~~ ~~1\cr \cr
                           {\rm flux} &:~~ $2\hbar c \theta_s$  \cr}~~~.
        \eqn\AnyonChargeFlux  $$
It's not exactly a Maxwell interaction, however, since there is no
charge-charge
interaction.  We shall see in the next section  that the Chern-Simons
gauge theory precisely describes the anyon interaction.

We would like to note that Aharonov-Bohm effects have wider applications.
Let us consider the motion of a particle in an arbitrary multiply-connected
manifold, ${\cal M}$.  We can imagine that the space itself has nontrivial
topology like $T^2 \times R^1$ etc., or that the three-dimensional Eucledian
space is obstructed by the presence of closed strings.  Further we suppose that
field strengths, or more specifically magnetic fields, $F_{jk}=-\d_j A^k +\d_k
A^j$,  identically vanish in ${\cal M}$.  Schr\"odinger equation is given by
$$ - {\hbar^2\over 2m} \Big( \nabla   - i{e \over \hbar c}{\bf A} \Big)^2  ~
      \Phi_0  = E ~\Phi_0  \qquad {\rm in} ~{\cal M}   \eqn\generalAB $$
where $\d_j A^k - \d_k A^j =0$.  In general $A_j$ is not gauge equivalent to
$A_j=0$ in a multiply-connected space.

We define a new wave function by
$$\Phi_1(\x;C) = \exp \bigg\{ - {i e\over \hbar c}
  \int_{C(\x_0,\x)} d\y \cdot {\bf A}(\y) \bigg\} ~ \Phi_0(\x)
      \eqn\ABnewPhi    $$
where $C(\x_0,\x)$ starts at $\x_0$ and ends at $\x$.
For two paths, $C_1(\x_0,\x)$ and $C_2(\x_0,\x)$,
 continuously deformable to each other,  $\Phi_1$ assumes the
same value,  since $F_{jk}=0$;
$$\Phi_1(\x;C_1) = \Phi_1(\x;C_2) \qquad {\rm if}
{}~~ C_1(\x_0,\x) \sim C_2(\x_0,\x) ~~.
      \eqn\equivalentPhi $$
Hence derivatives of $\Phi_1$ with respect to $\x$ is well defined.

$\Phi_1$ satisfies a free equation
$$ - {\hbar^2\over 2m} \nabla^2  ~ \Phi_1  = E ~\Phi_1  ~,  \eqn\newABone $$
but is not single-valued.    Let $\Gamma (\x)$ denote a transport
of \x along a closed path $\Gamma$.  Then
$$\eqalign{
&\Phi_1[\,\Gamma (\x)\, ] = W(\,\Gamma\,) ~\Phi_1(\x) \cr
&W(\,\Gamma\,) = \exp \bigg\{ -{ie\over \hbar c}
  \int_\Gamma d\y \cdot {\bf A}(\y) \bigg\}   \cr}   \eqn\multiPhi  $$
If $\Gamma$ can be continuosly shrunk to a point, then $W=1$.   $W(\,\Gamma\,)$
depends on only homology of $\Gamma$ with respect to ${\cal M}$.
It is a non-integrable phase factor, often called a Wilson
line integral in the particle physics literature.\myref{\ABHosotani}
We have seen that the general Aharonov-Bohm problem is traded for a free system
with  nontrivial boundary conditions.

\sectionskip

\secno=4 \meqno=1

\line{\tenbf 4.  Chern-Simons Gauge Theory \hfil}
\vglue 5pt
Consider a quantum field theory described by a Lagrangian
$$\eqalign{
\L_0 =& - {N\over 4\pi} \eps a_\mu \d_\nu a_\rho + i \psi^\dagger  D_0 \psi
    - {1\over 2m} |D_k \psi|^2 ~~~, \cr
&D_0 = \d_0 +i a_0 \next D_k = \d_k - ia^k ~~~. \cr}
      \eqn\LagZero $$
$a_\mu$ ($a_0=a^0, a_k=-a^k$)
is a gauge field whose motion is described by the Chern-Simons
term ($\propto \eps a_\mu \d_\nu a_\rho$). (Refs. \CStheory -- \JackiwPi)
$\psi$ is a non-relativistic
matter field, obeying either bose or fermi statistics.  $N$ determines the
magnitude of the gauge coupling.  A large $|N|$ corresponds to a weak
coupling, as can be seen by rescaling  $a_\mu$.   We show that the system
defined by (\LagZero) is equivalent to a neutral anyon fluid described by
(\EqNeutral), and therefore by (\SchEq).

Euler equations derived from (\LagZero) are
$$\eqalign{
&- {N\over 4\pi} ~ \eps f_{\nu\rho} = j^\mu \cr
&i \d_0 \psi = \Big\{ -{1\over 2m} D_k^2 + a_0 \Big\} ~\psi \cr}
   \eqn\EulerEq $$
where $f_{\mu\nu}=\d_\mu a_\nu - \d_\nu a_\mu$ and
$$\eqalign{
j^0 &= \psi^\dagger \psi ~~~, \cr
j^k &= - {i\over 2m} \Big( \psi^\dagger D_k \psi - (D_k \psi)^\dagger \psi
         \Big) ~~~.  \cr}
   \eqn\current $$
The strengths $f_{\mu\nu}$ of the Chern-Simons gauge fields are determined
by the current.  There is no physical degree of freedom for the gauge field.
Hence the Chern-Simons field can be eliminated in favor of the matter field.

It is convenient to take the radiation gauge div{\hskip .1cm}${\bf a}=0$.  The
Chern-Simons field equation in (\EulerEq) becomes
$$\eqalign{
{N\over 2\pi} ~ \Delta a^k &= - \ep^{kl} \d_l j^0 \cr
{N\over 2\pi} ~ \Delta a_0 &= \d_1 j^2 - \d_2 j^1 ~~~. \cr}
   \eqn\CSone $$
With an appropriate boundary condition, the equations can be solved to
express $a_\mu$ in terms of $j^\mu$.

In the case of a plane ($R^2$)
the boundary condition at infinity is subtle.  A safe and rigorous
derivation is obtained on a torus ($T^2$). (Refs. \CStorus -- \Ho)  We quote
the
result on a torus, taking the infinite volume limit.  The argument presented
here is very close to those in ref.~\JackiwPi  ~and in ref.~\Ho.
$$\eqalign{
a_0(x) &= \int d\y ~ h_k(\x-\y)  j^k(y) ~~~, \cr
a^j(x) &= \abar^j (x) + \ahat^j (x) ~~~, \cr
&\abar^j(x) = - {\pi n_e \over N} ~ \ep^{jk} x_k \next
    {N\over 2\pi} (\d_1 \abar^2 - \d_2 \abar^1) = n_e  \cr
&\ahat^j(x) = \int d\y ~ h_j(\x-\y) \big( j^0(y) - n_e \big)  ~~~. \cr}
   \eqn\expressCS $$
Here $n_e$ is the average matter density, which generates $\abar^j(x)$.
$h_j(\x)$ is related to the two dimensional Green's function
$G(\x)= (2\pi)^{-1} \ln r$ by
$$h_j(\x) = - {2\pi\over N} ~\ep^{jk} \d_k G(\x) =
           - {\ep^{jk} x_k\over N r^2} ~~.  \eqn\hj $$

If a finite number of particles on an infintely large plane ($R^2$) are
considered, one can set $n_e=0$ in the above formulas.  In applying
to the superconductivity it is convenient to deal with a system with a finite
density on $R^2$.

$j^k(y)$ in the expression for $a^0(x)$ in (\expressCS) contains $a^k(y)$
which is expressed in terms of $\psi$.  Hence the Chern-Simons
gauge fields are completely expressed in terms of the matter field.
The substitution of (\expressCS) into (\LagZero) gives a Lagrangian which
involves only $\psi$ and $\psi^\dagger$:
$$\eqalign{
\L_1 &= i \psi^\dagger \dot \psi - \H_1  \cr
\H_1 &= {1\over 2m} (D_k \psi)^\dagger (D_k \psi)  \cr}
    \eqn\LagOne $$
where $a^k(x)$ in $D_k = \d_k - ia^k$ is given by (\expressCS).  As it stands,
the resultant Hamiltonian $H_1 = \int d\x ~\H_1$ involves four- and six-fermi
interactions.  There arises an ambiguity in  ordering operators.  It can be
shown that the system defined by the Hamiltonian with the ordering
adopted in  (\LagOne) is equivalent to the system described by the
Schr\"odinger
equation (\EqNeutral).

The Hamiltonian is not completely normal-ordered.  It is given by
$$\eqalign{
H_1 &= {1\over 2m}\int d\x  ~
    \big( \d_k \psi^\dagger + i \psi^\dagger a^k \big)
    \big( \d_k \psi - i a^k \psi \big)  \cr
&= H_1^{(1)} + H_1^{(2)} + H_1^{(3)}  \cr}     \eqn\HamOne $$
where
$$\eqalign{
H_1^{(1)} &= {1\over 2m}\int d\x  ~ (\Dbar_k \psi)^\dagger (\Dbar_k \psi) \cr
H_1^{(2)} &= {i\over 2m}\int d\x d\y ~ h_k(\x-\y)  \Big\{
\psi^\dagger(x) ( \psi^\dagger \psi(y) - n_e ) \Dbar_k \psi(x) \cr
  &\hskip 3.7cm - (\Dbar_k \psi)^\dagger (x)
  (\psi^\dagger \psi(y) - n_e ) \psi(x) \Big\}  \cr
H_1^{(3)} &=  {1\over 2m} \int d\x d\y d\z ~h_k(\x-\y) h_k(\x-\z) \cr
&\hskip 2cm   \times\psi^\dagger(x) (\psi^\dagger\psi(y)-n_e)
   (\psi^\dagger\psi(z) - n_e) \psi(x)    \cr}
    \eqn\HamOneOne  $$
and $\Dbar_k \psi(x) = \big(\d_k - i \abar^k(x) \big) \psi(x)$.
Upon making use of $\int d\y ~h_k(\x-\y)=0$, one can write $H_1^{(3)}$ as
$$\eqalign{
H_1^{(3)} &=  {1\over 2m} \int d\x d\y d\z ~h_k(\x-\y) h_k(\x-\z)
  \psi^\dagger(x) \psi^\dagger(y) \psi^\dagger(z) \psi(z) \psi(y) \psi(x) \cr
& \hskip 1.cm
+ {1\over 2m} \int d\x d\y ~[h_k(\x-\y)]^2
\psi^\dagger(x)  \psi^\dagger(y) \psi(y) \psi(x) ~~~. \cr}  \eqn\normalHamOne
$$

The equation derived from $H_1$ is
$$\eqalign{
i & \dot \psi(x) = [\psi(x) , H_1] = \K_0 \psi(x) \cr
&\K_0 =  - {1\over 2m} D_k^2 + a_0(x) + g(x)   \cr}
  \eqn\PsiEq $$
where $a^k(x)$ in $D_k$ and $a_0(x)$ are given by (\expressCS) and
$$
g(x) = {1\over 2m} \int d\y ~  [h_k(\x-\y)]^2 ~\psi^\dagger\psi(y)
= {1\over 2mN^2} \int d\y~  {1\over (\x-\y)^2} ~\psi^\dagger\psi(y) ~~.
   \eqn\gx  $$
Eq. (\PsiEq) differs from the classical Euler equation (\EulerEq) by the
$g(x)$ term.  The additional term is important to establish the
equivalence between the anyon quantum mechanics and Chern-Simons gauge
theory.

Currents are given by
$$\eqalign{
J^0(x) &= \psi^\dagger \psi(x) \cr
J^k(x) &= -i \bigg\{
\psi^\dagger(x) ~{\delta H_1 \over \delta \nabla_k \psi^\dagger(x) }
-{\delta H_1 \over \delta \nabla_k \psi(x) } ~ \psi(x) \bigg\} \cr
&= -{i\over 2m} \big\{ \psi^\dagger D_k \psi - (D_k\psi)^\dagger \psi \big\}
\cr
&=-{i\over 2m}\big\{ \psi^\dagger(x) \Dbar_k\psi(x)
    - (\Dbar_k\psi)^\dagger(x) \psi(x) \big\} \cr
&\hskip 2.5cm -{1\over m} \int d\y \, h_k(\x-\y) \, \psi^\dagger(x)
\psi^\dagger(y) \psi(y) \psi(x) ~~~. \cr}   \eqn\bigCurrent  $$
They are conserved:  $\d_0 J^0 + \nabla_k J^k =0$.

The Schr\"odinger wave function $\fPhi$ in quantum mechnics for a $q$-particle
system is related to the field operator $\psi$ in (\LagOne) or
(\HamOne) by
$$\fPhi(1,\cdots,q) = \la 0 | \psi(1) \cdots \psi(q) | \Psi_q \ra  ~~.
   \eqn\waveF $$
Here $|0\ra$ and $|\Psi_q\ra$ are vacuum and $q$-particle states, respectively,
and $\psi(a)=\psi(x_a)$ where $x_a=(t,\x_a)$.  For a system with a finite
number
of particles on a plane ($R^2$), one can put $n_e=0$ and $\abar^k(x)=0$ in the
above formulas. To obtain the Schr\"odinger equation, we
differentiate $\fPhi$ with respect to $t$ and make use of (\PsiEq):
$$\eqalign{
i {\d\over \d t}\, \fPhi(1, \cdots,q) &= \sum_{a=1}^q
  \la 0 | \psi(1) \cdots i \dot \psi(a) \cdots \psi(q) | \Psi_q \ra \cr
&=\sum_{a=1}^q
  \la 0 | \psi(1) \cdots \K_0 \psi(a) \cdots \psi(q) | \Psi_q \ra ~~. \cr}
    \eqn\SEqOne $$

The following definitions
and identities facilitate futher manipulations.  First we define
$$
\vcenter{
\hbox{ \vbox{ \halign{ $\big #$\hfil &$\big #$\hfil  \cr
\K_1(a,b) &= {1\over m}  h_k(\x_a-\x_b) \big\{ \ahat^k(x_a) -
    \ahat^k(x_b) \big\}    + 2g(x_a,x_b) ~~,  \cr
\noalign{\kern 4pt}
V_2(a,b) &= {i\over m} ( \nabla^a_k - \nabla^b_k) h_k(\x_a-\x_b)
   + {1\over m} h_k(\x_a-\x_b)^2  ~~, \cr
\noalign{\kern 4pt}
V_3(a,b,c) &= {1\over m} \big\{ h_k(\x_a-\x_b) h_k(\x_a-\x_c)  \cr
\noalign{\kern 4pt}
&\hskip .7cm   +h_k(\x_b-\x_a)h_k(\x_b-\x_c) +h_k(\x_c-\x_a) h_k(\x_c-\x_b)
   \big\}  ~~. \cr}    } } }
   \eqn\stepOne $$
Here
$$g(x,y) = {1\over 2m} \int d\z~ h_k(\x-\z) h_k(\y-\z) ~ \psi^\dagger \psi(z)
  ~~~. \eqn\gxy $$
$\K_1(a,b)$ is an operator, whereas $V_2(a,b)$ and $V_3(a,b,c)$ are c-number
functions.

These operators with $\K_0(a)=\K_0(x_a)$ satisfy
$$
\vcenter{
\hbox{ \vbox{ \halign{ $\big #$\hfil &$\big #$\hfil  \cr
\psi(b) \K_0(a) &= \big\{ \K_0(a) + \K_1(a,b) + V_2(a,b) \big\} \psi(b) \cr
\noalign{\kern 4pt}
\psi(c) \K_1(a,b) &= \big\{ \K_1(a,b) + V_3(a,b,c) \big\} \psi(c)  \cr}
              } } }   \eqn\stepTwo $$
and
$$\eqalign{
&\la 0 | \K_0(a) \cdots = - {1\over 2m} ( \nabla^a_k )^2 ~\la 0| \cdots \cr
&\la 0 | \K_1(a,b) \cdots = 0 ~~~.  \cr}
     \eqn\stepThree  $$

Applications of (\stepTwo) and (\stepThree) lead to
$$\eqalign{
\la 0 | &\psi(1) \cdots \K_0 \psi(a) \cdots \psi(q) | \Psi_q \ra \cr
&=\la 0| \Big\{ \K_0(a) + \sum_{b=1}^{a-1} \big( \K_1(a,b)+V_2(a,b) \big)
 +\sum_{b=2}^{a-1} \sum_{c=1}^{b-1}V_3(a,b,c) \Big\} \cr
&\hskip 6.8cm \times \psi(1) \cdots \psi(q) | \Psi_q \ra \cr
&= \bigg\{ - {1\over 2m} (\nabla^a_k)^2 + \sum_{b=1}^{a-1}  V_2(a,b)
  + \sum_{b=2}^{a-1} \sum_{c=1}^{b-1}V_3(a,b,c)  \bigg\} ~\fPhi ~~. \cr}
   \eqn\stepFour $$
Therefore the Schr\"odinger equation for $\fPhi$ is given by
$$ i{\d\over \d t} ~ \fPhi(1, \cdots, q) =
  H^{(q)} ~ \fPhi(1, \cdots, q)   \eqn\SEqTwo $$
where
$$\eqalign{
H^{(q)} &= - {1\over 2m} \sum_{a=1}^q (\nabla^a_k)^2
+ \sum_{a=2}^q \sum_{b=1}^{a-1} V_2(a,b)
  + \sum_{a=3}^q \sum_{b=2}^{a-1} \sum_{c=1}^{b-1} V_3(a,b,c) \cr
&= - {1\over 2m} \sum_{a=1}^q \Big\{ \nabla^a_k -
  i \sum_{b\not= a} h_k(\x_a-\x_b)  \Big\}^2 ~. \cr}  \eqn\fHam $$
This is exactly Eq. (\EqNeutral), provided that
$$\theta_s= {\pi \over N}   \eqn\thetaN $$
and that $\psi(x)$ satisfies anti-commutation relations.  This establishes
the equivalence of the anyon quantum mechanics and Chern-Simons gauge
theory.   The fermion representation is convenient to incorpolate the
Pauli principle (\restrict) for anyons.

\sectionskip


\secno=5  \meqno=1

\line{\tenbf 5.  Charged anyon fluid \hfil}
\vglue 5pt

The anyon fluid described in the previous section is neutral.  Anyons may
be charged, interacting with each other electromagnetically.
In the application to superconductivity, one needs to consider
a charged anyon fluid.

We have in mind material which has a layered structure as
in newly discovered high $T_c$ superconductors.  The motion of  electrons are
mostly confined in two-dimensional layers.  The probability of the hopping of
electrons from one layer to adjacent layers is very
small.   In many high $T_c$ superconductors the resistivity of electrons in
the direction perpendicular to $CuO$ planes above $T_c$ is $10^2$ to $10^5$
times bigger than the in-plane resistivity.
To the first approximation we may neglect the hopping
interaction.\myref{\Burns}

We shall adopt the ``holon'' picture originally advocated by
Anderson.\myref{\holon-\Wiegmann}
In this picuture collective modes created by electron holes are spinless
and charged.  They are called holons, and are supposed to obey half-fermion
statistics ($\theta_{\rm statistics} = \pm {1\over 2} \pi$).  The matter
field denoted by $\psi(x)$ corresponds to holon excitations.  In our
language $\psi(x)$ satisfies anti-commutation relations, interacting through
Chern-Simons gague fields with the coefficient $N=\pm 2$ and through
Maxwell fields with charge $e$.

The electromagnetic interaction is not two-dimensional, however.  Certainly
there is a Coulomb interactions among electrons in two different layers.
Electromagnetic waves can propagate in three dimensional space.

In this article we consider two extreme limits.  In one limit one can imagine
an ultra-thin film which has a couple of, or just one, superconducting layers.
Further we idealize the situation such that the only interaction
among holons, other than the Chern-Simons or fractional statistics
interaction, is the Coulomb interaction with a potential $1/r$.
  We call it {\it the ultra-thin film approximation}.

In the other limit we suppose material of an infinitely many layers
(in the $x_1$-$x_2$ plane)  which are evenly separated with a distance $d$.
Further we suppose that (1) electromagnetic fields $E_3$, $B_1$, and
$B_2$ identically vanish, and (2) all fields $E_1$, $E_2$, and $B_3$
are uniform in the $x_3$ direction.  Consequently all physical quantities
such as the expectation values of currents $\la J^\mu \ra$ are independent
of $x_3$.  One can mimic it by considering a system in the
idealized two-dimensional space, suppressing the third coordinate $x_3$.
It is called {\it the two-dimensional approximation}.  Fluctuations
of the $E_1$, $E_2$, and $B_3$ fields are retained, in addition to the
Coulomb interaction.

We note that the two-dimensional approximation incorporates
three-dimensional interactions in a specific way.  Both the three-dimensional
Coulomb interaction among electrons in  distinct layers and the
electron hopping between adjacent layers affect the three-dimensional motion
of electrons or holons.  When the electron or holon field is expanded
in Fourier series in the $x_3$ direction with a momentum $k_3$,
the two-dimensional approximation amounts to retaining only the $k_3=0$
component.

Real high $T_c$ superconductors lie somewhere between the
two approximation.  It is necessary and important to have more thorough
examinations of effects of the three-dimensional motion.  No such
analysis is available at the moment.

In the ultra-thin film approximation the Lagrangian is given by
$$\eqalign{
\L_\film &[\psi,\psi^\dagger,a_\mu ; \phi, A^\ext_\mu]  \cr
&= - {N\over 4\pi} \eps a_\mu \d_\nu a_\rho + i \psi^\dagger  D_0 \psi
    - {1\over 2m} |D_k \psi|^2 + \L_{\rm Coulomb}^{3D} ~~~, \cr
&\L_{\rm Coulomb}^{3D} = - {1\over 2} \phi \sqrt{ \nabla^2 } \phi
    + e \phi  (\psi^\dagger \psi - n_e) \cr
&D_0 = \d_0 +i (a_0 + e A^\ext_0)  \next
     D_k = \d_k - i(a^k + e A^k_\ext) ~~~. \cr}
      \eqn\filmLag $$
Here $\phi(\x)$ is an auxiliary field generating a $1/r$ Coulomb potential.
$A^\ext_\mu$ is an external electromagnetic field.  $\L_\film$ is bilinear
in $\psi$ and $\psi^\dagger$.
After eliminating $a_\mu$ and $\phi$, one obtains a Hamiltonian solely in terms
of $\psi$ and $\psi^\dagger$.
$$\eqalign{
H_\film &=  \int d\x~ \Big\{ {1\over 2m} (D_k \psi)^\dagger (D_k \psi)
  + e A^\ext_0 \psi^\dagger \psi \Big\} + H_{\rm Coulomb}^{3D} ~~, \cr
&H_{\rm Coulomb}^{3D} = {1\over 2} \int d\x d\y ~\psi^\dagger(x)
\psi^\dagger(y)
 ~{e^2\over |\x - \y| } ~\psi(y)\psi(x) ~~, \cr}   \eqn\filmHam $$
and $a_k$ in $D_k$ is given by (\expressCS).

Equations of motion derived from $\L_\film$ are
$$\eqalign{
&- {N\over 4\pi} ~ \eps f_{\nu\rho} = j^\mu \cr
&\sqrt{ \nabla^2} \phi = e( j^0 - n_e) \cr
&i \d_0 \psi = \Big\{ -{1\over 2m} D_k^2 + a_0+ e(\phi +A_0^\ext)  \Big\}
   ~\psi   \cr}     \eqn\filmEuler $$
where $j^\mu$ is given by (\current) with the covariant derivatives in
(\filmLag).

In the two-dimensional approximation the Lagrangian is given, instead, by
$$\eqalign{
\L_\twoD &[\psi,\psi^\dagger, a_\mu , A_\mu ]  \cr
&= - {1\over 4} F_{\mu\nu}^2 - {N\over 4\pi} \eps a_\mu \d_\nu a_\rho
 + en_e A_0  + i\psi^\dagger  D_0 \psi
    - {1\over 2m} |D_k \psi|^2  ~~~, \cr
&D_0 = \d_0 +i (a_0 + e A_0)  \next
     D_k = \d_k - i(a^k + e A^k) ~~~. \cr}
      \eqn\twoDLag $$
Here the electromagnetic field $A_\mu$ contains both external
and dynamical fields:  $A_\mu = A_\mu^\ext + A_\mu^{\rm dyn}$.
The corresponding Hamiltonian obtained by eliminating $a_\mu$ is
$$
H_\twoD =
\int d\x~ \Big\{ {1\over 2m}  (D_k \psi)^\dagger (D_k \psi)
  + e A_0 \psi^\dagger \psi  +
   {1\over 2} (E_k^2 + B^2) \Big\}
   \eqn\twoDHam $$
where $E_k=F_{0k}$ and $B=-F_{12}$.

Equations of motion derived from (\twoDLag) are
$$\eqalign{
&- {N\over 4\pi} ~ \eps f_{\nu\rho} = j^\mu \cr
&\d_\nu F^{\nu\mu} = ej^\mu - en_e \delta^{\mu 0} \cr
&i \d_0 \psi = \Big\{ -{1\over 2m} D_k^2 + (a_0 + eA_0) \Big\} ~\psi \cr}
   \eqn\twoDEuler $$
Again $j^\mu$ is given by (\current) with $D_k$ in (\twoDLag).

The Lagrangian forms $\L_0$ in (\LagZero) (for neutral fluids)
and $\L_\film \, / \,\L_\twoD$ (for charged fluids) are convenient to develop a
perturbation scheme.  They are bilinear in $\psi$ or $\psi^\dagger$, and
in the charged case the gauge invariance can be easily implemented in the
perturbation scheme.      On the other hand, the
Hamiltonian forms $H_1$ in (\HamOne) and $H_\film\, /\,H_\twoD$
has an advantage that they involve only physical fields, and
are particularly suited to
the evaluation of physical quantities beyond the perturbation scheme.  We shall
make use of both in subsequent sections.

As we shall see, there is a subtle but important difference between
neutral and charged anyon fluids.  It seems that charged fluids
(with a neutralizing background charge) are more stable than
neutral fluids.

\sectionskip

\secno=6 \meqno=1

\line{\tenbf 6.  Mean field ground state \hfil}
\vglue 5pt
We consider an anyon system with a finite density $n_e\not= 0$ on a
plane.  First we ask what would be the average Chern-Simons fields
in the ground state.  In the
equation (\EulerEq), (\filmEuler) or (\twoDEuler), we replace the
operator $j^\mu$ by its expectation value $\la j^\mu \ra$ in the ground state.
We expect that $\la j^0 \ra = n_e$ and $\la j^k \ra=0$ and all Maxwell fields
vanish.  In both neural and charged fluids we have
$$ b = {2\pi n_e\over N} \equiv b^{(0)}   \eqn\averageField  $$
and $f_{0k}=0$.  In other words, particles, or holons in high $T_c$
superconductors, move in a uniform Chern-Simons magnetic field on the average.

In the mean field approximation all gauge fields in
(\EulerEq), (\filmEuler), or (\twoDEuler) are replaced by the average fields.
The equation for the field operator $\psi$ is, in all cases,
$$\eqalign{
i \d_0 \psi &= -{1\over 2m} \Dsquare  ~\psi \cr
\Dbar_k &= \d_k - i \bar a^k    \cr}     $$
where the average vector potential is
$$
\hbox{$\bar a^k=$}
\left\{ \vcenter{
\hbox{ \vbox{ \halign{ $\big #$\hfil &$\big #$\hfil &\hskip 1cm #\hfil \cr
   -\ep^{kj} ~{\pi n_e x_j\over N}
       &= -\ep(N) ~\ep^{kj} ~ {x_j\over 2l^2}
           &in the symmetric gauge;\cr \cr
   -\delta^{k1} ~{2\pi n_e x_2\over N}
       &= -\ep(N) ~\delta^{k1} ~{x_2\over l^2}
              &in the Landau gauge.    \cr }      }}      }\right.
   \eqn\averagePotential $$
Here $\ep(N)=+1 ~(-1)$ for $N>0 ~(<0)$, and the magnetic length,
$l$, is defined by
$$l^2 = {|N|\over 2\pi n_e}    \eqn\magneticLength $$

The corresponding one-particle Schr\"odinger equation
$$ -{1\over 2m} \Dsquare  ~u_\alpha(\x) = \ep_\alpha ~u_\alpha(\x)
  \eqn\averageEq  $$
is easily solved in either gauge.  The energy spectrum is characterized
by Landau levels:
$$\eqalign{
\alpha &= (n, p)  \hskip 2cm n=0,1,2,\cdots \cr
\ep_\alpha &= \Big( n+{1\over 2} \Big) {1\over ml^2} \equiv \ep_n ~~~.  \cr}
    \eqn\spectrum $$
The first integer index $n$ labels Landau levels.
The second index $p$ is either a momentum in the $x_1$-direction in the
Landau gauge, or an orbital angular momentum in the symmetric gauge.

In the symmetric gauge
$$\eqalign{
&u_{sp}^{\rm sym}(r,\phi)
   = \bigg[ { s!\over (s+|p|)!} {1\over  2\pi l^2} \bigg]^{{1\over 2}} ~
  e^{-i \ep(N) p\phi}~ w^{|p|/2} e^{-w/2}~ L_s^{|p|}(w)  \next
          w={r^2\over 2l^2} \cr
&\ep_{sp} = \Big( s+ {1\over 2} + \theta(-p) |p| \Big) {1\over ml^2}  \cr
&\hskip 6.cm (q =0,1,2,\cdots ~~{\rm and}~~ p \in \zz) . \cr}
     \eqn\symmetricGauge  $$
The Landau level index is $n= s+ \theta(-p) |p|$.  Here $L_s^\alpha(w)$ is the
associated Laguerre polynomial
$$L_s^\alpha(w) = {1\over s!} ~w^{-\alpha} e^w ~{d^s\over dw^s}
(w^{s+\alpha} e^{-w} ) ~~~.   \eqn\Laguerre $$

In the Landau gauge we impose a periodic boundary condition in the
$x_1$-direction: $u_\alpha(x_1+L_1,x_2)=u_\alpha (x_1,x_2)$.  Then
$$\eqalign{
&u_{np}^{\rm Landau}(\x) = {1\over \sqrt{\, lL_1} } ~ e^{-ikx_1}
   v_n [  (x_2-\bar x_2 )/l ]  \next
     k={2\pi p\over L_1} \next \bar x_2 =  \ep(N)~kl^2 \cr
&\ep_{np} = \Big( n+ {1\over 2} \Big) {1\over ml^2}   \cr
&\hskip 6.cm (n=0,1,2,\cdots ~~{\rm and}~~ p \in \zz). \cr}
    \eqn\LandauGauge  $$
Here $v_n(x)$ is related to the Hermite polynomial:
$$\eqalign{
&v_n(x) = {(-1)^n\over 2^{n/2} \pi^{1/4} (n!)^{1/2} } ~
   e^{x^2/2} {d^n\over dx^n} e^{-x^2}  \cr
&\int_{-\infty}^\infty dx ~v_n(x) v_m(x) = \delta_{nm} ~~~. \cr}
        \eqn\Hermite $$
$u_{np}^{\rm Landau}(\x)$ is a plane wave in the $x_1$-direction, but
is localized around $\bar x_2$ in the $x_2$-direction.

In a box ($0\le x_2 \le L_2$), $\bar x_2$ must satisfy
$0\le \bar x_2 \le L_2$, or equivalently $0 \le p \le L_1L_2/2\pi l^2$.
Hence
the number of states per area for each Landau level, $n_L$, is given by
$$n_L = {1\over 2\pi l^2}   ~~~.  \eqn\levelDensity  $$

Combinig (\levelDensity) with (\magneticLength), one finds that the
filling factor, $\nu$,  is given by
$$\nu \equiv {n_e\over n_L} = |N| ~~~.  \eqn\fillingFactor $$
In other words, for an integer $N$, Landau levels are completely filled
at least in the mean field approximation.   It has to be stressed that
this property holds irrespective of the density $n_e$.

If the $\psi$ field has spin ${1\over 2}$ as electrons, the filling factor
is given by $\nu={1\over 2} |N|$ so that the complete filling of Landau
levels holds for an even integer $N$, provided that the magnetic moment
interaction is sufficiently small as in the usual cases.  We are mostly
interested in the case $|N|=2$, corresponding to semions or half-fermions.
There arises no qualitative change in physics
properties,\myref{\HosoChak,\RDSS,\HHL} and we shall
restrict ourselves in this article to spinless $\psi$.

We expand $\psi$ in terms of $\{ u_\alpha(\x) \}$.
$$
\psi(x) = \sum_\alpha c_\alpha u_\alpha(\x) \next
\{ c_\alpha^{}, c_\beta^{\dagger} \} = \delta_{\alpha\beta} ~~~.
                  \eqn\expandPsi $$
The mean field ground state is given by, for $|N|=2$,
$$\eqalign{
&\st \Psi_{\rm mean} \ra = \prod_{\alpha \in G} c_\alpha^\dagger ~\st 0 \ra \cr
&G = \{ \alpha=(n,p) ~;~ n=0,1 \}  ~~. \cr}   \eqn\meanGround  $$
The corresponding mean field energy is
$$E_{\rm mean} = \la \Psi_{\rm mean} \st H_1^{(1)} \st \Psi_{\rm mean} \ra
  = \sum_{\alpha \in G} \ep_\alpha  ~~~. \eqn\meanEnergy $$
Here $H_1^{(1)}$ is given in (\HamOneOne).
Again putting  the system  in a box in the Landau gauge,
one finds  the energy density to be
$$\E_{\rm mean} = {1\over L_1L_2} \sum_{n=0,1} \sum_{p=1}^{L_1L_2 /2\pi l^2}
   \Big( n+{1\over 2} \Big) {1\over ml^2} = {\pi n_e^2 \over m} ~~.
   \eqn\meanEnergyDensity $$
We recall that the energy density of a free spinless fermion fluid is
exactly $\pi n_e^2/m$.

It is straightforward to check
$$\eqalign{
\la j^0(x) \ra_{\rm mean}
&= \la \Psi_{\rm mean} \st \psi^\dagger \psi \st \Psi_{\rm mean} \ra  \cr
&=\sum_{\alpha\in G} u_\alpha(\x)^\dagger u_\alpha(\x) \cr
&=n_e  \cr
\la j^k(x) \ra_{\rm mean}
 &= -{i\over 2m} ~\la \Psi_{\rm mean} \st
  \big\{  \psi^\dagger \Dbar_k\psi - (\Dbar_k \psi)^\dagger \psi \big\}
                     \st \Psi_{\rm mean} \ra  \cr
&= -{i\over 2m} \sum_{\alpha\in G}
 \big\{  u_\alpha^\dagger \Dbar_ku_\alpha
            - (\Dbar_k u_\alpha)^\dagger u_\alpha \big\} \cr
&=0  \cr}    \eqn\meanCurrent  $$
where $\Dbar_k=\d_k - i \bar a^k(\x)$.

\sectionskip

\secno=7 \meqno=1

\line{\tenbf 7.  Hartree-Fock ground state \hfil}
\vglue 5pt
In this section we incorporate many-particle correlations in the
Hartree-Fock (HF) approximation.   The mean field approximation retains only
$H_1^{(1)}$, defined in (\HamOneOne), in the total Hamiltonian $H_1$,
(\HamOne).
In the Hartree-Fock approximation ``diagonal parts'' of $H_1^{(2)}$ and
$H_1^{(3)}$ in (\HamOneOne) are taken into account self-consistently, or
equivalently to say, the ground state is determined to satisfy the
Hartree-Fock equation.

It was shown by Hanna, Laughlin, and Fetter\myref{\HLF}
 that the Hartree-Fock ground state is exactly the same as
the mean field ground state $\st \Psi_{\rm mean} \ra$, (\meanGround).
$$\st \Psi_\HF \ra =
\st \Psi_{\rm mean} \ra = \prod_{\alpha \in G} c_\alpha^\dagger ~\st 0 \ra
  ~~.  \eqn\HFground  $$
However, it has a different energy.
$$E_\HF = \la H_1 \ra_\HF \equiv \la \Psi_\HF \st H_1 \st \Psi_\HF \ra
    \not= E_{\rm mean} ~~.    \eqn\HFenergyOne $$

We first compute $E_\HF$.  The original computation of Hanna, Laughlin, and
Fetter was given in the first quantized theory.   We present the
evaluation in the second quantized theory.

To facilitate the computations, we define the following sums.
$$
\vcenter{
\hbox{ \vbox{ \halign{ $\big #$\hfil &$\big #$\hfil  &$\big #$\hfil \cr
f(\x,\y) &= \sum_{\alpha\in G} u_\alpha(\x)^* u_\alpha(\y)
                          &= f(\y,\x)^*  \cr
\noalign{\kern 3pt}
f_k(\x,\y) &= \sum_{\alpha\in G} u_\alpha(\x)^*  i \Dbar_k u_\alpha(\y)
                   &= i \Dbarky f(\x,\y)   \cr}   } } }
    \eqn\FunctionF  $$

We need to evaluate expectation values of various products of $\psi$ and
$\psi^\dagger$.
We denote $\la {\cal Q} \ra_\HF = \la \Psi_\HF \st
{\cal Q} \st \Psi_\HF \ra$.  Then

{\baselineskip=22pt plus 1pt  \noindent
$$\eqalign{
\noalign{\kern 7pt}
&\la \psi^\dagger(x)\psi(x) \ra_\HF = f(\x,\x) \cr
&\la \psi^\dagger(x) \psi^\dagger(y) \psi(y) \psi(x) \ra_\HF
  =f(\x,\x) f(\y,\y) - f(\x,\y) f(\y,\x)  \cr
&\la \psi^\dagger(x) \psi^\dagger(y) \psi^\dagger(z)
   \psi(z) \psi(y) \psi(x) \ra_\HF  \cr
&\hskip .7cm =f(\x,\x) f(\y,\y) f(\z,\z) + f(\x,\y) f(\y,\z) f(\z,\x)
       + f(\x,\z) f(\y,\x) f(\z,\y)  \cr
&\hskip .7cm  - f(\x,\x) f(\y,\z) f(\z,\y) - f(\x,\z) f(\y,\y) f(\z,\x)
       - f(\x,\y) f(\y,\x) f(\z,\z)  \cr}
   \eqn\expectHFone  $$
Secondly
$$\eqalign{
&i \la \psi^\dagger(x) \Dbar_k \psi(x)
     - \big( \Dbar_k \psi(x) \big)^\dagger \psi(x) \ra_\HF
      = f_k(\x,\x) + f_k(\x,\x)^*  \cr
&i \la \psi^\dagger(x) \psi^\dagger(y) \psi(y) \Dbar_k \psi(x)
    - \big( \Dbar_k \psi(x) \big)^\dagger \psi^\dagger(y) \psi(y)
    \psi(x) \ra_\HF   \cr
&\hskip .3cm = \big\{ f_k(\x,\x) + f_k(\x,\x)^* \big\} f(\y,\y)
   - \big\{ f(\x,\y) f_k(\y,\x) + f(\y,\x) f_k(\y,\x)^* \big\} ~. \cr}
  \eqn\expectHFtwo $$
}

Making use of (\HamOneOne), (\normalHamOne), (\expectHFone), and
(\expectHFtwo),
one finds
$$\E_\HF = {1\over \vol }
       \la (H_1^{(1)} + H_1^{(2)} + H_1^{(3)}) \ra_\HF
  \equiv \E^{(1)} +  \E^{(2)} + \E^{(3)}  \eqn\HFenergy $$
where $\vol$ is the volume.
$\E^{(1)}$ is the same as the mean field energy density
$$\E^{(1)} = \E_{\rm mean} = {\pi n_e^2 \over m}  \eqn\HFenergyOne  $$

{\baselineskip=22pt plus 1pt  \noindent
and
$$\eqalign{
\E^{(2)} &= {1\over \vol} {1\over 2m} \int d\x d\y ~h_k(\x-\y)
\Big[ ~ \big\{ f_k(\x,\x) + f_k(\x,\x)^* \big\} \big\{ f(\y,\y) - n_e\big\} \cr
&\hskip 4.cm - \big\{ f(\x,\y) f_k(\y,\x)
     + f(\y,\x) f_k(\y,\x)^* \big\} ~ \Big]  \cr
\E^{(3)} &= {1\over \vol} {1\over 2m} \int d\x d\y d\z ~
  h_k(\x-\y) h_k(\x-\z)  \cr
&\hskip .3cm \times \Big[ ~ 2f(\x,\y) f(\y,\z) f(\z,\x)
               - f(\x,\x) f(\y,\z) f(\z,\y) \cr
&\hskip .3cm -2 |f(\x,\y)|^2 \big\{ f(\z,\z)-n_e \big\}
+f(\x,\x) \big\{ f(\y,\y)-n_e \big\} \big\{ f(\z,\z)-n_e \big\} ~ \Big]  \cr
&+ {1\over \vol}{1\over 2m} \int d\x d\y ~[h_k(\x-\y)]^2
   ~\big\{ f(\x,\x) f(\y,\y) - |f(\x,\y)|^2 \big\}  \cr}
     \eqn\HFenergyDensity $$
}

The quantities $f(\x,\y)$ and $f_k(\x,\y)$ in (\FunctionF) depend on the gauge
chosen.   The symmetric and Landau gauge defined in (\averagePotential) are
related by
$$\eqalign{
&\bar a^k_{\rm sym}(\x) = \bar a^k_{\rm Landau} - \nabla_k \Lambda(\x) \cr
&\Lambda(\x) = \ep(N) ~ {x_1x_2 \over 2l^2} ~~~.  \cr}
    \eqn\symmetricLandau  $$
If $u_{sp}^{\rm sym}(\x)$ is a solution to the Schr\"odinger
equation in the symmetric gauge, then $e^{i\Lambda(\x)} ~u_{sq}^{\rm sym}(\x)
\equiv u_{sq}^{\rm Landau}(\x)$ is a solution in the Landau gauge with the
same energy eigenvalue.  It is a linear combination of
$u_{np}^{\rm Landau}(\x)$ in (\LandauGauge).  With a given Landau level
(energy eigenvalue),  the sets $\{ u_{sq}^{\rm Landau}(\x) \}$ and
$\{ u_{np}^{\rm Landau}(\x) \}$ are related by a unitary transformation.

Hence, if the set $G$ in (\meanGround) represents completely filled Landau
levels as in the case under consideration, then
$$\sum_{(s,q) \in G^{\rm sym}} u_{sq}^{\rm Landau}(\x)^* u_{sq}^{\rm
Landau}(\y)
= \sum_{(n,p) \in G^{\rm Landau}} u_{np}^{\rm Landau}(\x)^*
   u_{np}^{\rm Landau}(\y)  $$
so that
$$\eqalign{
f(\x,\y)^{\rm sym}
   &= e^{i\{\Lambda(\x)-\Lambda(\y)\} } ~ f(\x,\y)^{\rm Landau}  \cr
f_k(\x,\y)^{\rm sym}
   &= e^{i\{\Lambda(\x)-\Lambda(\y)\} }   ~ f_k(\x,\y)^{\rm Landau} ~~. \cr}
    \eqn\symmetricLandauF  $$
Thanks to the relation (\symmetricLandauF), every term in (\HFenergyDensity)
is separately gauge independent.

It is easiest to evaluate $f(\x,\y)$ in the Landau gauge.  Making use of
(\LandauGauge), one finds
$$\eqalign{
f(\x,&\y)^{\rm Landau} = \sum_{n=0}^{|N|-1} \sum_{p=-\infty}^\infty
{1\over lL_1} ~e^{ik(x_1-y_1)}
    ~v_n[(x_2-\bar x_2)/l] ~v_n[(y_2-\bar y_2)/l] \cr
&={1\over 2\pi l} \sum_{n=0}^{|N|-1} \int_{-\infty}^\infty dk
  ~e^{ik(x_1-y_1)} ~v_n[(x_2-\bar x_2)/l] ~v_n[(y_2-\bar y_2)/l] ~~. \cr}
   \eqn\generalLandauF $$
Employing explicit forms of $v_n$'s and integrating over $k$, one obtains
$$\eqalign{
N=\pm 1 ~:~  f(\x,\y)^{\rm Landau} &=
n_e \cdot \exp \bigg\{ -{(\x-\y)^2 \over 4l^2}
      \pm i {(x_1-y_1)(x_2+y_2) \over 2l^2} \bigg\} \cr
          &\equiv f_0^\pm(\x,\y) \cr
N=\pm 2 ~:~  f(\x,\y)^{\rm Landau} &=
\bigg\{ 1 - {(\x-\y)^2\over 4l^2} \bigg\}  \cdot f_0^\pm(\x,\y) \cr}
   \eqn\OneTwoLandauF  $$
Note that $l^2 = |N|/2\pi n_e$.  $f_k(\x,\y)$ is obtained from (\FunctionF).
$$\eqalign{
N=\pm 1 ~:~ f_1(\x,\y)^{\rm Landau} &=
{i\over 2l^2} \big\{ (x_1-y_1) \mp i (x_2-y_2) \big\} ~ f_0^\pm(\x,\y) \cr
f_2(\x,\y)^{\rm Landau} &= \pm i f_1(\x,\y)^{\rm Landau} \cr
\noalign{\kern 8pt}
N=\pm 2 ~:~ f_k(\x,\y)^{\rm Landau} &=
  \bigg\{ 1 - {(\x-\y)^2\over 4l^2} \bigg\} \cdot
             f_k(\x,\y)^{\rm Landau}_{N=\pm 1}  \cr
&\hskip 3cm    + {i\over 2l^2} ~(x_k-y_k) f_0^\pm(\x,\y)     \cr}
       \eqn\OneTwoLandauFk  $$

We are ready to evaluate (\HFenergyDensity).
We drop the superscripts `Landau' in $f(\x,\y)$ and $f_k(\x,\y)$.
Note that $f(\x,\x)=n_e$ and $f_k(\x,\x)=0$.  Furthermore,
$$h_k(\x-\y) f_k(\x,\y) = -{1\over 2 l^2 |N|} ~f(\x,\y) ~~.
   \eqn\usefulHFk  $$
Therefore $\E^{(2)}$ becomes
$$
\E^{(2)} = - {1\over \vol} {1\over 2|N|ml^2} \int d\x d\y ~ |f(\x,\y)|^2
= \cases{ \big - {\pi n_e^2\over m} &for $N= \pm 1$\cr \cr
           \big -{\pi n_e^2\over 4m} &for $N= \pm 2$.\cr}
  \eqn\HFeTwo$$

For $\E^{(3)}$ we have
$$\eqalign{
\E^{(3)} &= {1\over \vol} {1\over 2m} \int d\x d\y d\z ~
  {1\over N^2} {(\x-\y)(\x-\z) \over (\x-\y)^2 (\x-\z)^2}  \cr
&\hskip 3.cm \times \Big[ ~ 2f(\x,\y) f(\y,\z) f(\z,\x)
               - n_e |f(\y,\z) |^2 ~ \Big]  \cr
&+ {1\over \vol}{1\over 2m} \int d\x d\y ~{1\over (\x-\y)^2}
   ~\big\{ n_e^2 - |f(\x,\y)|^2 \big\}  \cr
&\equiv \Big[~ \E^{(3)}_1 + \E^{(3)}_2 ~\Big] + \E^{(3)}_3   ~~.\cr}
     \eqn\HFeThree $$
As will been seen below, $\E^{(3)}_2$ and $\E^{(3)}_3$ have infra-red
divergences, which cancell each other.

In evaluating $\E^{(3)}_1$, we rename $\y-\x$ and $\z-\x$ to be new $\y$ and
$\z$, respectively.  Then
$$\eqalign{
\E^{(3)}_1 &= {n_e^3 \over N^2 m} \int d\y d\z ~ {\y \z \over \y^2 ~\z^2} ~
   g_N(\y) g_N(\z) g_N(\y-\z)  \cr
 &\hskip 1cm \times \exp \Big\{ - {1\over 4l^2} [\y^2 + \z^2 + (\y-\z)^2 ]
  \pm {i\over 2l^2} (y_1 z_2 - y_2 z_1) \Big\}  \cr
\noalign{\kern 5pt}
&g_N(\x) = \cases{ 1 &for $N=\pm 1$\cr
      \big 1-{\x^2\over 4l^2} &for $N=\pm 2$ \cr}  \cr}
  \eqn\eThreeOneOne  $$
We introduce polar coordinates by
$$y_1+iy_2 = \sqrt{2} \, l \, r e^{i\theta} \next
   z_1+iz_2 = \sqrt{2} \,l \,\rho e^{i(\theta+\phi)} ~~. $$
The integrand is independent of $\theta$, and
$\y\z/2l^2 = r\rho \cos \phi$ and $(y_1z_2-y_2 z_1)/2l^2 = r\rho \sin \phi$.
(\eThreeOneOne) is transformed to
$$\eqalign{
\E^{(3)}_1 &= { n_e^3 \over N^2 m}\cdot 2\pi \cdot 2l^2
    \int_0^\infty dr d\rho \int_0^{2\pi} d\phi~ \cos \phi
\cdot \exp \Big\{ -r^2 -\rho^2 + r\rho e^{\pm i\phi} \Big\} \cr
\noalign{\kern 4pt}
&\hskip .5cm \times \cases{ 1 &for $N=\pm 1$\cr
\noalign{\kern 3pt}
 (1-{1\over 2} r^2) (1-{1\over 2} \rho^2)
 (1- {1\over 2} r^2 -{1\over 2} \rho^2 +r\rho \cos \phi ) &for $N=\pm 2$\cr}
            \cr}  $$
Integrating over $\phi$, one finds
$$\eqalign{
\E^{(3)}_1 &= { 4\pi l^2 n_e^3 \over N^2 m}    \int_0^\infty dr d\rho ~
    e^{-r^2 - \rho^2}  \cr
&\hskip 1.5cm \times \cases{ \pi r \rho \cr
\noalign{\kern 5pt}
  \pi r \rho  (1-{1\over 2}r^2) (1-{1\over 2}\rho^2)
[ 2- {1\over 2} (r^2 + \rho^2) + {1\over 4} r^2\rho^2 ]  \cr}
          \cr
\noalign{\kern 5pt}
&=\cases{ \big {\pi n_e^2\over 2m} &\hskip 1cm for $N=\pm 1$\cr \cr
          \big {\pi n_e^2 \over 8m} &\hskip 1cm for $N=\pm 2$\cr}  \cr}
    \eqn\eThreeOneTwo  $$

Similarly, the second term, $\E^{(3)}_2$, in (\HFeThree) becomes
$$\eqalign{
\E^{(3)}_2 &= - {n_e^3\over 2mN^2} \int d\y d\z ~
 {\y\z \over \y^2 ~\z^2} ~g_N(\y-\z)^2 ~e^{-(\y-\z)^2/2l^2} \cr
&== - {n_e^3\over 2mN^2} \int d\y d\u ~
 {\y(\y-\u) \over \y^2 (\y-\u)^2} ~g_N(\u)^2 ~e^{-\u^2/2l^2} \cr}
  \eqn\eThreeTwoOne $$
where we have introduced $\u=\y-\z$.  This time we define
$$y_1+iy_2 = \sqrt{2} \, l \, r e^{i\theta} \next
   u_1+iu_2 = \sqrt{2} \,l\, \rho e^{i(\theta+\phi)} ~~. $$
Then
$$
\E^{(3)}_2 = - {2\pi l^2n_e^3\over mN^2} \int dr d\rho d\phi ~
{\rho\over r} {r^2 - r\rho \cos \phi\over r^2 + \rho^2 - 2r \rho \cos \phi}
{}~ g_N(\u)^2 ~ e^{-\rho^2}  ~.
              \eqn\eThreeTwoTwo $$
The $\phi$-integral gives
$$\int_0^{2\pi} d\phi ~
   {r^2 - r\rho \cos \phi\over r^2 + \rho^2 - 2r \rho \cos \phi}
 = 2\pi \theta(r-\rho) ~~. $$
The rest of the computation is straightforward.  We find
$$\E^{(3)}_2 = - {\pi n_e^2\over mN^2} \bigg\{
 \int_0^\infty dr~ {1- e^{-r^2} \over r} + {1\over 4} ~\delta_{N,\pm 2} \bigg\}
  ~~. \eqn\eThreeTwoThree  $$
The integral in (\eThreeTwoThree) diverges in the upper limit.

The evaluation of $\E^{(3)}_3$ is easy.
$$\eqalign{
\E^{(3)}_3 &= {n_e^2 \over 2 mN^2} \int d\y~ {1\over \y^2}
 \Big\{ 1- g_N(\y)^2 ~e^{-\y^2/2l^2} \Big\} \cr
&= {\pi n_e^2\over mN^2} \bigg\{  \int_0^\infty dr ~
   {1-e^{-r^2} \over r} + {3\over 8} ~\delta_{N,\pm2} \bigg\}   \cr}
  \eqn\eThreeThree  $$
The divergent integrals in (\eThreeTwoThree) and (\eThreeThree) cancell
each other.

Adding (\eThreeOneTwo), (\eThreeTwoThree), and (\eThreeThree), one finds
$$\E^{(3)} = \cases{ \big {1\over \,2\,}{\pi n_e^2\over m} &for $N=\pm 1$\cr
  \noalign{\kern 5pt}
  \big {5\over 32}{ \pi n_e^2\over  m} &for $N=\pm 2$ ~~~. \cr}
   \eqn\HFeThreeTotal $$
Finally combining (\HFenergyOne), (\HFeTwo), and (\HFeThreeTotal), we
obtain
$$\E_\HF = \left\{  \vcenter{   \hbox{ \vbox{
\halign{ $\big #$\hfil &\hskip .3cm $\big #$ \hfil &\hskip 1cm for $#$ \cr
\Big( 1 - \,1\, + {1\over \,2\,} \Big) {\pi n_e^2\over m}
    &= {1\over \,2\,} {\pi n_e^2 \over m} &N=\pm 1 \cr
\noalign{\kern 5pt}
\Big( 1 - {1\over 4} + {5\over 32} \Big) {\pi n_e^2\over m}
 &= {29\over 32} {\pi n_e^2 \over m} &N=\pm 2  \cr}
     } } }   \right.
   \eqn\HFenergyFinal  $$
The correction to the mean field energy is large for $N=\pm 1$, but
is relatively small for $N=\pm2$.

There are various ways of showing that $\st \Psi_\HF \ra$ in (\HFground) is
the Hartree-Fock ground state.  The Hartree-Fock approximation amounts
to finding an approximate ground state for a many-particle system in
the form of a Slater determinant formed from one-particle wave functions.
In the language of the second quantized theory we expand the field
operator $\psi(x)$ in a complete orthonormal set
$\{ u_\alpha(\x) \}$ to be yet determined, and write a trial ground state as
$$\eqalign{
&\psi(x) = \sum c_\alpha u_\alpha(\x) ~e^{-i \ep_{\alpha,\HF} t}  \cr
&\st \Psi_G \ra = \prod_{\alpha \in G} c_\alpha^\dagger \st 0 \ra  \cr
&\int d\x ~\la  \psi^\dagger \psi(x)  \ra
     =  n_e \cdot vol  ~~~.   \cr}
    \eqn\trialHF $$
where $\la {\cal Q} \ra = \la \Psi_G \st {\cal Q} \st \Psi_G \ra$.
$\{ u_\alpha(\x) \}$ and the set $G$ are determined to minimize the
expectation
value of the Hamiltonian  $\la  H  \ra$.

In Eq. (\PsiEq)
$$i  \dot \psi(x) =\bigg\{ - {1\over 2m} D_k^2 + a_0(x) + g(x) \bigg\}
   ~\psi(x)      \eqno(\PsiEq) $$
the approximation amounts to retaining only diagonal pieces on the
right hand side.   For instance
$$\eqalign{
&\psi^\dagger(y) \Dbar_k \psi(y) \psi(x)  \cr
&\hskip .3cm \longrightarrow
\la \psi^\dagger(y) \Dbar_k \psi(y) \ra \, \psi(x)
 - \la \psi^\dagger(y) \psi(x) \ra \, \Dbar_k  \psi(y) \cr
\noalign{\kern 5pt}
&\psi^\dagger(y) \psi^\dagger(z)  \psi(z) \psi(y) \psi(x) \cr
&\hskip .3cm \longrightarrow
\la \psi^\dagger(y)\psi(y) \ra \, \la \psi^\dagger(z)\psi(z) \ra \, \psi(x)
- \la\psi^\dagger(y)  \psi(z)\ra \, \la\psi^\dagger(z)  \psi(y)\ra \,\psi(x)
\cr
&\hskip .6cm
-\la \psi^\dagger(y)\psi(x) \ra \,\la \psi^\dagger(z)\psi(z) \ra \, \psi(y)
+\la \psi^\dagger(y)\psi(z) \ra \,\la \psi^\dagger(z)\psi(x) \ra \, \psi(y) \cr
&\hskip .6cm
-\la \psi^\dagger(y)\psi(y) \ra \,\la \psi^\dagger(z)\psi(x) \ra \, \psi(z)
+\la \psi^\dagger(y)\psi(x) \ra \,\la \psi^\dagger(z)\psi(y) \ra  \,\psi(z) ~.
           \cr}
  \eqn\HFapprox  $$
The set $\{ u_\alpha(\x) \, ;\, \alpha \in G \}$ is determined to satisfy
Eq. (\PsiEq) with the above substitution made.  The equation thus obtained
is called the Hartree-Fock equation.

It was shown by Hanna, Laughlin, and Fetter\myref{\HLF} that $|N|$ completely
filled Landau levels formed from the mean field eigenstates (\symmetricGauge)
or
(\LandauGauge) satisfy the Hartree-Fock equation, and therefore
$\st \Psi_\HF \ra$ in (\HFground) is the Hartree-Fock ground state.
The computation is similar to, but more complicated than, that of
$\la H_1 \ra_\HF$ presented above.  Readers should refer to the original
paper for details.

The Hartree-Fock equation can be written solely in terms of
the function $f(\x,\y)$ defined in (\FunctionF).  With the ansatz (\trialHF),
the expectation value $\la H \ra$ can be viewed as a function of
$\la \psi^\dagger(\x) \psi(\y) \ra = f(\x,\y)$.
$$
\la H_1 \ra = {\cal H}[f(\x,\y)] \next  f(\x,\y)^* = f(\y,\x)  ~~.
    \eqn\HFfunctional $$
$f(\x,\y)$ is determined by the extremum condition:
$$ {\delta {\cal H} \over \delta f(\x,\y) } =0
 \hskip .6cm {\rm subject ~to} ~~~ \int d\x \, f(\x,\x) = n_e \cdot vol ~~.
     \eqn\alternativeHFeq  $$
$f(\x,\y)$ is determined by the equation up to an arbitrariness due to
gauge degrees of freedom.   Eq. (\alternativeHFeq) is equivalent to
the Hartree-Fock equation.

\sectionskip


\secno=8  \meqno=1

\line{\bf 8.  RPA and SCF \hfil}
\vglue 5pt
There are approximation schemes which lie between the mean filed and
Hartree-Fock approximations, and are useful to investigate various physical
quantities such as the excitation spectrum,  current-current correlation
functions, response to external perturbations, and so on.
They are the random phase approximation (RPA), self-consistent
field method (SCF), and hydrodynamic description.  In the
anyon model under consideration, all these three are equivalent to each other.

In a neutral anyon fluid RPA is formulated from the Hamiltonian
$H_1[\psi,\psi^\dagger]$ in (\HamOne), whereas SCF from
the Lagrangian  $\L_0[\psi,\psi^\dagger,a_\mu]$ in (\LagZero).  The
difference lies in whether Chern-Simons fields are integrated out first, or
retained.  A similar statement is valid for a charged anyon fluid.

For definiteness we shall restrict ourselves to a neutral anyon fluid in this
section.  One can establish the diagram method, or Feynman rules, from the
Hamiltonian $H_1$.  $H_1^{(1)}$ in (\HamOneOne) defines a bare propagator
for the $\psi$-field.  It is a propagator in the mean field, depicted
by a solid arrowed line.   $H_1^{(2)}$ and $H_1^{(3)}$ define interaction
vertices.

The two-body interaction generated by $H_1^{(2)}$ is given by $V_a$ in
Fig. 3.   A dashed line corresponds to $h_k(\x-\y)$, representing
virtual ``propagation'' of Chern-Simons fields.  A crossed circle at the vertex
$x$ indicates the derivative factor $i \Dbar_k$.

For the purpose of establishing Feynman rules,
it is convenient to start with the normal-ordered form (\normalHamOne) for
$H_1^{(3)}$.   It yields  three- and two-body
interactions $V_b$ and $V_c$ in Fig. 3, respectively.
Both  involve two dashed lines.

{\figure
\vglue 6cm
\centerline{Fig. 3 \quad
 \vtop{\hsize=9 cm \noindent
 Feynman rules derived from (\HamOne).  The rules are simplified by retaining
Chern-Simons fields as independent variables.  See Fig.~7 in Section 10.} }
}
\bigskip

There are important rules resulting from the form of the Hamiltonian.
Recall that $\la \psi^\dagger(x) \psi(x) \ra=n_e$,
 $\la \psi^\dagger(x) \Dbar_k \psi(x) \ra=0$, and $\int d\x \, h_k(\x)=0$.
Therefore, contraction of two solid lines at the same vertex in $V_a$ yield a
vanishing result.  Similarly, contraction of two solid lines at either
$y$ or $z$ vertex in $V_b$ yields a vanishing result.  See Fig. 4a. However,
contraction of two solid lines at the $x$ vertex in $V_b$, and at the $x$ and
$y$ vertices in $V_c$ yield a non-vanishing result.  See Fig. 4b.

{\figure
\vglue 7cm
\centerline{Fig. 4  \quad Feynman rules -- constraints.}
}
\bigskip

Let's consider  correlation functions of the currents $J^\mu(x)$ defined in
(\bigCurrent).   Note that $J^k(x)$  defines two- and four-point vertices.
(See Fig. 5.)
To all order in perturbation theory
$$\la J^\mu(x) \ra = n_e ~\delta^{\mu 0}  ~~~.  \eqn\expectJ $$

{\figure
\vglue 6cm
\centerline{Fig. 5  \quad Currents {\eightit J}$^{\sevenmu}$({\eightit x}) .}
}
\bigskip

For the correlation function for a neutral fluid
$$ D_n^{\mu\nu}(x,y) = -i \la T[ \tilde J^\mu(x) \tilde J^\nu(y)] \ra
 \next  \tilde J^\mu(x) = J^\mu(x) - n_e~ \delta^{\mu 0}
    \eqn\defineDn  $$
RPA constitutes in keeping only daigrams in which no dashed line is a part
of closed loops involved.\myref{\FHL-\CWWH}  Typical diagrams involved are
depicted in  Fig. 6.  In RPA we have only sums of bubble diagrams.  Further,
the
four-point vertex in $J^k(x)$ does not enter at this level.

{\figure
\vglue 10cm
\centerline{Fig. 6  \quad
{\eightit D}$_{\hbox{\sevenmit n}}^{\sevenmu\sevennu}$({\eightit x,y})\ in
RPA.}
}
\bigskip

One can write down  Schwinger-Dyson equations for
$\Dn^{\mu\nu}$, which was done  by Fetter, Hanna, and Laughlin,\myref{\FHL}
and by Chen, Wilczek, Witten, and Halperin.\myref{\CWWH}  Due to the presence
of
the off-diagonal couplings, the equation takes the form of  a $3 \times 3$
matrix.  In the following section we shall show an alternative method
to evaluate   $\Dn^{\mu\nu}$.

The self-consistent field method starts with the field equation (\EulerEq),
iretaining the Chern-Simons gauge fields:\myref{\HosoChak,\HHL}
$$\eqalign{
&- {N\over 4\pi} ~ \eps f_{\nu\rho} = j^\mu \cr
&i \d_0 \psi = \Big\{ -{1\over 2m} D_k^2 + a_0 \Big\} ~\psi  ~~. \cr}
   \eqno(\EulerEq) $$
We suppose first that there is a consistent gauge field configuration
$a_\mu(x)$ which is treated classically.  Secondly, with this given $a_\mu(x)$
we determine a quantum mechanical state vector, $\st \Psi[a_\mu] \ra$,
for the matter field $\psi(x)$, by solving the second equation of (\EulerEq).
Thirdly, we replace  the current $j^\mu$ on the right side of the first
equation
of (\EulerEq) by its expectation value:
$$- {N\over 4\pi} ~ \eps f_{\nu\rho}
 = \la \Psi[a_\mu] \st j^\mu \st \Psi[a_\mu] \ra
 \equiv J^\mu[x;a_\mu] ~~.    \eqn\SCFeq  $$
Eq. (\SCFeq) is solved for $a_\mu(x)$ to find a self-consistent field
configuration.

We have known that the configuration, (\averagePotential), of the  mean field
ground state, which is the same as the Hartree-Fock
ground state, solves Eq.$\,$(\SCFeq).  We are seeking for more general,
$x$-dependent solutions.  If a deviation of $a_\mu(x)$ from (\averagePotential)
is small, one can employ a perturbation theory to find a solution.

For time-independent configurations the procedure is particularly simple,
as was first performed by Hosotani and Chakravarty.\myref{\HosoChak}   The
generalization to finite temperature is  done by Randjbar-Daemi,
Salam, and Strathdee,\myref{\RDSS} and by Hetrick, Hosotani, and
Lee.\myref{\HHL}  Time-dependent configurations have been recently analysed by
this method by Chakravarty.\myref{\Sumantra}

Consider  a time-independent, small fluctuation of Chern-Simons gauge fields,
$a_\mu^{(1)}(x)= a_\mu(x) - \bar a_\mu(x)$.  We first solve the one-particle
Schr\"odinger equation with this $a_\mu(x)$:
$$\bigg\{ - {1\over 2m} \, D_k^2 + a_0(x) \bigg\} ~ u_\alpha (\x;a_\mu^{(1)})
  = \ep_\alpha (a_\mu^{(1)}) ~ u_\alpha (\x;a_\mu^{(1)}) ~~.
    \eqn\SCFSchroEq  $$
Both $\ep_\alpha (a_\mu^{(1)})$ and $u_\alpha (\x;a_\mu^{(1)})$ are determined
perturbatively.  $\{ u_\alpha (\x;a_\mu^{(1)}) \}$ defines a complete,
orthonormal basis, with which we expand $\psi(x)$ as
$$
\psi(x) = \sum_\alpha c_\alpha(a^{(1)}) ~ u_\alpha(\x;a^{(1)}) \next
\{ c_\alpha^{}, c_\beta^{\dagger} \} = \delta_{\alpha\beta} ~~~.
                  \eqn\SCFexpandPsi $$

So long as $a_\mu^{(1)}(x)$ is sufficiently small and smooth, the spectrum
retains the structure of Landau levels, although they are not degenerate any
more in general.   With the same set as $G$ in (\meanGround), we define
a state
$$\Psi[a^{(1)}] = \prod_{\alpha \in G}
    c_\alpha^\dagger (a^{(1)}) \st 0 \ra ~~,  \eqn\SCFstate  $$
from which the current in (\SCFeq) is determined as
$$J^\mu[x;a_\mu] = n_e \, \delta^{\mu 0}
 - \int d^3y ~\Gamma^{\mu \nu} (x,y) \, a_\nu^{(1)}(y) + \cdots ~~.
   \eqn\SCFcurrent  $$

At finite temperature $T (= \beta^{-1})$ we evaluate the matter part of
the free energy with a given $a_\mu(x)$ by
$$e^{-\beta F[a_\mu]} = \tr e^{-\beta H_0[\psi,a_\mu]}
   \eqn\SCFfreeEnergy  $$
where
$$H_0[\psi,a_\mu] = \int d\x ~ \bigg\{ {1\over 2m} \, (D_k \psi)^\dagger
  (D_k \psi) + a_0 \psi^\dagger \psi \bigg\}  ~~.
   \eqn\SCFHam  $$
The current is given by
$$\eqalign{
J^\nu[x;a_\mu] &= \tr j^\nu(x) ~ e^{\beta(F[a_\mu]-H_0[\psi,a_\mu])}
= \tr {\delta H_0 \over \delta a_\nu(x) }
  ~ e^{\beta(F[a_\mu]-H_0[\psi,a_\mu])}  \cr
 &= {\delta F[a_\mu] \over \delta a_\nu(x) } ~~,  \cr}
   \eqn\SCFcurrentTwo $$
which leads to an expression similar to (\SCFcurrent).
Incorporation of electromagnetic interactions is straightforward.

In this section we have explained the two approximation methods, RPA and SCF.
These two look quite different from each other.  RPA is defined for Green's
functions for matter fields in terms of Feynman diagrams, whereas SCF is
written
in the form of  gauge field equations.  We shall show in the next section
that these two are indeed the same, and are equivalent.

\sectionskip

\secno=9 \meqno=1

\line{\bf 9.  Path integral representation \hfil}
\vglue 5pt
A bridge between RPA and SCF becomes most transparent in the path integral
formalism.\myref{\PathIntegral}  One can deal with both neutral and charged
anyon fluids at once. The key step is to consider the transition amplitude or
partition function in the presence of external gauge potentials.

To simplify notations, we write as
$$\eqalign{
&\L_0^\CS [a] = - {N\over 4\pi} \, \eps a_\mu \d_\nu a_\rho  \next
\L_0^\EM [A]= -{1\over 4} \, F^2_{\mu\nu} + en_e A_0\cr
\noalign{\kern 5pt}
&\L_f[\psi;a] = i \psi^\dagger D_0 \psi
   - {1\over 2m} (D_k \psi)^\dagger (D_k \psi) \cr}  \eqn\simplifyL $$
where $D_0=\d_0 + ia_0$ and $D_k=\d_k - ia^k$.  In terms of these definitions
$$\eqalign{
\L_0[\psi,a] &= \L_0^\CS [a] + \L_f[\psi;a] \cr
\L_\twoD[\psi,a,A] &= \L_0^\CS [a] + \L_0^\EM [A] + \L_f[\psi;a + eA]  \cr} $$

Consider a charged anyon fluid described by (\twoDLag) or (\twoDHam).
We introduce an external gauge potential $a^\ext_\mu(x)$ in a gauge invariant
way, making a  replacement
$$\L_f[\psi;a + eA] \Longrightarrow \L_f[\psi;a + eA + a^\ext] ~~~.
  \eqn\externalPotential $$
In other words, the external Lagrangian is given by
$$\eqalign{
\L_\ext &= \L_f[\psi;a + eA + a^\ext] - \L_f[\psi;a + eA ]  \cr
&=  - a_0^\ext j^0 + a^k_\ext j^k - {1\over 2m} \, (a^k_\ext)^2 \, j^0 \cr}
   \eqn\externalL $$
where
$$
\cases{
j^0(x) = \psi^\dagger \psi \cr \noalign{\kern 3pt}
\big j^k(x) = - {i\over 2m} \big( \psi^\dagger \nabla_k \psi
 - \nabla_k \psi^\dagger \cdot \psi \big)
   -{1\over m} \psi^\dagger \psi \cdot (a^k + eA^k) \cr   }
    \eqn\currentsReal  $$
In the presence of an external potential a total gauge-invariant
current is
$$\eqalign{
&j_\tot^\mu(x)= -{ \delta \over \delta a^\ext_\mu(x) }  \int dy ~
            \L_f[\psi;a+eA+a^\ext](y) \cr
&j_\tot^0 =j^0 \next
   j_\tot^k = j^k - {1\over m} \, a^k_\ext \psi^\dagger \psi ~~. \cr}
    \eqn\realCurrent $$

The transition amplitude at $T=0$ is given, in the path integral
representation,
by
$$\eqalign{
\exp \Big( \, i I_\twoD [a^\ext] \,\Big)
      = \int &\D A \D a \D\psi^\dagger \D\psi ~
  \exp \bigg\{ i\int dx ~ \Big( \L_0^\CS[a] + \L_{\rm g.f.}^\CS[a] \cr
&  + \L_0^\EM[A] + \L_{\rm g.f.}^\EM[A]
 +\L_f[\psi;a+eA+a^\ext] \, \Big) \,\bigg\} ~~. \cr}
   \eqn\amplitudeI $$
Here $\L_{\rm g.f.}^\CS[a]$ and $\L_{\rm g.f.}^\EM[A]$ are
 the gauge-fixing terms for $a_\mu$ and $A_\mu$, respectively,
and shall be specified shortly.
The initial and final  state wave functions have been absorbed
in the definition of the path integration measure.  Their explicit forms
are irrelevant to compute various quantities described
below.

Several identities follow from (\amplitudeI).  First
$$ -{\delta I_\twoD \over \delta a_\mu^\ext (x) }
= e^{-iI_\twoD} ~i  {\delta \over \delta a_\mu^\ext (x)} ~  e^{iI_\twoD}
=\la \, j_\tot^\mu(x) \, \ra_{a_\ext}  ~~. \eqn\FirstDiffI $$
Recalling that $\la\, j_\tot^\mu \,\ra_{a_\ext=0}= n_e \delta^{\mu 0}$, one
finds a current dynamically induced to be
$$\eqalign{
\indJ^\mu(x; a^\ext)
   &= \la j_\tot^\mu(x) \ra_{a_\ext} - n_e \, \delta^{\mu 0}  \cr
&= -{\delta I_\twoD \over \delta a_\mu^\ext (x) }
   - n_e \, \delta^{\mu 0} ~~. \cr}   \eqn\inducedJ  $$

Similarly the second derivative leads to
$$\eqalign{
&e^{-iI_\twoD} ~  i^2 {\delta^2\over \delta a_\mu^\ext (x)
  \delta a_\nu^\ext (y)} \, e^{iI_\twoD}  \cr
\noalign{\kern 4pt}
&\quad =\la T[j_\tot^\mu(x) j_\tot^\nu(y) ] \ra_{a_\ext}
 +i (1-\delta^{\mu 0}) \delta^{\mu\nu} ~
  {1\over m} \, \la j^0(x) \ra_{a_\ext} ~ \delta(x-y) \cr
\noalign{\kern 4pt}
&\quad = - i {\delta^2 I_\twoD \over \delta a_\mu^\ext(x)
   \delta a_\nu^\ext (y) }
+ {\delta I_\twoD \over \delta a_\mu^\ext(x)}
     {\delta I_\twoD \over \delta a_\nu^\ext(y)}  \cr}
  \eqn\JJgeneralCorrelation $$
In the $a^\ext_\mu(x)=0$ limit one has
$$\eqalign{
\la T &[j^\mu (x) j^\nu(y) ] \ra_{a_\ext=0}
 -  \delta^{\mu 0} \delta^{\nu 0} ~ n_e^2 \cr
\noalign{\kern 4pt}
&=- i {\delta^2 I_\twoD \over \delta a_\mu^\ext(x)
   \delta a_\nu^\ext (y) } ~\bigg|_{a_\ext=0}
 -i (1-\delta^{\mu 0}) \delta^{\mu\nu} ~  {n_e\over m}  ~ \delta(x-y)~~.  \cr}
  \eqn\JJzeroCorrelation $$

Formulas for a neutral anyon fluid are obtained from the expressions above,
dropping electromagnetic fields $A_\mu$ entirely.  The amplitude becomes
$$\eqalign{
\exp \Big( \, &i I_{\rm n} [a^\ext] \,\Big)
      = \int  \D a \D\psi^\dagger \D\psi \cr
&\times \exp \bigg\{ i\int dx ~ \Big( \L_0^\CS [a]
  + \L_{\rm g.f.}^\CS [a] + \L_f[\psi;a+a^\ext] \Big) \,\bigg\} ~. \cr}
   \eqn\amplitudeNeutral $$
The induced current and correlation function are given by
$$\eqalign{
& \indJ^\mu(x; a^\ext)^{\rm neutral}
= - {\delta I_{\rm n} \over \delta a_\mu^\ext (x) }
  - n_e \, \delta^{\mu 0} \cr
\noalign{\kern 4pt}
&\la T[j^\mu(x) j^\nu(y) ] \ra_{a^\ext=0}^{\rm neutral}
- \delta^{\mu 0} \delta^{\nu 0} ~ n_e^2  \cr
\noalign{\kern 4pt}
&\hskip 1.cm =-i {\delta^2 I_{\rm n} \over \delta a_\mu^\ext (x)
   \delta a_\nu^\ext (y) } ~\bigg|_{a_\ext =0}
-i (1-\delta^{\mu 0}) \delta^{\mu\nu} ~  {n_e\over m}  ~ \delta(x-y) ~~.  \cr}
  \eqn\NeutralFormula   $$

The first equation of (\NeutralFormula) is related to
(\SCFcurrent) in SCF, whereas the second equation gives the correlation
function (\defineDn) in RPA.  Both are derived from the effective action
$I_{\rm n}[a^\ext]$ or $I_\twoD[a^\ext]$ for a neutral or charged anyon fluid,
respectively.  The connection between RPA and SCF is established by
evaluating the effective action.

As for gauge-fixing, it is most convenient to take
$$\eqalign{
\L_{\rm g.f.}^\CS[\,a\,]&= {1\over 2\alpha} \, (\nabla_k a^k )^2   \cr
\L_{\rm g.f.}^\EM [A] &= {1\over 2} \, (\d_\mu A^\mu)^2  \cr}
   \eqn\gaugeFix  $$
$\L_{\rm g.f.}^\EM [A]$ gives the standard Lorentz covariant Feynman gauge
for electromagnetic fields.  We have retained the gauge parameter $\alpha$
for Chern-Simons fields.    For $\alpha=1$,  $\L_{\rm g.f.}^\CS[\,a\,]$ gives
 a spatial Feynman gauge.  In the
$\alpha=0$ limit it reproduces the radiation gauge  $\nabla_k a^k =0$.

For Chern-Simons gauge fields we have
$$\eqalign{
\L_0^\CS [a] + \L_{\rm g.f.}^\CS [a]
 &= {N\over 2\pi} \, a_0^{(1)} b^{(0)}
   -{N\over 4\pi} \, \eps \, a_\mu^{(1)} \d_\nu a_\rho^{(1)}
  + {1\over 2\alpha} \, (\nabla_k a^{(1)k} )^2  \cr
\noalign{\kern 5pt}
&= n_e \, a_0^{(1)}
 +{1\over 2} \, a^{(1)}_\mu \, ( \Lambda_0
 + \Lambda_{\rm g.f.} )^{\mu\nu} \, a^{(1)}_\nu  \cr}
   \eqn\CSeffective  $$
where the kernels $\Lambda_0$ and $\Lambda_{\rm g.f.}$ are given by
$$
\Lambda_0 = - {N\over 2\pi}
    \left(  \matrix{ 0 & -\d_2 &+\d_1 \cr
                    +\d_2 & 0 & -\d_0 \cr
                     -\d_1 & +\d_0 & 0 \cr} \right) \next
\Lambda_{\rm g.f.} = -{1\over \alpha}
 \left( \matrix{ 0 & 0 & 0 \cr
                 0 & \d_1^2 & \d_1\d_2 \cr
                 0 & \d_1\d_2 & \d_2^2 \cr}  \right) ~~. \eqn\DefLambda $$
Note that $a_\mu^{(1)}= (a_0^{(1)} , a_1^{(1)} , a_2^{(1)})
 = (a^{(1)0}, -a^{(1)1} , -a^{(1)2})$.  Integration by parts has been made
in the above formulas.
The Chern-Simons term itself is singular, as $det \,\Lambda_0 =0$.  However
the total kernel is regular:
$$\eqalign{
\Lambda &\equiv \Lambda_0 + \Lambda_{\rm g.f.} \cr
det~\Lambda
  &= - {1\over \alpha} \, \Big( {N\over 2\pi} \Big)^2 ~ (\nabla^2)^2
    \not=  0  ~. \cr}
    \eqn\detLambda $$
For electromagnetic fields we have
$$\L_0^\EM [A] + \L_{\rm g.f.}^\EM [A]
 = {1\over 2} \, A_\mu \, g^{\mu\nu} \d^2 \, A_\nu + en_e A_0
   \eqn\EMeffective  $$
where the metric is $g^{\mu\nu} = {\rm diag} \, (1,-1,-1)$.

The integration over various fields in the formula (\amplitudeI) can be
done in any order.  Integrating first over the Chern-Simons fields $a_\mu(x)$
in the limit $\alpha=0$
is equivalent to eliminating them to get the Hamiltonian (\HamOne)
or (\twoDHam).   An  alternative is to integrate the matter fields $\psi(x)$
and
$\psi^\dagger(x)$ first, maintaining the symmetry of the CS and EM gauge
couplings.

To see that the Hamiltonian (\HamOne) is reproduced in the neutral case
by integrating $a_\mu$, we write
$$
{1\over 2} \, a^{(1)}_\mu  \Lambda^{\mu\nu}  a^{(1)}_\nu  + \L_f [\psi;a]
={1\over 2} \, a^{(1)}_\mu  ( \Lambda + \Xi )^{\mu\nu}  a^{(1)}_\nu
  + a_\mu^{(1)} \bar j^\mu + \L_f [\psi; a^{(0)}]
  \eqn\effectiveCSlag $$
where
$$\eqalign{
\Xi^{\mu\nu} &=  - (1-\delta^{\mu 0})
   \delta^{\mu\nu} {1\over m} \, \psi^\dagger \psi  \cr
\bar j^0 ~&= j^0 = \psi^\dagger \psi \cr
\bar j^k ~&= - {i\over 2m} \big\{ \psi^\dagger \Dbar_k \psi - (\Dbar_k
         \psi)^\dagger \psi \big\}  ~~~.\cr}
  \eqn\effectiveLambda   $$
Further we note that
$$\eqalign{
 \Lambda^{-1} &= {2\pi\over N} {1\over \nabla^2}
 \left(  \matrix{
         0 & +\d_2 & -\d_1 \cr
       -\d_2 & 0 & 0 \cr
        +\d_1 & 0 & 0 \cr} \right)
         + {\rm O}(\alpha)   \cr
\noalign{\kern 6pt}
(\Lambda + \Xi)^{-1} &= \Lambda^{-1} + \Lambda^{-1} \Xi \Lambda^{-1}
 + \Lambda^{-1} \Xi \Lambda^{-1} \Xi \Lambda^{-1} + \cdots \cr}
   \eqn\usefulLambda $$
In the $\alpha \go 0$ limit
$$\eqalign{
&\Lambda^{-1} \Xi \Lambda^{-1} \big|_{\alpha=0} =
 - \Big( {2\pi\over N} \Big)^2
   {1\over m} ~ {\d_k \over \nabla^2}~ j^0 {\d_k \over \nabla^2} ~
   \left( \matrix{ 1&0&0\cr 0&0&0\cr 0&0&0\cr } \right)  \cr
&\Lambda^{-1} \Xi \Lambda^{-1} \Xi \Lambda^{-1} \big|_{\alpha=0}
  =0 \hskip 5cm  {\rm etc.}  \cr} $$
so that the integration over $a_\mu^{(1)}$ yields a Lagrangian
$$\eqalign{
\L_f&[\psi;a^{(0)}] -
{1\over 2}\,\bar j^\mu ~(\Lambda+\Xi)^{-1}_{\mu\nu} ~ \bar j^\nu \cr
\noalign{\kern 5pt}
&= {1\over 2m} \Big( {2\pi\over N} \Big)^2  j^0 \cdot
 {\d_k \over \nabla^2} ~ j^0 \cdot {\d_k \over \nabla^2} ~ j^0
- {2\pi \over N} ~\ep^{kl} ~ \bar j^k ~ {\d_l \over \nabla^2} ~ j^0 ~~~. \cr}
   \eqn\almostRadiation $$
Noticing that
$$\int d\y\, h_j(\x-\y) j^0(y)
    = - {2\pi \over N} \ep^{jk} ~ {\d_k \over \nabla^2}\, j^0 (x)  ~~~, $$
we observe that (\almostRadiation) yields  the same Hamiltonian as in
(\HamOne).   The path integral formalism is equivalent to the operator
formalism.
\sectionskip

\secno=10 \meqno=1

\line{\bf 10.  RPA = linearized SCF  \hfil}
\vglue 5pt
In practice it is more convenient  to integrate the fermion fields $\psi$
and $\psi^\dagger$ first to evaluate $I_{\rm n}[a^\ext]$ in (\NeutralFormula)
or  $I_\twoD[a^\ext]$ in (\amplitudeI).   Let us define
$$\exp \big\{  iS_f[a] \big\} = \int \D \psi^\dagger \D\psi \,
 \exp \bigg\{ i \int dx \, \L_f[\psi; a] \bigg\} ~~.  \eqn\defineSf  $$
It gives the transition amplitude for the fermion fields in the presence
of gauge fields $a_\mu(x)$.  Then we have
$$\eqalign{
\exp \big\{ i I_{\rm n}[a^\ext] \big\} &= \int \D a ~
 \exp \bigg\{ i S_f[a+a^\ext] +
     i \int dx ~ \big( \L_0^\CS[a] + \L_{\rm g.f.}^\CS[a] \big) \bigg\} \cr
\exp \big\{ i I_\twoD[a^\ext] \big\} &= \int \D A  ~
    \exp \bigg\{ iI_{\rm n}[\,eA + a^\ext]  \cr
&\hskip 3cm + i \int dx ~
   \big( \L_0^\EM[A] + \L_{\rm g.f.}^\EM[A] \big) \bigg\} \cr}
   \eqn\formulaIone $$

To evaluate $S_f[a]$ we decompose $\L_f[\psi; a]$
into the zeroth order and interaction parts:
$$\L_f[\psi; a] = \L_f[\psi; a^{(0)}]   + \L_f^{int}  ~~.
    \eqn\decomposeM  $$
Here $a_\mu^{(0)}(x)$ is the average field configuration given in
(\averagePotential) and
$$\eqalign{
\L_f^{int} = &- a_0^{(1)} \psi^\dagger \psi \cr
& -a^{(1)k} \, {i\over 2m}
\big\{ \psi^\dagger  \Dbar_k \psi -  (\Dbar_k \psi)^\dagger \psi \big\} \cr
& - \big( a^{(1)k} \big)^2  \, {1\over 2m} \, \psi^\dagger \psi   \cr}
  \eqn\vertexM  $$
where $a_\mu^{(1)} = a_\mu - a_\mu^{(0)}$.

$\L_f[\psi; a^{(0)}]$ defines a propagator of the $\psi$ field in the
background potential $a_\mu^{(0)}$, whereas $\L_f^{int}$ defines
interaction vertices containing
$a_\mu^{(1)}(x)$.   The first and second terms in (\vertexM)
give one gauge field leg, whereas the last term gives two legs,
as  depicted in Fig. 7.

$S_f[a]$ is nothing but the effective action for $a_\mu(x)$ generated by
dynamics of $\psi$ and $\psi^\dagger$ fields.  The standard diagram technique
 can be employed.  Since $\psi$ and $\psi^\dagger$ are integrated,
fermion lines must be closed.  $S_f[a]-S_f[a^{(0)}]$ is the sum of
connected diagrams.  Further, since the interaction $\L_f^{int}$ is
bilinear in $\psi$ and $\psi^\dagger$, diagrams thus generated are all
one-loop.  One can arrange them according to the number of legs of gauge
field  $a_\mu^{(1)}$ as in Fig. 8.

{\figure
\vglue 5cm
\centerline{Fig. 7. ~
  \vtop{ \hsize=9cm \noindent
  Vertices generated by {\smallL}$_{\fivemit f}^{\fivemit int}$ in (\vertexM).
``0'' and ``k''   at the ends of dashed lines indicate
{\eightit a}$_{\fiverm 0}^{\fiverm (1)}$ and
{\eightit a}$^{\fiverm (1)k}$,  respectively. } }
}
\bigskip

{\figure
\vglue 8cm
\centerline{Fig. 8.  The effective action
    {\eightit S}$_{\fivemit f}$[{\eightit a}]. }
}
\bigskip

Contributions coming from diagrams (a) and (b) in Fig. 8 are easy to evaluate.
$$\eqalign{
\hbox{diagram (a)}
 &= - a_0^{(1)} \, \la \psi^\dagger \psi \ra = - a_0^{(1)} \, n_e \cr
\hbox{diagram (b)} &= - {1\over 2m} \big( a^{(1)k} \big)^2 \, n_e ~~. \cr}
   \eqn\SfOne  $$
Note that in the nonrelativistic system under consideration we always have
$\la \psi^\dagger \psi (x) \ra$, instead of $\lim_{x\go y} (-1)\la T[\psi(x)
\psi^\dagger(y) ]\ra$, in the first-order perturbation.  A similar diagram
containing the second vertex in (\vertexM) vanishes, since
$\la\, \bar j^k \,\ra =0$.

In general one has an expansion
$$\eqalign{
S_f[a] = S_f &[a^{(0)}]  - \int dx \, n_e \, a_0^{(1)}(x)   \cr
&+  \int dx dy ~ {1\over 2} \, a_\mu^{(1)}(x) \, \Gamma^{\mu\nu} (x-y)
\, a_\nu^{(1)}(y)  + \cdots ~~.  \cr}
  \eqn\SfExpansion  $$
In writing the $\Gamma^{\mu\nu}$ term,
 we have employed the translation invariance of the zeroth-order system
described by $\L_f[\psi, a^{(0)}]$.  In the momentum space it becomes
$$\int {d\omega d\q \over (2\pi)^3} ~
{1\over 2} \, a^{(1)}_\mu (-\omega,-\q) \, \Gamma^{\mu\nu} (\omega,\q) \,
a^{(1)}_\nu (\omega,\q) ~~.  \eqn\momentumGamma  $$
Contributions from diagrams (c), (d), and (e) to $\Gamma^{\mu\nu}$ will be
evaluated in the following sections.

The next step is to integrate over Chern-Simons fields $a_\mu(x)$ in
(\formulaIone).  Again a diagram method may be developed for the integral.
Higher order terms in (\SfExpansion), namely terms involving
three or more $a_\mu^{(1)}$'s, give ``interaction'' vertices.

By droping all these higher order terms the system is linearlized.
We call it  ``{\it the linear approximation}''.  The resulting integral is a
simple Gaussian integral, whose   evaluation is straightforward.
As we shall see shortly,
RPA and the linearlized SCF are nothing but the linear approximation.

Let us consider a neutral anyon fluid.  We observe
$$\eqalign{
S_f&[a+a^\ext] + \int dx \, \big( \L_0^\CS[a] + \L_{\rm g.f.}^\CS[a] \big) \cr
&= S_f[a^{(0)}] - \int n_e \, (a_0^{(1)} + a_0^\ext)
 + \int {1\over 2} (a_\mu^{(1)} +a_\mu^\ext ) \Gamma^{\mu\nu}
   (a_\nu^{(1)} +a_\nu^\ext ) + \cdots \cr
&\hskip 2.cm + \Big( \int n_e \, a_0^{(1)}
 +\int {1\over 2} \, a^{(1)}_\mu \,  \Lambda^{\mu\nu} \, a^{(1)}_\nu \Big) \cr
&=  S_f[a^{(0)}] - \int n_e \,  a_0^\ext  \cr
&\hskip .5cm  + \int
{1\over 2} \big\{ a^{(1)} + a^\ext \,\Gamma \, (\Lambda+ \Gamma)^{-1} \big\}
\, (\Lambda+\Gamma) \, \big\{  a^{(1)}
    + (\Lambda+ \Gamma)^{-1} \Gamma \, a^\ext \big\}  \cr
&\hskip .5cm - \int
{1\over 2} a^\ext \,\Gamma \,(\Lambda+ \Gamma)^{-1} \Gamma \,a^\ext
 + \int{1\over 2} a^\ext \,\Gamma \, a^\ext  ~~~.  \cr}  \eqn\aIntegralOne  $$
We have suppressed a measure $dx$ or $d\omega d\q$ in the
experssion. The integration over $a_\mu^{(1)}$ immediately leads to
$$
I_{\rm n} [a^\ext] =
- \int dx\, n_e \,  a_0^\ext(x)
- \int dxdy \, {1\over 2} a_\mu^\ext(x) \, Q_{\rm n}^{\mu\nu}(x-y) \,
    a_\nu^\ext(y) + \cdots
   \eqn\finalIn   $$
where
$$\eqalign{
\Qn &=   \Gamma \, (\Lambda+ \Gamma)^{-1} \Gamma -\Gamma  \cr
\noalign{\kern 3pt}
&= - \,\Gamma \, {1\over 1+ \Lambda^{-1} \Gamma }
= - \,{1\over 1+\Gamma \, \Lambda^{-1} } \, \Gamma  \cr
\noalign{\kern 5pt}
&= - \,\Gamma + \Gamma \, \Lambda^{-1} \Gamma
 - \Gamma \, \Lambda^{-1} \Gamma  \, \Lambda^{-1} \Gamma + \cdots \cr}
   \eqn\QnSeries  $$
$\Lambda^{-1}$ represents a propagator for $a_\mu^{(1)}$.  Therefore
$\Qn$ has a diagram representation given in Fig. 9.

{\figure
\vglue 8cm
\centerline{Fig. 9. ~
\vtop{ \hsize=10.5cm \noindent
Diagrams for {\eightit Q}$_{\fivemit n}$.  Diagrams for
{\eightit Q}$_{\fivemit c}$ are obtained by
replacing each Chern-Simons field propagator (dashed line) by the sum of
propagators of  Chern-Simons fields and electromagnetic fields.   Compare
(\QnSeries) and (10.20). } }
}
\bigskip

{}From (\NeutralFormula)  the $jj$-correlation function is given by
$$\eqalign{
{1\over i} \Big( \, \la &T[j^\mu(x) j^\nu(y) ] \ra_{a^\ext=0}^{\rm neutral}
- \delta^{\mu 0} \delta^{\nu 0} ~ n_e^2 \, \Big) \cr
&= Q_{\rm n}^{\mu\nu}(x-y) -
 (1-\delta^{\mu 0}) \delta^{\mu\nu} ~  {n_e\over m}  ~ \delta(x-y)
  + \cdots ~~.  \cr}  \eqn\jjNeutralOne $$
We recall that the contribution of diagram (b) in Fig. 8 to $\Gamma$ is
given by (\SfOne) so that
$$- \Gamma^{\rm (b)}_{jk} = \delta^{jk} \,{n_e \over m} \, \delta(x-y)~~,
   \eqn\diagramB $$
which is precisely the negative of the last term in (\jjNeutralOne).  In a
normal metal this is the end of  cancellation.  In anyon fluids something
special happens.  As shall be shown in  section 12, diagram (e) in Fig. 8,
in part, yields the same contribution as (\diagramB) with the opposite sign.
Hence the last term in (\jjNeutralOne) survives.

The series generated in (\jjNeutralOne) with (\QnSeries) substituted
exactly reproduces the diagrams  in Fig. 6 in Section 8.
Hence
$$\eqalign{
D_{\rm n}^{\mu\nu}(x,y)^{\rm RPA}
&= D_{\rm n}^{\mu\nu}(x,y)^{\rm linear}  \cr
&=  Q_{\rm n}^{\mu\nu}(x-y) -
 (1-\delta^{\mu 0}) \delta^{\mu\nu} ~  {n_e\over m}  ~ \delta(x-y) ~~.  \cr}
   \eqn\DnRPA  $$
RPA is equivalent to droping, in $I_{\rm n}[a^\ext]$, all terms
cubic or higher-order in $a^\ext_\mu(x)$.

How about SCF?  We return to the expression (\SCFcurrent).  In SCF one
determines the current in the presence of non-trivial gauge fields
$a_\mu(x)$, which is nothing but evaluating $S_f[a]$ in (\defineSf).   Since
$j^\mu(x) = - \delta \int \L_f[\psi;a]  / \delta a_\mu(x)$,  one immediately
finds  $$\eqalign{
J^\mu[x;a]^{\rm SCF} &= e^{-iS_f[a]} \, \int \D\psi^\dagger \D\psi ~
  j^\mu(x) \, e^{i\int dx \L_f}  \cr
&= - {\delta S_f[a] \over \delta a_\mu^{(1)}(x) } \cr
&= n_e \, \delta^{\mu 0} - \int dy ~ \Gamma^{\mu\nu}(x-y) \, a_\nu^{(1)}(y)
 + \cdots ~~.  \cr}  \eqn\SCFandGamma$$
We recognize that $\Gamma^{\mu\nu}$ appearing in (\SCFcurrent) is the same
as $\Gamma^{\mu\nu}$ defined in (\SfExpansion).

In SCF the current thus obtained is inserted into the Chern-Simons field
equation (\EulerEq), which is then solved to determine a self-consistent
nontrivial field configuration.  If higher-order terms in (\SCFandGamma)
are dropped, the resulting field equation becomes linear.  Solving the
equation is equivalent to
performing a Gaussian integral in the path integral formalism, since
the latter  amounts to picking a stationary path of the action.

An alternative way of seeing this is to examine a response to an
external field in SCF.  Since $J_\ext =  \Lambda \, a_\ext$, the field
equation in SCF, with the gauge fixing term added,  becomes
$$ \Lambda \, a^{(1)} = J_\ind + J_\ext
  = \big( - \Gamma \, a^{(1)}+ \cdots \big)  + \Lambda\, a_\ext ~~~, $$
from which it follows that
$$\eqalign{
{J_\ind}^{\rm linearized \,SCF}
&= -\Gamma \, (\Lambda+\Gamma)^{-1} \Lambda \, a_\ext
= - \Gamma\, (1+\Lambda^{-1} \Gamma)^{-1} \, a_\ext \cr
&= \Qn \, a_\ext \cr
&= {J_\ind}^{\rm linear}~~.  \cr}   \eqn\responseZero $$
We have stressed in the last equality that
the expression is exactly what one obtains from the first equation of
(\NeutralFormula) combined with (\finalIn).
The relations (\DnRPA) and (\responseZero)
together establish the equivalence between  RPA and the linearized SCF.

The generalization to a charged anyon fluid is easy.   In the two-dimensional
approximation one needs to do one more integration over $A_\mu(x)$ in
(\formulaIone).
With the expressions (\finalIn) and (\EMeffective) inserted,
the exponent of  the integrand becomes
$$\eqalign{
- n_e (e&A_0 + a_0^\ext)
 - {1\over 2} ~ (eA + a^\ext) \, \Qn \, (eA + a^\ext) + \cdots
     + \,{1\over 2}\, A \, P \, A + en_e A_0 \cr
\noalign{\kern 5pt}
= &{1\over 2} \,\big\{ A - a^\ext \, e\Qn \,(P - e^2\Qn)^{-1} \big\}
  \, (P - e^2\Qn)
\big\{ A - (P - e^2\Qn)^{-1}\, e\Qn \, a^\ext \big\} \cr
\noalign{\kern 3pt}
& -{1\over 2} \, a^\ext \, e\Qn \, (P - e^2\Qn)^{-1}\, e\Qn \, a^\ext
 - {1\over 2} \, a^\ext\, \Qn \, a^\ext - n_e a_0^\ext + \cdots \cr}
  \eqn\chargeFormulaTwo $$
where
$$P^{\mu\nu} = g^{\mu\nu} \, \d^2  ~~~.
  \eqn\EMpropagatorOne  $$

Therefore we have
$$
I_\twoD [a^\ext] = - \int dx~ n_e a_0^\ext(x)
- \int dx dy~ {1\over 2} \, a^\ext_\mu(x) Q_{\rm c}^{\mu\nu}(x-y)
 a^\ext_\nu(y) + \cdots   \eqn\finalItwoD $$
where
$$\eqalign{
\Qc &=   \Qn ( e^{-2}P - \Qn )^{-1} \Qn + \Qn \cr
\noalign{\kern 3pt}
 &=  {1\over 1-  \Qn \, e^2 P^{-1} } \, \Qn
  = \Qn \, {1\over 1  - e^2 P^{-1} \,\Qn}  \cr
\noalign{\kern 5pt}
&=\Qn + \Qn \, e^2P^{-1}\, \Qn
     + \Qn \, e^2P^{-1}\, \Qn  e^2P^{-1}\, \Qn + \cdots  \cr}
   \eqn\QcSeries $$
Combining (\QnSeries) and (\QcSeries), one finds
$$\eqalign{
\Qc &= - \,\Gamma \, {1\over 1 + (\Lambda^{-1} + e^2 P^{-1})  \Gamma }
=- \, {1\over 1 + \Gamma (\Lambda^{-1} + e^2 P^{-1}) }\, \Gamma \cr
\noalign{\kern 5pt}
&= -\, \Gamma + \Gamma \,\Big({1\over \Lambda} + {e^2\over P} \Big) \,\Gamma
 - \Gamma \,\Big({1\over \Lambda} + {e^2\over P} \Big)\, \Gamma \,
\Big({1\over \Lambda} + {e^2\over P} \Big) \,\Gamma + \cdots ~. \cr}
  \eqn\QcTaotalSeries  $$
The proper vertex $-\Gamma$, which summarizes one-loop fermion interactions,
is connected to the next one
by a propagator of either Chern-Simons fields ($\Lambda^{-1}$)
or electromagnetic fields ($e^2P^{-1}$).  There is only one proper vertex,
since both Chern-Simons and electromagnetic fields minimally couple to
the fermions.   The final expression for $\Qc$ above is obvious
from the viewpoint of the diagram method.  (See Fig. 9.)

To summarize, in the linear approximation, which is equivalent to RPA and
the linearized SCF,
$$\eqalign{
D_{\rm c}^{\mu\nu}(x,y)^{\rm linear}
&=  Q_{\rm c}^{\mu\nu}(x-y) -
 (1-\delta^{\mu 0}) \delta^{\mu\nu} ~  {n_e\over m}  ~ \delta(x-y)\cr
{J_\ind}^{\rm linear} &= \Qc \, a^\ext ~~~.  \cr}
  \eqn\linearCharged  $$

\sectionskip


\secno=11  \meqno=1
\line{\bf 11.  Response function \hfil}
\vglue 5pt
As it stands from (\DnRPA), (\responseZero), and (\linearCharged),
$\Qn$ or $\Qc$ determines a linear response to an external perturbation.
It defines a response function.  Location of poles in the response
function $\Qn$ or $\Qc$ gives an energy spectrum of particle-hole
excitations.  In passing, location of poles in the fermion propagator
yields a spectrum for fermionic excitations, which is quite different
from that of particle-hole excitations in anyon fluids under consideration.

One can also examine a response
to an external magnetic field in a charged anyon fluid, from which
the exsitence or non-existence of a Meissner effect is checked.  Conductivity
or resistivity tensors can be computed from $\Qn$ or $\Qc$.  Examinig
a response to external density perturbation $J^0_\ext(x)$ gives information
on muon spin relaxation.  Many other consequences can be drawn from
$\Qn$ or $\Qc$.

We have seen in the previous section that $\Qn$ and $\Qc$ are determined
by the proper vertex $\Gamma$ defined in (\SfExpansion).
There are many restrictions resulting from the definition (\SfExpansion) and
the conservation law.

First, from the definition we have
$\Gamma^{\mu\nu}(x-y) = \Gamma^{\nu\mu}(y-x)$, or
$$\Gamma^{\mu\nu}(\omega,\q) = \Gamma^{\nu\mu}(-\omega,-\q) ~~.
  \eqn\GammaOne  $$
Secondly, the gauge invariance or current conservation implies
$\d_\mu^x \Gamma^{\mu\nu}(x-y) = 0$ and  $\d_\nu^y \Gamma^{\mu\nu}(x-y)=0$, or
$$q_\mu \, \Gamma^{\mu\nu}(\omega,\q) =0
 = q_\nu \,\Gamma^{\mu\nu}(\omega,\q)
  \eqn\GammaTwo $$
where $q_\mu = (\omega, -\q)$.  Thirdly, the rotational invariance implies
that
$$\eqalign{
\Gamma^{00} &= A\cr
\Gamma^{0j} &= q_j B + \ep_{jk} q_k C \cr
\Gamma^{j0} &= q_j B' + \ep_{jk} q_k C' \cr
\Gamma^{jk} &= \delta_{jk} D + \ep_{jk} E + q_jq_k F \cr}
   \eqn\GammaThree  $$
where $A \sim F$ are functions of $\omega$ and $q= |\q|$.

Relations (\GammaOne), (\GammaTwo), and (\GammaThree) lead to a
decomposition\myref{\FHL,\RDSS}
$$\eqalign{
\Gamma^{00}(\omega,\q) &= q^2 \Pi_0 \cr
\Gamma^{0j}(\omega,\q) &= \omega q_j \Pi_0 - i \ep_{jk} q_k \Pi_1 \cr
\Gamma^{j0}(\omega,\q) &= \omega q_j \Pi_0 + i \ep_{jk} q_k \Pi_1  \cr
\Gamma^{jk}(\omega,\q) &= \delta_{jk} \,\omega^2 \Pi_0
  + i \ep_{jk}\omega \Pi_1  - (q^2 \delta_{jk} -q_jq_k)\, \Pi_2 \cr}
  \eqn\GammaDecomposition $$
where all $\Pi_j$'s are functions of $\omega^2$ and $\q^2$ only.  If the
perturbative ground state is stable, $S_f[a]$ defined in (\defineSf) is real,
and therefore from (\SfExpansion) $\Gamma^{\mu\nu}(x)^*=\Gamma^{\mu\nu}(x)$,
or $\Gamma^{\mu\nu}(-\omega, -\q)^*= \Gamma^{\mu\nu}(\omega,\q)$.  In other
words, $\Pi_k$'s ($k=0,1,2$) are real:   $\Pi^*_k=\Pi^{}_k$.
In a frame $\q= (q, 0)$,
$$\Gamma^{\mu\nu} = \left( \matrix{
q^2 \Pi_0 & \omega q \Pi_0 & +iq \Pi_1 \cr
\noalign{\kern 4pt}
\omega q \Pi_0 & \omega^2 \Pi_0 & +i \omega \Pi_1 \cr
\noalign{\kern 4pt}
- iq \Pi_1 & -i \omega \Pi_1 & \omega^2 \Pi_0 - q^2 \Pi_2 \cr}  \right) ~~.
  \eqn\GammaMatrix $$

To evaluate $\Qn$, we recall Eq. (\QnSeries):
$$\eqalign{
\Qn &= \Gamma \, (\Lambda + \Gamma)^{-1} \, \Gamma - \Gamma  \cr
&= - \Lambda \,  (\Lambda + \Gamma)^{-1} \, \Gamma \cr
&= - \Gamma \,  (\Lambda + \Gamma)^{-1} \, \Lambda ~~. \cr}
        \eqn\evaluateQn  $$
Take a frame in which $\q=(q,0)$.  $\Gamma$ is given by (\GammaMatrix), and
$$\Lambda =
 - {N\over 2\pi}
    \left(  \matrix{ 0 & 0 & iq \cr
                    0 & 0 & i\omega \cr
                  -iq & -i\omega & 0 \cr} \right)
 +{q^2\over \alpha}
 \left( \matrix{ 0 & 0 & 0 \cr
                 0 & 1 & 0\cr
                 0 & 0 & 0 \cr}  \right) ~~. \eqn\NewLambda $$
Combinig (\GammaMatrix) and (\NewLambda), one finds
$$\eqalign{
\Lambda + \Gamma =&
 \left( \matrix{
q^2 \Pi_0 & \omega q \Pi_0 & +iq \Pibar_1 \cr
\noalign{\kern 4pt}
\omega q \Pi_0 & \omega^2 \Pi_0 + \alpha^{-1} q^2 & +i \omega \Pibar_1 \cr
\noalign{\kern 4pt}
- iq \Pibar_1 & -i \omega \Pibar_1 &  \Pibar_2 \cr}  \right) ~~,\cr
\noalign{\kern 10pt}
&\Pibar_1 = \Pi_1 - {N\over 2\pi}  ~~~, \cr
&\Pibar_2 = \omega^2 \Pi_0 - q^2 \Pi_2 ~~~. \cr}
    \eqn\LamGam $$

We need to evaluate $(\Lambda + \Gamma)^{-1}$.  First we note
$$det ~ (\Lambda + \Gamma) = {q^4\over \alpha} \,
 (\Pi_0 \Pibar_2 - \squarePibar ) ~~~.  \eqn\detLamGam  $$
A straightforward manipulation leads to
$$\eqalign{
&(\Lambda + \Gamma)^{-1} = \cr
& {1\over q^2  (\Pi_0 \Pibar_2 - \squarePibar ) } ~
 \left( \matrix{
 \Pibar_2 & 0 & -iq \Pibar_1 \cr \noalign{\kern 3pt}
 0 & 0 & 0 \cr \noalign{\kern 3pt}
 +iq \Pibar_1 & 0 &  q^2 \Pi_0 \cr}  \right)
+{\alpha\over q^4} \left( \matrix{
\omega^2 & - \omega q  & 0 \cr \noalign{\kern 3pt}
-\omega q  & q^2  & 0 \cr \noalign{\kern 3pt}
0 & 0 &  0 \cr}  \right) ~ .  \cr}
          \eqn\inverseLamGam $$
Both $\Lambda$ and $(\Lambda + \Gamma)^{-1}$ depend on $\alpha$, but
the product $\Lambda\,(\Lambda + \Gamma)^{-1}$
or $(\Lambda + \Gamma)^{-1}\Lambda$ is independent of $\alpha$.
(We shall see shortly that the current conservation guarantees the
$\alpha$-independence.)  $\Qn$ is given by
$$\eqalign{
\Qn &=
\Big( {N\over 2\pi} \Big)^2 \,
   {1\over \Pi_0 \Pibar_2 - \squarePibar } ~
 \left( \matrix{
q^2 \Pi_0 & \omega q \Pi_0 & -iq \Pibar_1 \cr
\noalign{\kern 4pt}
\omega q \Pi_0 & \omega^2 \Pi_0 & -i \omega \Pibar_1 \cr
\noalign{\kern 4pt}
+ iq \Pibar_1 & +i \omega \Pibar_1 & \Pibar_2         \cr}  \right) \cr
\noalign{\kern 8pt}
&\hskip 3.8cm
 - {N\over 2\pi} ~ \left( \matrix{
0 & 0 & -iq \cr \noalign{\kern 6pt}
0 & 0 & -i\omega \cr \noalign{\kern 6pt}
+iq & +i\omega &  0 \cr}  \right) ~~ . \cr}  \eqn\finalQn   $$
This expression was first given by Aronov and Mirlin.\myref{\Aronov}

For a charged fluid
it is easiest and most convenient to evaluate $\Qc$ in the form
(\QcTaotalSeries):
$$
\eqalign{
\Qc &= - \Gamma (1+ \Lambda_c^{-1} \Gamma )^{-1}
  = - (1+ \Gamma\, \Lambda_c^{-1} )^{-1} \Gamma ~~,\cr
{1\over \,\Lambda_c} &= {1\over \Lambda} + {e^2\over \, P\,}  \cr}
     \eqn\QcNewFormula $$
Here $P^{-1}$ is the propagator for electromagnetic fields.  In the
two-dimensional approximation it is given by (\EMpropagatorOne).
Its form in the ultra-thin film approximation can be deduced from
(\filmLag).
$$
{e^2\over \,P\, }= \cases{
\hskip .6cm \big {e^2_{\rm 3D}\over q} ~\left( \matrix{
1 & 0 & 0 \cr
0 & 0 & 0 \cr
0 & 0 & 0 \cr}  \right) &(ultra-thin film approx.) \cr
\cr\cr
\big {e^2\over q^2-\omega^2} ~\left( \matrix{
1 & ~0 & ~0 \cr
0 & -1 & ~0 \cr
0 & ~0 & -1 \cr}  \right) &(two-dimensional approx.)\cr}
   \eqn\EMpropagatorP  $$
Here $e_{\rm 3D}$ is the coupling constant in three dimensions.

We evaluate $\Qc$ in the two-dimensional approximation.    We introduce
$$ A= {e^2\over q^2 - \omega^2}  \next B= {2\pi\over N}\,{1\over q} ~~~.
    \eqn\ABdefinition $$
Then
$$\Lambda_c^{-1} =
\left( \matrix{
A & 0 & -iB \cr
0 & -A & 0 \cr
iB & 0 & -A \cr}  \right)
+ {\alpha\over q^4} \, \left( \matrix{
\omega^2 & -\omega q & 0 \cr
-\omega q  & q^2 & 0 \cr
0 & 0 & 0 \cr}  \right) ~~.  \eqn\LamCinverse $$

The substitution of (\GammaMatrix) and (\LamCinverse) into (\QcNewFormula)
immediately confirms that the $\alpha$-dependence entirely drops in $\Qc$
(or $\Qn$),  since both $(\Gamma^{0\nu},\Gamma^{1\nu})$ and
 $(\Gamma^{\mu 0},\Gamma^{\mu 1})$ are proportional to
$(q,\omega)$.  It is a consequence of the current conservation
(\GammaTwo).

The computation of $\Qc$ is rather involved.  We record intermediate steps
for readers' convenience.  $\Lambda_c^{-1} \Gamma$ is found to be
$$\Lambda_c^{-1} \Gamma =
\left( \matrix{
q(q A \Pi_0 - B\Pi_1) & \omega (q A\Pi_0 - B\Pi_1)& i(qA\Pi_1 - B\Pibar_2) \cr
\cr
-\omega q A\Pi_0 & -\omega^2 A\Pi_0 & -i\omega  A\Pi_1 \cr
\cr
iq(qB\Pi_0 + A\Pi_1) & i\omega(qB\Pi_0 + A\Pi_1) & -qB\Pi_1-A\Pibar_2 \cr}
\right) ~~. \eqn\LambdacGamma $$
Further the determinant of $(1+\Lambda_c^{-1} \Gamma)$ is evaluated to be
$$\eqalign{
\Delta_c &\equiv det ~(1+ \Lambda_c^{-1} \Gamma) \cr
\noalign{\kern 6pt}
&= 1 + (q^2-\omega^2) A \Pi_0 - 2qB \Pi_1 - A \Pibar_2 \cr
&\hskip 3cm + \big\{ q^2 (A^2+B^2) -\omega^2 A^2 \big\}
    ( \Pi_1^2 - \Pi_0 \Pibar_2) \,    \cr
\noalign{\kern 6pt}
&=\Big( {2\pi\over N} \Big)^2 (\squarePibar - \Pi_0 \Pibar_2 )
+ e^2 \Pi_0 -{e^2\over q^2-\omega^2} \Pibar_2
  +{e^4\over q^2-\omega^2}  (\Pi_1^2 - \Pi_0 \Pibar_2) \cr}
   \eqn\defineDeltaC $$

Straightforward manipulations lead to
$$\eqalign{
 (1+ \Lambda_c^{-1} &\Gamma)^{-1}
= {1\over ~\Delta_c} \cdot \big( 1 + \hat M_1 + \hat M_2 \big) \cr
\noalign{\kern 12pt}
\hat M_1 &=
\left( \matrix{
0 & 0 & 0 \cr
0 & 0 & 0 \cr
-iq^2 B & -i\omega q B
              & (q^2-\omega^2) A  \cr}  \right) \Pi_0 \cr
\noalign{\kern 5pt}
&\hskip .5cm + \left( \matrix{
-qB & 0 & -iq A \cr
0 & -qB &+i\omega A\cr
-iq A & -i\omega A & -qB \cr}  \right) \Pi_1
+ \left( \matrix{
-A& 0 & iB \cr
0 & -A & 0 \cr
0 & 0 & 0 \cr}  \right)  \Pibar_2  \cr
\noalign{\kern 12pt}
\hat M_2 &= \left( \matrix{
-\omega^2 & -\omega q & 0 \cr
\omega q & q^2 & 0 \cr
0 & 0 & 0 \cr}  \right) ~ A\Pi_0
+ \left( \matrix{
0 & \omega & 0 \cr
0 & -q & 0 \cr
0 & 0 & 0 \cr}  \right) ~B \Pi_1\cr
\noalign{\kern 5pt}
&\hskip .5cm + \left( \matrix{
-\omega^2 A^2 & -\omega q(A^2+B^2) & i\omega^2 AB \cr
\omega q A^2& q^2(A^2+B^2) & -i\omega q AB \cr
0 & 0 & 0 \cr}  \right) ~ (\Pi_1^2 - \Pi_0 \Pibar_2)  \cr}
   \eqn\MoneMtwo  $$

In computing $\Gamma \, (1+ \Lambda_c^{-1} \Gamma )^{-1}$, the contribution
from the $\hat M_2$ part vanishes thanks to the current conservation
$\Gamma \,\hat M_2 =0$.  The final result takes a simple form.
$$\eqalign{
\Qc &= - {1\over \Delta_c} \, \Bigg\{ ~\Gamma
+\left( \matrix{
q^2 A & \omega qA & -iq^2 B \cr
\omega qA  & \omega^2 A & -i\omega q B \cr
iq^2 B & i\omega q B & (\omega^2-q^2)A \cr}  \right)
   ~ (\Pi_1^2 - \Pi_0 \Pibar_2) ~ \Bigg\} \cr
\noalign{\kern 12pt}
&= - {1\over \Delta_c} \, \Bigg\{ ~\Gamma
+  \Bigg[ ~{e^2\over q^2-\omega^2} \,\left( \matrix{
q^2  & \omega q & 0 \cr
\omega q   & \omega^2  & 0 \cr
0 & 0 & 0 \cr}  \right)
- e^2 \,\left( \matrix{
0 & 0 & 0 \cr
0 & 0 & 0 \cr
0 & 0 & 1 \cr}  \right)  \cr
\noalign{\kern 6pt}
&\hskip 3cm  + {2\pi\over N} \,\left( \matrix{
0 & 0 & -iq \cr
0 & 0 & -i\omega \cr
iq & i\omega & 0 \cr}  \right) ~ \Bigg]
   ~ (\Pi_1^2 - \Pi_0 \Pibar_2) ~ \Bigg\} \cr}
   \eqn\finalQc  $$

It is easy to check that the formula (\finalQc) reduces to (\finalQn) in the
neutral limit $e^2 \go 0$.   Indeed one sees that
$$\eqalign{
&\Delta_c \go \Delta_n =
\Big( {2\pi\over N} \Big)^2 (\squarePibar - \Pi_0 \Pibar_2 )   \cr
\noalign{\kern 5pt}
&\Pi_1 - {2\pi\over N}  \, (\Pi_1^2 - \Pi_0 \Pibar_2 )
= - \Pibar_1 -{N\over 2\pi} \,\Delta_n   ~~~.  \cr}
   \eqn\neutralLimit  $$
As we shall see, the difference between $\Qn$ and
$\Qc$ is important in discussing superconductivity.

\sectionskip

\secno=12  \meqno=1

\line{\bf 12.  Evaluation of the kernel \hfil}
\vglue 5pt
We evaluate the kernel $\Gamma^{\mu\nu}(\omega,\q)$ defined in
(\momentumGamma),
or equivalently the invariant functions $\Pi_a$'s in (\GammaDecomposition).
We need to evaluate the four diagrams (b), (c), (d), and (e) in Fig. 8.
For $\q=(q,0)$ we have
$$\eqalign{
 q^2 \, \Pi_0 &= \Gamma^{00}(\omega,\q)
     = \Gamma^{{\rm (c)}00}(\omega,\q) \cr
iq\, \Pi_1 &= \Gamma^{02}(\omega,\q) = \Gamma^{{\rm (d)}02}(\omega,\q)   \cr
\Pibar_2 &= \Gamma^{22}(\omega,\q) = \Gamma^{{\rm (b,e)}22}(\omega,\q) ~~.\cr}
          \eqn\findPi  $$
The diagram (b) has been already evaluated in (\diagramB):
$$\Gamma^{{\rm (b)}jk} (\omega,\q) = - {n_e\over m} \, \delta^{jk} ~~.
      \eqn\diagramBB $$

To evaluate other diagrams, we first examine the zeroth order propagator
$$\eqalign{
G(x,y) &= -i \la {\rm T} [ \psi(x) \psi^\dagger (y) ] \ra \cr
\bigg( \, i{\d\over \d x_0} + {1\over 2m} \, &\Dsquare ~ \bigg) \, G(x,y) =
\delta^3 (x-y) ~~. \cr}
  \eqn\propagator  $$
Here the expectation value is taken with the ground state (\meanGround) and
$\psi(x)$ satisfies the mean field equation.

More explicitly
$$\eqalign{
\noalign{\kern 5pt}
&G(x,y) = -i \bigg\{ \theta(x_0-y_0) \sum_{\alpha \notin G}
 - \theta(y_0-x_0) \sum_{\alpha \in G} ~ \bigg\} ~
  u_\alpha(\x) u_\alpha(\y)^* e^{-i \ep_\alpha (x_0-y_0)}  \cr
\noalign{\kern 5pt}   }
               \eqn\GreenOne $$
where $u_\alpha(\x)$ is given by either (\symmetricGauge) or (\LandauGauge).
In the Landau gauge
$$\eqalign{
G(x,y) &=-i \bigg\{ \theta(x_0-y_0) \sum_{n=|N|}^\infty
 - \theta(y_0-x_0) \sum_{n=0}^{|N|-1} ~ \bigg\} \cr
\noalign{\kern 5pt}
&\hskip 1.5cm  \times {1\over l L_1} \sum_p e^{-2\pi ip(x_1-y_1)/L_1}~
v_n\Big({x_2-\bar x_2 \over l}\Big) v_n\Big({y_2-\bar y_2 \over l}\Big) \cr
\noalign{\kern 5pt}
&=e^{i \phi(x,y) }  \,\cdot \, G_0(x-y) ~~~~. \cr
\noalign{\kern 5pt}   }
   \eqn\GreenTwo $$
Here
$$\eqalign{
\phi(x,y) &= - \ep(N) \, {1\over 2l^2} \,(x_1-y_1)(x_2+y_2)  \cr
\noalign{\kern 5pt}
G_0(x) &=-i \bigg\{ \theta(x_0) \sum_{n=|N|}^\infty
 - \theta(-x_0) \sum_{n=0}^{|N|-1} ~ \bigg\} ~ e^{-i \ep_n x_0} \cr
\noalign{\kern 5pt}
&\hskip 1.5cm \times {1\over 2\pi l^2}
\int_{-\infty}^\infty dz ~ e^{-izx_1/l} ~
  v_n[z-\bar z(x_2) ] \, v_n [ z+\bar z(x_2) ] \cr
\noalign{\kern 7pt}~~
 &\hskip 3cm \bar z(x_2) =  \ep(N) ~{x_2\over 2l} ~~.  \cr}
    \eqn\GreenThree   $$
We have introduced a new integration vbariable
$z=(2\pi pl/L_1)- \ep(N) (x_2+y_2)/2l$, taking the limit $L_1 \go \infty$.
It should be noticed that the Green's function $G(x,y)$ is not manifestly
translation invariant, but is invariant
up to a gauge transformation, due to the presence of  a non-vanishing
magnetic field.
The Fourier transform of $G_0(x)=G_0(t,\x)$ is given by
$$\eqalign{
G_0&(p) = G_0(\omega,\p)
= \int dt d\x ~ G_0(t,\x) \, e^{i(\omega t - \p \,\x)}  \cr
\noalign{\kern 8pt}
&=\sum_{n=0}^\infty {1\over \omega - \ep_n + i \delta(n)}
{}~ \int {dx_2\over l} ~ e^{-ip_2x_2} ~
 v_n[p_1l+\bar z(x_2)] \, v_n[p_1l - \bar z(x_2)]  \cr
\noalign{\kern 6pt}
&\hskip 4cm  \hbox{where ~~}
  i \delta(n) = \cases{ + i\ep &for $n \ge |N|$\cr
                        - i\ep &for $n < |N|$~. \cr}  \cr}
   \eqn\FourierGreen  $$

We come back to evaluating the remaining diagrams.    Recalling $\L_f^{int}$
in (\vertexM), we see that the diagram (c) in Fig. 8
yields
$$\eqalign{
 {i^2\over 2i} &\int dxdy~ G(x,y) \,[-a_0(y)]\, G(y,x) \, [-a_0(x)] \cr
&= {i\over 2} \int dxdy ~ a_0(x) \, G_0(x-y) \, G_0(y-x) \, a_0(y) \cr}
    \eqn\diagramCone $$
so that
$$\Gamma^{{\rm (c)}00} (q) =
  i \int {d^3p\over (2\pi)^3} ~
  G_0(p) \, G_0(p-q) ~~. \eqn\diagramCtwo $$

The diagram (d) in Fig. 8 yields
$$\eqalign{
&{1\over 2} \int dxdy ~ a_0(x) \, \Gamma^{(d)0j}(x,y) \, a^{(1)j}(y) \cr
&={i^2\over 2i} \int dxdy~ [-a_0(x)] \Big[ - {i\over 2m} a^{(1)j} (y) \Big] \cr
&\hskip .1cm \times \Big( \,G(x,y) \cdot [ \d^y_j - i\bar a^j(y) ] G(y,x)
 - [ \d^y_j + i\bar a^j(y) ] \, G(x,y)
  \cdot G(y,x)   \, \Big) ~. \cr}  \eqn\diagramDone $$
With the aid of
$$\eqalign{
&[ \d^x_j - i\bar a^j(x) ] \, G(x,y)
= + e^{i\phi(x,y) } \, D^-_j G_0(x-y)  \cr
&[ \d^y_j + i\bar a^j(y) ] \, G(x,y)
=  - e^{i\phi(x,y) } \, D^+_j G_0(x-y)  \cr
\noalign{\kern 8pt}
&D^\pm_j G_0(x-y)  = \Big( \d^x_j \mp i \ep(N) \,
\ep^{jk} \, {x_k-y_k\over 2l^2} \Big) \, G_0(x-y)  ~~, \cr}
   \eqn\delGzero $$
eq. (\diagramDone) becomes
$$\eqalign{
&- {1\over 4m} \int dxdy ~a_0(x) \,a^{(1)j}(y) \cr
&\hskip .5cm \times \Big\{ G_0(x-y) \cdot D_j^- G_0(y-x)
  + D^+_j G_0(x-y) \cdot G_0(y-x)  \Big\} ~~. \cr}  \eqn\diagramDtwo $$
Note that $\Gamma^{(d)0j}(x,y)$ is a function of $x-y$ only.
Hence, in the Fourier space
$$
\Gamma^{(d)0j}(q) = -{1\over 2m} \int {d^3p\over (2\pi)^3} ~
\Big\{ G_0(p) \cdot D_j^- G_0(p-q) + D_j^+ G_0(p) \cdot G_0(p-q) \Big\}
  \eqn\diagramDthree $$

Similar manipulations lead to
$$\eqalign{
{1\over 2} \int &dxdy  ~ a^{(1)j}(x) \, \Gamma^{(e)jk}(x,y) \, a^{(1)k}(y) \cr
=- &{i\over 8m^2} \int dxdy ~ a^{(1)j}(x) \, a^{(1)k}(y) \cr
 \Big\{
&D^-_k G_0(y-x) \cdot D^-_j G_0(x-y) + D^+_j G_0(y-x) \cdot D^+_k G_0(x-y) \cr
+&D^+_jD^-_k G_0(y-x) \cdot  G_0(x-y) + G_0(y-x) \cdot  D^-_jD^+_k G_0(x-y)
  \Big\} ~, \cr}  \eqn\diagramEone $$
and therefore
$$\eqalign{
\Gamma^{(e)jk}&(q) = - {i\over 4m^2} \int {d^3p\over (2\pi)^3}\cr
\Big\{
&D^-_k G_0(p) \cdot D^-_j G_0(p-q) + D^+_j G_0(p) \cdot D^+_k G_0(p-q) \cr
+&D^+_jD^-_k G_0(p) \cdot  G_0(p-q) + G_0(p) \cdot  D^-_jD^+_k G_0(p-q) \Big\}
  ~~. \cr}  \eqn\diagramEtwo $$

We need to evaluate $\Gamma^{(c)00}(q)$, $\Gamma^{(d)02}(q)$, and
$\Gamma^{(e)22}(q)$ in the frame $\q=(q,0)$ to find $\Pi_0$, $\Pi_1$,
and $\Pibar_2$.  Upon inserting (\FourierGreen) into (\diagramCtwo),
(\diagramDthree), or (\diagramEtwo), we encounter an $\omega'$-integral
$$\eqalign{
{i\over 2\pi} \int_{-\infty}^\infty &d\omega' ~
{1\over [\omega'-\ep_n + i\delta(n) ]
\,          [\omega' - \omega - \ep_m + i\delta(m)]}  \cr
\noalign{\kern 8pt}
&=\cases{\big  {1\over \ep_n-\ep_m - \omega -i\ep}
          &for $n\ge |N|, m<|N|$\cr
\noalign{\kern 6pt}
         \big  {1\over \ep_m-\ep_n + \omega -i\ep}
          &for $n< |N|, m\ge |N|$\cr
\noalign{\kern 15pt}
         ~~~0 &otherwise.   \cr}       \cr}
   \eqn\omegaIntegral  $$
Further, since
$$\eqalign{
D^\pm_2 & G_0(p)
  =\Big( ip_2 \pm \ep(N)\, {1\over 2l^2} {\d\over \d p_1} \Big) ~ G_0(p) \cr
\noalign{\kern 10pt}
&=\sum_{n=0}^\infty {1\over \omega - \ep_n + i \delta(n)}
{}~ \int {dx_2\over l} ~ e^{-ip_2x_2}  \cr
&\hskip 3.cm \times
\Big( {\d\over \d x_2} \pm \ep(N) {1\over 2l^2}{\d\over \d p_1} \Big)
\Big\{  v_n(p_1l+\bar z) \, v_n(p_1l - \bar z) \Big\} ~, \cr}  $$
one finds that
$$\eqalign{
D^\pm_2  G_0(p)
&={\ep(N)\over l} \sum_{n=0}^\infty {1\over \omega - \ep_n + i \delta(n)}
{}~ \int {dx_2\over l} ~ e^{-ip_2x_2} ~ \cr
&\hskip 4.cm \times \cases{
            +v_n^{(1)}(p_1l+\bar z) \, v_n(p_1l - \bar z)  \cr
         \noalign{\kern 4pt}
            -v_n(p_1l+\bar z) \, v_n^{(1)}(p_1l - \bar z)  \cr} \cr}
   \eqn\derivativeGzero $$
where $v_n^{(p)}(z)$ is the $p$-derivative of $v_n(z)$.
Another  integral which frequently appears is
$$
C^{(p)}_{nm} (a) \equiv
 \int_{-\infty}^\infty dx ~ v_n^{(p)}(x) \, v_m^{}(x-a)
= (-1)^p \int_{-\infty}^\infty dx ~ v_n^{}(x) \, v_m^{(p)}(x-a) ~~.
  \eqn\CintegralOne $$
It satisfies that
$$\eqalign{
C^{(p)}_{mn} (a) &= (-1)^{n+m} \, C^{(p)}_{nm} (a) \cr
C^{(p)}_{nm} (-a) &= (-1)^{p+n+m} \, C^{(p)}_{nm} (a)~~~.  \cr}
    \eqn\CintegralTwo $$

We start with computing $\Pi_0$:
$$\eqalign{
&q^2 \, \Pi_0 = \Gamma^{(c)00}(\omega,q,0) \cr
\noalign{\kern 8pt}
&= i \int {d\omega' d\p\over (2\pi)^3} ~
 \sum_{n=0}^\infty {1\over \omega' - \ep_n + i \delta(n)}
{}~ \int {dx\over l} ~ e^{-ip_2x} ~
 v_n[p_1l+\bar z(x)] \, v_n[p_1l - \bar z(x)]  \cr
\noalign{\kern 2pt}
&\hskip .1cm
\times \sum_{m=0}^\infty {1\over \omega' -\omega - \ep_m + i \delta(m)}
{}~ \int {dy\over l} ~ e^{-ip_2y} ~
 v_m[p_1l-ql +\bar z(y)] \, v_m[p_1l-ql  - \bar z(y)]  \cr
\noalign{\kern 10pt}
&= {1\over 2\pi l^2} \bigg\{ \sum_{n=|N|}^\infty \sum_{m=0}^{|N|-1}
 {1\over \ep_n-\ep_m -\omega - i\ep} +
\sum_{m=|N|}^\infty \sum_{n=0}^{|N|-1}
  {1\over \ep_m-\ep_n +\omega - i\ep}  \bigg\} \cr
\noalign{\kern 2pt}
&\hskip 7.8cm \times C^{(0)}_{nm}(ql) C^{(0)}_{nm}(ql)   \cr}
    \eqn\PiZeroOne  $$
Therefore
$$\eqalign{
&q^2 \, \Pi_0
= {1\over 2\pi l^2} \sum_{n=|N|}^\infty \sum_{m=0}^{|N|-1} \cr
&\hskip 2cm \Big\{ {1\over \ep_n-\ep_m -\omega - i\ep} +
  {1\over \ep_n-\ep_m +\omega - i\ep}  \Big\} ~
C^{(0)}_{nm}(ql)^2  ~. \cr}  \eqn\PiZeroTwo $$

Evaluation of $\Pi_1$ proceeds similarly.  In view of
of (\diagramDthree) and (\derivativeGzero), one needs to make small
modifications to (\PiZeroTwo)
$$\eqalign{
&\hskip 1cm \times ~ {i\over 2m}{\ep(N)\over l} \cr
\noalign{\kern 8pt}
&C^{(0)}_{nm}(ql)^2 \Longrightarrow
2 C^{(1)}_{nm}(ql) C^{(0)}_{nm}(ql) \cr}
  \eqn\PiOneOne  $$
to find
$$\eqalign{
&q  \Pi_1 =   {\ep(N) \over 2\pi m l^3}
\sum_{n=|N|}^\infty \sum_{m=0}^{|N|-1}  \cr
&\hskip 2cm \Big\{ {1\over \ep_n-\ep_m -\omega - i\ep} +
  {1\over \ep_n-\ep_m +\omega - i\ep}  \Big\} ~
C^{(1)}_{nm}(ql) \, C^{(0)}_{nm}(ql) ~. \cr}  \eqn\PiOneTwo $$

To find $\Pibar_2$, we first note that  partial integrations in (\diagramEtwo)
lead to
$$\Gamma^{(e)22}(\omega,q,0)
 = - {i\over m^2} \int {d^3p\over (2\pi)^3} ~D^-_2 G_0(p) \cdot D^-_2 G_0(p-q)
   ~~~.\eqn\PiTwoOne $$
This time we modify (\PiZeroTwo) such that
$$\eqalign{
&\hskip .2cm \times
  ~\Big(- {1\over m^2}\Big) \Big( {\ep(N)\over l} \Big)^2  \cr
\noalign{\kern 8pt}
&C^{(0)}_{nm}(ql)^2 \Longrightarrow
- \, C^{(1)}_{nm}(ql)^2  ~~.  \cr}   \eqn\PiTwoTwo $$
Since $\Pibar_2=\Gamma^{(b)22}(\omega,q,0)+\Gamma^{(e)22}(\omega,q,0)$,
we find that
$$\eqalign{
\Pibar_2 &= - {n_e\over m} \cr
&+ {1\over 2\pi m^2l^4}
 \sum_{n=|N|}^\infty \sum_{m=0}^{|N|-1} \cr
&\hskip 2cm \Big\{ {1\over \ep_n-\ep_m -\omega - i\ep} +
  {1\over \ep_n-\ep_m +\omega - i\ep}  \Big\} ~
C^{(1)}_{nm}(ql)^2  ~.  \cr}   \eqn\PiTwoThree $$

(\PiZeroTwo), (\PiOneTwo), and (\PiTwoThree) are the results in the RPA and
linearized SCF.  These $\Pi_k$'s, through the formulas (\responseZero),
(\linearCharged), (\finalQn), and (\finalQc), determine the response to
external perturbations.  As is evident from the discussions in Sections 10 and
11, it describes a response to harmonic perturbations $a_\ext(x) \propto
e^{i\p\x-i\omega t}$ introduced from $t=-\infty$ to $t=+\infty$.   One may
introduce a perturbation adiabatically  from $t=-\infty$ to the
present (but not to $t=+\infty$).  In this case the response function is
related to the retarded, but not time-ordered,  Green's function of
currents.  This amounts to making a change\myref{\FTFT,\Abrikosov}
$$\eqalign{
&{1\over \ep_n-\ep_m -\omega - i\ep} +
  {1\over \ep_n-\ep_m +\omega - i\ep}   \cr
&\hskip .5cm \Longrightarrow
{1\over \ep_n-\ep_m -\omega - i\ep} +
  {1\over \ep_n-\ep_m +\omega + i\ep}  \cr}
   \eqn\retarded  $$
in all formulas.  Hence, for instance,
$$\eqalign{
&q^2 \, \Pi_0^R =
 {1\over 2\pi l^2} \sum_{n=|N|}^\infty \sum_{m=0}^{|N|-1} \cr
&\hskip 2cm \Big\{ {1\over \ep_n-\ep_m -\omega - i\ep} +
  {1\over \ep_n-\ep_m +\omega + i\ep}  \Big\} ~
C^{(0)}_{nm}(ql)^2  ~. \cr}  \eqn\retardedPiZero $$

The $x$-integral in
$C^{(p)}_{nm}(a)$,  (\CintegralOne), can be done to yield
$$\eqalign{
C^{(0)}_{nm} (a) &= \Big( {m!\over n!} \Big)^{1/2}
  \Big( {a\over \sqrt{2} } \Big)^{n-m}  e^{-a^2/4} ~
 L_m^{n-m} ( \hbox{${1\over 2}$}a^2) \hskip 1.cm (n \ge m) \cr
&\hskip 3cm L_m^{n-m}(z) = {1\over m!} z^{m-n} \, e^z \,
  {d^m\over dz^m} \,(z^n\, e^{-z})  \cr
\noalign{\kern 8pt}
C^{(1)}_{nm} (a) &= {1\over \sqrt{2}} ~
 \Big\{ - \sqrt{n+1} ~ C^{(0)}_{n+1,m} (a) +
 \sqrt{n} ~ C^{(0)}_{n-1,m} (a)\Big\} \hskip 1.cm
  {\rm etc.}   \cr}   \eqn\CintegralTwo $$
$L_n^\alpha(x)$ is the Laguerre polynomial.
This expression is useful to investigate the response function at finite
$ql$.

For a small momentum $ql \ll 1$, one can expand $v_m(x-ql)$ in a
Taylor series to find
$$
C^{(p)}_{nm}(a) = \sum_{k=0}^\infty (-1)^{k+p} \,{1\over k!}  \,
                       d^{k+p}_{nm} ~ a^k  \next
d^k_{nm} = \int_{-\infty}^\infty dx~ v_n(x) \, v_m^{(k)}(x) ~~~,
  \eqn\CformulaOne  $$
where the coefficients $d^k_{nm}$'s are given by
$$\eqalign{
d^0_{nm} & = \delta_{n,m} \cr
\noalign{\kern 6pt}
d^1_{nm} &= {1\over \sqrt{2}} \, \big\{ \sqrt{m}\, \delta_{n,m-1} - \sqrt{m+1}
    \, \delta_{n,m+1} \big\} \cr
\noalign{\kern 6pt}
d^2_{nm} &=  {1\over 2}
   \Big\{ \sqrt{m(m-1)} \delta_{n,m-2} - (2m+1) \delta_{n,m}
+ \sqrt{(m+1)(m+2)} \, \delta_{n,m+2} \Big\} \cr
\noalign{\kern 6pt}
d^3_{nm} &=  {1\over  2\sqrt{2}} \Big\{ \sqrt{m(m-1)(m-2)} \,
              \delta_{n,m-3}  - 3 m^{3/2} \, \delta_{n,m-1}  \cr
&\hskip 2cm + 3 (m+1)^{3/2} \, \delta_{n,m+1} -
  \sqrt{(m+1)(m+2)(m+3)} \, \delta_{n,m+3} \Big\} \cr
\noalign{\kern 6pt}
d^4_{nm} &=  {1\over  4} \Big\{ \sqrt{m\cdots (m-3)} \, \delta_{n,m-4}
 - 2(2m-1)\sqrt{m(m-1)} \, \delta_{n,m-2}  \cr
&\hskip 1cm + 3 (2m^2+2m+1) \, \delta_{n,m}
-  2(2m+3) \sqrt{(m+1)(m+2)} \, \delta_{n,m+2}  \cr
&\hskip 6cm  + \sqrt{(m+1)\cdots (m+4)} \, \delta_{n,m+4} \Big\} \cr
\noalign{\kern 6pt}
d^5_{nm} &= {1\over  4\sqrt{2}}
  \Big\{ \sqrt{m\cdots (m-4)} \, \delta_{n,m-5}
 - 5(m-1)^{3/2} \sqrt{m(m-2)} \, \delta_{n,m-3}  \cr
&\hskip .3cm  + 5(2m^2+1) \sqrt{m} \, \delta_{n,m-1}
 -5(2m^2+4m+3) \sqrt{m+1} \, \delta_{n,m+1}  \cr
&\hskip .3cm + 5(m+2)^{3/2} \sqrt{(m+1)(m+3)} \, \delta_{n,m+3}
 - \sqrt{(m+1)\cdots (m+5)} \, \delta_{n,m+5} \Big\} \cr
\noalign{\kern 6pt}
&  \cdots ~~~~. \cr}
   \eqn\dknmFormula $$
In particular, for $n>m$,
$$\eqalign{
C^{(0)}_{nm}(a)^2 &= a^2 \,{n\over 2} \, \delta_{n,m+1}
+ a^4 \Big\{ - {n^2\over 4} \, \delta_{n,m+1} +
{n(n-1)\over 16} \, \delta_{n,m+2} \Big\} +\cdots ~, \cr
\noalign{\kern 6pt}
C^{(0)}_{nm}(a)\, C^{(1)}_{nm}(a)
&= a\, {n\over 2}\, \delta_{n,m+1} + a^3 \Big\{ - {n^2\over 2} \,
\delta_{n,m+1}
  + {n(n-1)\over 8} \, \delta_{n,m+2} \Big\} + \cdots ~, \cr
\noalign{\kern 6pt}
C^{(1)}_{nm}(a)^2 &=
{}~~{n\over 2} \,\delta_{n,m+1} + a^2\,\Big\{  -{ 3 n^2\over 4} \,
\delta_{n,m+1}
+{n(n-1)\over 4}  \, \delta_{n,m+2}  \Big\} \cr
&\hskip .5cm + a^4 \, \Big\{ {n(37n^2+5)\over 96} \, \delta_{n,m+1}
 - {n(n-1)(2n-1)\over 12} \, \delta_{n,m+2} \cr
&\hskip 4cm + {n(n-1)(n-2)\over 32} \, \delta_{n,m+3}  \Big\} + \cdots~. \cr}
   \eqn\CformulaTwo $$

Applying (\CformulaTwo) in (\PiZeroTwo), (\PiOneTwo), and (\PiTwoThree),
one finds that for small frequency and momentum
$$\eqalign{
\Pi_0 &= \Big( {N\over 2\pi} \Big)^2 {m\over n_e} \,
  \bigg\{ 1 + \Big( {\omega\over \omega_c} \Big)^2 - {3\over 8} \, |N| (ql)^2
  + \cdots \bigg\} \cr
\noalign{\kern 4pt}
\Pi_1 &= {N\over 2\pi} \, \Big\{ 1 + \Big( {\omega\over \omega_c} \Big)^2
 - {3\over 4} \, |N| (ql)^2 + \cdots \Big\}  \cr
\noalign{\kern 4pt}
\Pibar_2 &= {n_e\over m} \, \Big\{\Big( {\omega\over \omega_c} \Big)^2
 -  |N| (ql)^2 + \cdots \Big\}  \cr
\noalign{\kern 4pt}
&\hskip .6cm      l^2 = {|N|\over 2\pi n_e}~~~, ~~~
\omega_c = {1\over ml^2} = {2\pi n_e\over |N| m}  ~~~.  \cr}
      \eqn\PiExpansion $$
We also note that
$$
\Pi_2 = {1\over q^2} \, (\omega^2 \Pi_0 - \Pibar_2)
= {N^2\over 2\pi m} + \cdots ~~~. \eqn\PiTwoExpansion $$

There are two notable cancellations in the above formulas.  First, in $\Pi_1$,
the dominant term is exactly $N/2\pi$ so that  $\Pibar_1=\Pi_1-
(N/2\pi)$ vanishes at $q=\omega=0$.   This fact is phrased in the literatrure
that the bare Chern-Simons term is exactly cancelled by the one-loop
correction.  Secondly, $\Pibar_2$ vanishes at $q=\omega=0$.  In other words,
the first term (diagram (b)) in (\PiTwoThree) is cancelled by the second term
(diagram (e)).  We have mentioned about it just below eq. (\diagramB).

$\Pi(\omega, q=0)$'s can be evaluated in a closed form.  It is straightforward
to find
$$\eqalign{
\Pi_0(\omega,q=0) &= {|N|\over 2\pi} ~{\omega_c \over \omega_c^2 - \omega^2}
\cr
\Pi_1(\omega,q=0) &= {N\over \,2\pi\,}
  ~{\omega_c^2 \over \omega_c^2 - \omega^2}  \cr
\Pibar_1(\omega,q=0) &= {N\over \,2\pi\,}
  ~{\omega^2 \over \omega_c^2 -  \omega^2} \cr
\Pibar_2(\omega,q=0) &= {|N|\over 2\pi} ~
  {\omega_c \omega^2 \over \omega_c^2 - \omega^2} ~~. \cr}
   \eqn\staticPi $$

In the literature these kernels appear in different notations.
For the sake of readers' convenience we have summarized them in Table 1
below.

\vskip 17pt


\def\smallGamma{{\char'0}}

\def\smallPi{{\char'5}}

\def\smallmu{\hbox{\sevenmit\char'26}}
\def\smallnu{\hbox{\sevenmit\char'27}}

\def\nineneq{ {\ninerm = \kern-1.em\hbox{/}} }

{ \ninerm \baselineskip=12pt
\centerline{ \hbox{ Table 1. ~
 \vtop{ \hsize=8.5cm \noindent
The comparison of various references. Relations among
{\smallPi}$_{\sevenit k}$'s,
{\smallGamma},
{\nineit Q}$_{\sevenit n}$, and
{\nineit Q}$_{\sevenit c}$ are given by (\GammaDecomposition),
(\finalQn), and (\finalQc). At {\nineit T}{\nineneq}0, the notation
{\smallPi}$^{\sevenit E}_{\sevenit k}$ and
{\smallGamma}$^{\smallmu\smallnu}_{\sevenit E}$ has been adopted in this
article
and in ref.\ \RDSS. }
        } }
}
\vglue 13pt

{
\def\space{height7pt &\omit  &&\omit
           &&\omit &&\omit &&\omit \cr}
\def\bigspace{height10pt &\omit  &&\omit
           &&\omit &&\omit &&\omit \cr}
\def\hh{\hskip .15cm}

\centerline{ \hbox{
\vtop{\offinterlineskip
\hrule
\halign{&\vrule# &\strut\hh \hfil $\big #$\hfil \hh &#\vrule\hskip 1pt \vrule
             &\strut\hh \hfil $\big #$\hfil \hh &#\vrule
             &\strut\hh \hfil $\big #$\hfil \hh &#\vrule
             &\strut\hh \hfil $\big #$\hfil \hh &#\vrule
             &\strut\hh \hfil $\big #$\hfil \hh &#\vrule\cr
\space
&{\hbox{\ninerm This}\atop \hbox{\ninerm article}}
  &&{\hbox{\ninerm ref.\ \FHL}\atop \hbox{\ninerm FHL}}
      ~~{\hbox{\ninerm ref.\ \CWWH}\atop \hbox{\ninerm CWWH}}
      ~~{\hbox{\ninerm ref.\ \Fetter}\atop \hbox{\ninerm FH}}
  &&{\hbox{\ninerm ref.\ \HosoChak}\atop \hbox{\ninerm HC}}
      ~~{\hbox{\ninerm ref.\ \HHL}\atop \hbox{\ninerm HHL}}
  &&{\hbox{\ninerm ref.\ \RDSS}\atop \hbox{\ninerm RSS}} & \cr
\space
\noalign{\hrule}
\space
&{T=0\atop \omega\not=0} ~~{T\not=0\atop \omega\not=0}
  &&{T=0\atop \omega\not=0}
  ~~~{T=0\atop \omega\not=0}  ~~~{T\not=0\atop \omega\not=0}
  &&{T=0\atop \omega=0} ~~{T\not=0\atop \omega=0}
  &&{T=0\atop \omega\not=0} ~~{T\not=0\atop \omega=0}    &\cr
\space
\noalign{\hrule}
\space
&\Pi_0
  &&- \big \Big( {N\over 2\pi} \Big)^2 {m\over n_e} \, \Sigma_0
  && &&\Pi_0 &\cr
\space
&\Pi_1
  &&\big -{N\over 2\pi} \, \Sigma_1 && &&\Pi_1 &\cr
\space
&\Pibar_2
  &&- \big {n_e\over m} \, (1+ \Sigma_2) && && &\cr
\space
&\Pi_2
  && && &&-\Pi_2~~ &\cr
\bigspace
\noalign{\hrule}
\bigspace
& \Gamma^{\mu\nu} && - D_0^{\mu\nu} + {n_e\over m}
\delta^{\mu\nu}(1-\delta^{\mu0})
  &&K^{\mu\nu} && \Gamma^{\mu\nu}  &\cr
\bigspace
\noalign{\hrule}
\bigspace
& Q_n^{\mu\nu} &&e^{-2} \, K^{\mu\nu}
  && e^{-2} \, R_n^{\mu\nu} &&   &\cr
\bigspace
\noalign{\hrule}
\bigspace
& Q_c^{\mu\nu} &&  && R_c^{\mu\nu} &&   &\cr
\bigspace
}
\hrule}     }
}

\vskip 26pt


\secno=13  \meqno=1
\line{\bf 13.  Phonons and plasmons  \hfil}
\vglue 5pt
It follows from (\jjNeutralOne) or (\linearCharged) that
the location of poles of the response function $Q_n$ or $Q_c$
in the Fourier space determines
the spectrum of excitations which couple to currents $J^\mu$.
In SCF they appear as self-consistent configurations in the absence of
external fields.  Indeed,
$$\eqalign{
Q_n^{-1} \, J_\ind &= a_\ext =0   \cr
&{\rm or} \cr
Q_c^{-1} \, J_\ind &= a_\ext =0   \cr}
  \eqn\SCFspectrum  $$
has a non-trivial solution ($J_\ind \not= 0$) only at $(\omega,\q)$
for which $det\, Q_n^{-1}=0$ or $det\, Q_c^{-1}=0$.

{}From (\finalQn), (\defineDeltaC), (\finalQc), and (\neutralLimit) one finds
that the location of poles are determined by
$$\Delta_n =
\Big( {2\pi\over N} \Big)^2 (\squarePibar - \Pi_0 \Pibar_2 )  =0
  \eqn\neutralPole $$
for a neutral anyon fluid, and by
$$
\Delta_c=\Big( {2\pi\over N} \Big)^2 (\squarePibar - \Pi_0 \Pibar_2 )
+ e^2 \Pi_0 -{e^2\over q^2-\omega^2} \Pibar_2
  +{e^4\over q^2-\omega^2}  (\Pi_1^2 - \Pi_0 \Pibar_2) =0
  \eqn\chargedPole $$
for a charged anyon fluid.

Solving (\neutralPole) for a small momentum, Fetter, Hanna, and Laughlin
first showed\myref{\FHL} that a neutral anyon fluid admits a phonon excitation
and is a superfluid.  Later the equation was numerically solved for a finite
momentum in refs. (\Wang) and (\Dai).  The spectrum in the charged case has
been examined in ref. (\Sumantra).

In this article we confine ourselves to the spectra at  small momenta
$ql \ll 1$.  For a neutral fluid one can employ the formula (\PiExpansion),
as is justified a posteriori.  Since
$$\eqalign{
\Pi_0 &= {\rm O}(1) \cr
\Pibar_1 &= \Pi_1 - {N\over 2\pi} = {\rm O}( \omega^2, q^2) \cr
\Pibar_2 &= {\rm O}(\omega^2, q^2) ~~~, \cr} $$
Eq. (\neutralPole) is solved for a small momentum by  $\Pibar_2=0$, or
$$\eqalign{
&\omega^2 = c_s^2 \, q^2 \cr
&c_s = \sqrt{|N|} ~\omega_c l = \hbar \,{\sqrt{2\pi n_e}\over m}   ~~. \cr}
    \eqn\phononOne $$
We have recovered $\hbar$ in the last relation.  It is a phonon excitation.
The velocity $c_s$ does not
depend on $N$ in this approximation (RPA, linearized SCF).

In the charged case $\omega$ approaches a finite value as $q\go 0$.   To find
$\omega(q=0)$, we insert (\staticPi) into (\chargedPole).  Many cancellations
take place.  One finds, at $q=0$,
$$\eqalign{
\Big( {2\pi\over N} \Big)^2 (\squarePibar - \Pi_0 \Pibar_2 )
&= ~~ - {\omega^2\over \omega_c^2 - \omega^2} \cr
e^2 \Pi_0 +{e^2\over \omega^2} \Pibar_2 ~~
 &= ~ 2 ~{|N| e^2\over 2\pi} ~{\omega_c \over \omega_c^2 - \omega^2} \cr
-{e^4\over \omega^2}  (\Pi_1^2 - \Pi_0 \Pibar_2)
 &= - \Big( {Ne^2\over 2\pi} \Big)^2 ~
 {\omega_c^2 \over \omega^2 (\omega_c^2-\omega^2)} \cr} $$
Hence Eq. (\chargedPole) reduces to a polynomial equation for $\omega^2$:
$$\bigg( ~{\omega^2\over \omega_c^2} - {|N|e^2\over 2\pi \omega_c} ~\bigg)^2
 =0  ~~~.  \eqn\plasmonOne $$
In other words, the dispersion relation is
$$\omega(q=0) = \sqrt{ {|N|e^2\omega_c\over 2\pi} }
 = \sqrt{ {e^2n_e\over m} }  ~~~.  \eqn\plasmonTwo $$
This is nothing but a plasmon, representing a plasma oscillation.
There is only one solution for $\omega$.  [The two solutions
in ref. (\RDSS) are the result of  $\Pi_k(\omega,0)$'s
being approximated by $\Pi_k(0,0)$'s.]

\sectionskip

\secno=14  \meqno=1
\line{\bf 14.  Hydrodynamic description   \hfil}
\vglue 5pt
In the preceeding sections we have integrated the matter field first,
to obtaine the effective theory for the gauge fields.  The kernel
$\Gamma^{\mu\nu}$ which appears in the effective action (\formulaIone) or
(\SfExpansion) has an important physical meaning.  It appears in
(\SCFandGamma) as the
coefficient relating gauge field configurations to the induced current.
In SCF, for a given gauge field confuguration $a_\mu(x)$ (in the
neutral case),
$$\eqalign{
J^\mu(x) &= \la j^\mu(x) \ra \cr
&= n_e \, \delta^{\mu 0} - \int dy ~ \Gamma^{\mu\nu}(x-y) \, a_\nu^{(1)}(y)
 + \cdots ~~, \cr}  \eqn\SCFinducedCurrent $$
with which the field equation
$$- {N\over 4\pi} ~ \eps f_{\nu\rho}
=  J^\mu(x)    \eqn\SCFequation $$
has been solved self-consistently.

One can read eq. (\SCFequation) differently.  It says that the Chern-Simons
field strengths are nothing but the currents.  The roles of the
two equations (\SCFinducedCurrent) and (\SCFequation) are interchanged.
Now the latter gives a relation, while the former yields  equations
for the currents $J^\mu(x)$.

The time component is the density field $J^0(x) = n(x)$, while the spatial
components define the velocity fields ${\bf v}(x)$ by $J^k = n\, v^k$.
The equations for $n(x)$ and ${\bf v}(x)$ give the hydrodynamic
description of the system.  This is the viewpoint originally adopted by
Wen and Zee.\myref{\WenZee}   The advantage of this approach is that everything
is expressed in terms of macroscopic physical quantities so that physics
is clearest.  In particular, as we shall see, it gives an interpretation
of a phonon excitation as a breathing mode of a density wave, the
picture first spelled out by Wen and Zee.  Anyon fluids are unique systems
in which the ``microscopic'' RPA or SCF is equivalent to the ``macroscopic''
hydrodynamic description.

Substitution of (\GammaDecomposition) into  eq. (\SCFinducedCurrent) yields,
in the linear approximation which drops O[$(J_\ind)^2$],
$$\eqalign{
J^0 &= n_e  -i q_k f_{0k} \, \Pi_0 + b^{(1)} \, \Pi_1  ~~,\cr
\noalign{\kern 5pt}
J^k &= - i\omega \, f_{0k} \, \Pi_0 + \ep^{kl} f_{0l}\, \Pi_1
   - i \ep^{kl} q_l \, b^{(1)} \, \Pi_2 ~~, \cr}
   \eqn\hydroOne  $$
where $b^{(1)}= - f^{(1)}_{12}$.
At $T=0$ all $\Pi(\omega,q)$'s have finite limits at $\omega=q=0$ so that
the right sides of (\hydroOne) are expressed solely in terms of the field
strengths $f_{\mu\nu}$.  Hence, with the aid of the fundamental identity
(\SCFequation), eq. (\SCFinducedCurrent) gives a set of differential
equations for $J^\mu(x)$.  We remark that at finite temperature $\Pi_0$
develops a $1/q^2$ pole so that eq. (\SCFinducedCurrent) becomes an
integral-differential equation.  We shall come back to this point later.

Eq. (\hydroOne) becomes
$$\eqalign{
- \Pibar_1 \, J^0_\ind&= \Pi_0 \, i(q_1 J^2_\ind - q_2 J^1_\ind) \cr
- \Pibar_1 \, \ep^{jk} J^k_\ind &= - \Pi_0 \, i \omega \, J^j_\ind
 +\Pi_2 \, i q_j \, J^0_\ind  \cr}
  \eqn\hydroTwo $$
where $\Pibar_1 = \Pi_1 - (N/ 2\pi)$.
A solution to (\hydroTwo) exsits only if
$ \Pibar_1 \, (\squarePibar - \Pi_0 \Pibar_2) =0  $
where $\Pibar_2 = \omega^2 \Pi_0 - q^2 \Pi_2$.
$\Pibar_1=0$ is not permissible, since the equations would imply $\Pibar_2=0$
as well, which is incompatible.  Hence we have
$$\squarePibar - \Pi_0 \Pibar_2 =0 ~~~.  \eqn\hydroExcitation $$
This is the same as eq. (\neutralPole), $\Delta_n=0$.  As was shown in
the previous section, it admits a phonon spectrum (\phononOne).
We also note that eq. (\hydroTwo) contains the continuity equation.  Indeed
it follows from (\hydroTwo) that
$$ \Pibar_1 \, (\omega  J^0_\ind - q_k J^k_\ind) =0~~~. $$
Here the fact $\Pibar_1 \not= 0$ (at finite $\omega$ and $\q$) is important.

To recognize that the phonon mode found in (\phononOne) represents a density
wave, we write the equations in terms of $n(x)$ and ${\bf v}(x)$.  In the
long wavelength limit, one may insert (\PiExpansion) into (\hydroTwo).
Since $\Pi_0$, $\Pi_2=$O(1), while $\Pibar_1={\rm O}(\omega^2)$, the left
sides of eq. (\hydroTwo) give negligible contributions.  Hence we have
$$\eqalign{
&\d_1 J_\ind^2(x) - \d_2 J_\ind^1(x) =0 \cr
&\d_0 J_\ind^k(x) + c_s^2 \, \d_k J^0_\ind(x) =0  \cr} \eqn\hydroThree $$
with the continuity equation
$$\d_0 J^0_\ind(x) + \d_k J^k_\ind(x) =0 ~~~. \eqn\continuity $$
The first of (\hydroThree) says that there is no circulation (vorticity).
The second of (\hydroThree) and (\continuity) give
$$(\d_0^2 - c_s^2 \d_k^2) \, J^0_\ind(x) =0 ~~~,  $$
implying a density wave.

It is easy to see that (\hydroThree) results by linearizing the hydrodynamic
equation.  The Euler equation for an ideal fluid (with no viscosity
and thermal conductivity) is
$${\d {\bf v}\over \d t} + ({\bf v} \cdot \nabla) \, {\bf v} =
  - {1\over m n} \, \nabla P   \eqn\NSequation $$
where $P(x)$ is the pressure
$$ P = - \Big( {\d F\over \d V} \Big)_T
  = - \Big( {\d E \over \d V} \Big)_S ~~.
   \eqn\pressure  $$

We have evaluated the energy density $E/V$ at zero temperature in Sections 6
and 7.  In the RPA and SCF
$${\rm SCF/RPA :} \qquad E= V \, {\pi n^2 \over m} =
  {\pi N_e^2 \over mV}  ~~~, \eqn\RPAenergy  $$
where $N_e$ is the total anyon number, and in the Hartree-Fock approximation
$${\rm HF :} \qquad E= \cases{
\big  {1\over \,2\,} {\pi N_e^2 \over mV} &for $N=\pm 1$\cr \cr
\big  {29\over 32} {\pi N_e^2 \over mV} &for $N=\pm 2$~.\cr}
   \eqn\HFenergy   $$
Hence
$$P= {\pi n^2 \over m}   \hskip 2.5cm \hbox{in RPA/SCF}
  \eqn\pressureRPA   $$
or
$$
P= \cases{
\big {1\over \,2\,} {\pi n^2 \over m} &for $N=\pm 1$\cr \cr
\big {29\over 32} {\pi n^2 \over m} &for $N=\pm 2$\cr}
 \hskip 1cm \hbox{in Hartree-Fock}
    \eqn\pressureHF  $$

Substituting (\pressureRPA) into (\NSequation) and keeping only linear terms
in $n$ and ${\bf v}$, one finds
$$\eqalign{
{\d {\bf v}\over \d t} &= - {2\pi\over m^2} \, \nabla n \cr
&\Longrightarrow \d_0 J^k = \d_0 (n v^k) \sim
 - {2\pi n_e\over m^2} \, \d_k J^0  ~~,   \cr}  \eqn\hydroFour $$
which is exactly the second equation in (\hydroThree) with $c_s$ given by
(\phononOne).  This derivation demonstrates that in the Hartree-Fock
approximation the sound velocity is modified to\myref{\HLF}
$$
c_s = \cases{
\big \phantom{ {\sqrt{29}\over 4} \, }
     {\hbar \sqrt{\pi n_e} \over m}     &for $N=\pm 1$\cr \cr \cr
\big  {\sqrt{29}\over 4} \,{\hbar\sqrt{\pi n_e }\over m}   &for $N=\pm
2$~~.\cr}
   \eqn\phononTwo  $$

Crucial in the above argument is the fact that the energy density is given
by (\RPAenergy) or (\HFenergy), independent of the number density $n$.
The energy density is proportional to $n^2$.  In the Hartree-Fock language
$|N|$ lowest Landau levels are completely filled even for slowly varying
density $n(x)$.  As $n(x)$ varies, the magnetic length $l$ also varies
such that the particles precisely fill the space.  If one looks at the
motion of each particle (in the half classical picture), the Larmor orbit
expands and shrinks periodically as the density changes.  It breathes.
Wen and Zee called it the breathing mode.\myref{\WenZee}

\sectionskip

\vfil\eject

\secno=15  \meqno=1

\line{\bf 15.  Effective theory \hfil}
\vglue 5pt
The effective theory of anyon fluids in terms of Chern-Simons and Maxwell
fields is obtained by integrating the matter field $\psi(x)$ first.
We have already encountered it in Sections 9 and 10.
It is given by
$$
\L_\eff (a,A)^{T=0} = \L_0^\CS(a) + \L_0^\EM(A) + \L_F(a + eA)
   \eqn\ourGLone  $$
where $\L_0^\CS(a)$ and  $\L_0^\EM(A)$ are defined in (\simplifyL):
$$\eqalign{
\L_0^\CS [a] &= - {N\over 4\pi} \, \eps a_\mu \d_\nu a_\rho  \cr
\L_0^\EM [A] &= -{1\over 4} \, F^2_{\mu\nu} + en_e A_0 ~~. \cr}   $$
$\L_F(a+eA)$ summarizes the effect of the matter field, and is given by
(\SfExpansion) with $a$ replaced by $a+eA$ :
$$\eqalign{
\L_F(a+eA) = &\L_F^{(1)}(a+eA) + \L_F^{(2)}(a+eA) + \cdots  \cr
\noalign{\kern 8pt}
\L_F^{(1)}(a+eA) &=  - n_e \, (a_0 +e A_0 )  \cr
\L_F^{(2)}(a+eA) &= + {1\over 2} \,( a_\mu^{(1)} + e A_\mu) \, \Gamma^{\mu\nu}
\,  (a_\nu^{(1)} + e A_\nu)  ~~~.  \cr}   \eqn\ourGLtwo  $$
In the linear approximation of SCF, or equivalently in RPA, higher order
terms in $a^{(1)}+eA$ are neglected.

The kernel $\Gamma^{\mu\nu}$ has been evaluated in Section 12.  If one is
interested in physics at a large length scale, $\Gamma^{\mu\nu}$ can be
expanded in a Taylor series in $\d_\mu$.  In this section we shall retain
only the most dominant terms.   From (\PiExpansion) and (\PiTwoExpansion)
$$
\left. \eqalign{
\Pi_0 &=  {N^2 m \over 4\pi^2 n_e}  \cr
\noalign{\kern 4pt}
\Pi_1 &= ~~ {N\over 2\pi}  \cr
\noalign{\kern 4pt}
\Pi_2 &= ~ {N^2\over 2\pi m}  \cr} \right.
\hskip 1cm
{\rm for} \hskip .2cm
\left( \eqalign{
\sqrt{|N|}\, ql &= {|N|\over \sqrt{2\pi n_e} } \cdot q \ll 1 \cr
\noalign{\kern 7pt}
{\omega\over \omega_c} &= {~|N| m~\over 2\pi n_e} \cdot \omega \ll 1    \cr}
    \right) ~.   \eqn\dominantPi   $$
We note that the linear approximation gets better for a larger $|N|$, but
the low energy approximation (\dominantPi) breaks down when $|N|$  becomes too
large.

Since
$$\L_\CS(a) = n_e \, a_0
   - {N\over 4\pi} \, \eps a^{(1)}_\mu \d_\nu a^{(1)}_\rho~~~, $$
all linear terms in $\L_\eff(a,A)$ cancell each other.   (\GammaDecomposition)
and (\dominantPi) immediately lead to
$$\eqalign{
\L_\eff (a,A)^{T=0} &=
   - {N\over 4\pi} \, \eps a^{(1)}_\mu \d_\nu a^{(1)}_\rho
  -{1\over 4} \, F^2_{\mu\nu} + \L_F^{(2)}(a+eA)  \cr
\noalign{\kern 7pt}
\L_F^{(2)}(a+eA)
&= + {N\over 4\pi}  \, \eps ( a_\mu^{(1)} + e A_\mu) \, \d_\nu
  ( a_\rho^{(1)} + e A_\rho)  \cr
&\hskip 1.5cm + {N^2m\over 8\pi^2 n_e} \, (f_{0j} + eF_{0j} )^2
 - {N^2\over 4\pi m} \, (b^{(1)} + eB )^2 ~~. \cr}
     \eqn\ourGLthree $$
Or, noticing the cancellation of the bare Chern-Simons term, one may write
$$\eqalign{
\L_\eff (a,A)^{T=0} &=
   -{1\over 4} \, F^2_{\mu\nu} +
 {eN\over 8\pi} \, \eps A_\mu\, (2 \,f^{(1)}_{\nu\rho} + e F_{\nu\rho} )\cr
\noalign{\kern 5pt}
&\hskip 1.5cm + {N^2m\over 8\pi^2 n_e} \, (f_{0j} + eF_{0j} )^2
 - {N^2\over 4\pi m} \, (b^{(1)} + eB )^2 ~~~.  \cr}
     \eqn\ourGLfour $$

(\ourGLthree) or (\ourGLfour) is the effective theory of a charged anyon
fluid, valid for slowly varying configurations.  It replaces the
Ginzburg-Landau (GL) free energy for BCS superconductors.  Instead of the
GL order parameter $\Psi_{\rm GL}(x)$ we have Chern-Simons gauge fields
$a^{(1)}_\mu(x)$.   Higher order terms, namely terms cubic or quartic in
$a^{(1)}+eA$, become important for large gauge field
configurations such as vortices, but have not been evaluated so far.

The effective theory (\ourGLthree) or (\ourGLfour) was first derived by
Hosotani and Chakravarty\myref{\HosoChak} for static configurations at $T=0$.
There is  an alternative way of writing an effective theory.
Introducing a scalar field $\phi(x)$ in place of the Chern-Simons field
$a_\mu(x)$, Chen et al. have written down\myref{\CWWH}
$$\eqalign{
\L_\eff (\phi,A)^{\rm CWWH} &=
   -{1\over 4} \, F^2_{\mu\nu} +  g_1 F_{12} ( \dot \phi - g_2 A_0) \cr
 &\hskip 2.cm + {1\over 2} (\dot \phi - g_2 A_0)^2
 - {1\over 2} \, c_s^2 \, (\d_j \phi - g_2 A_j)^2  \cr
&g_1 = {Ne\over \sqrt{ 32\pi m}} ~~~,~~~ g_2= e \sqrt{ {m\over 2\pi} } \cr}
    \eqn\ChenGL  $$
with the sound velocity $c_s$ defined in (\phononOne).   It has been known
that in spite of different forms both (\ourGLfour) and (\ChenGL) lead
to the same predictions for many physical quantities.  Similar effective
Lagrangians have been written by Fradkin\myref{\Fradkin} and
 by Banks and Lykken.\myref{\Banks}

As we shall see in
later sections, the effective theory $\L_\eff (a,A)^{T=0}$ can be easily
generalized to finite temperature.  There we shall find that not only the
coefficients in $\L_F^{(2)}(a+eA)$ become $T$-dependent, but also a new
important term proportional to $(a_0+ eA_0)^2$ will turn up.

We notice that the effective theory $\L_\eff (a,A)^{T=0}$ neatly
summarizes the self-consistent field (SCF) method.  Equations derived by
taking variations over $a^{(1)}_\mu(x)$ and $A_\mu(x)$ are
$$\eqalign{
-{N\over 4\pi} \, \eps \, f^{(1)}_{\nu\rho} &= J^\mu_\ind \cr
\noalign{\kern 5pt}
 \d_\nu \, F^{\nu\mu} &= e \, J^\mu_\ind  \cr
\noalign{\kern 5pt}
J^\mu_\ind(x) &= - {\delta  \over \delta a^{(1)}_\mu(x) }
  \, \Big\{ \L_F^{(2)} (a+eA) + \cdots \Big\} \cr
\noalign{\kern 5pt}
&= -\Gamma^{\mu\nu} \, (a^{(1)}_\nu +eA_\nu)  + {\rm O}[(a^{(1)}+eA)^2]  \cr}
     \eqn\ourGLeq $$
The last equation which expresses the induced current $J^\mu_\ind$ in
terms of the two gauge fields $a_\mu$ and $A_\mu$ may be viewed as a new
London equation.
The first and second equations lead to an identity
$$-{eN\over 4\pi} \, \eps \, f^{(1)}_{\nu\rho} = \d_\nu \, F^{\nu\mu} ,
        \eqn\relateCSEM  $$
or in the component form
$$\eqalign{
{eN\over 2\pi} \, b^{(1)} &= \, div \, {\bf E}   \qquad (E_k=F_{0k})\cr
{eN\over 2\pi} \, f_{0k} &= \ep^{kl} \, \d_0 E_l + \d_k B ~,  \cr}
    \eqn\relateCSEMtwo $$
with which Chern-Simons fields may be eliminated.  We note that the identity
(\relateCSEM) or (\relateCSEMtwo) is valid beyond the linear approximation.
It is indeed a direct consequence resulting from the general structure of
(\ourGLone),  the minimal gauge couplings.

For slowly varying configurations
$$\eqalign{
\indJ^0(x) &= + {N\over 2\pi} \, (b^{(1)} + eB)
   - {N^2 m\over 4\pi^2 n_e} \, \d_j (f_{0j} + eF_{0j}) + \cdots  \cr
\indJ^k(x) &= + {N\over 2\pi} \, \, \ep^{kl} \,(f_{0l} + e F_{0l})
   - {N^2\over 2\pi m} ~ \ep^{kl} \d_l ( b^{(1)} + eB)  + \cdots \cr}
   \eqn\Londonone $$
With the aid of (\relateCSEMtwo) we eliminate Chern-Simons fields to obtain
$$\eqalign{
e \indJ^0 &=  \Big\{ \,div\, {\bf E} \, \Big\}
  - {e^2N^2 m\over 4 \pi^2 n_e}  \, div\, {\bf E} \cr
&\hskip 1.cm  - {Nm\over 2\pi n_e} \, \d_0 \, (rot \,{\bf E})
+ {e^2N\over 2\pi} \Big( 1 - {m\over e^2 n_e} \nabla^2 \Big)
 \, B + \cdots \cr
\noalign{\kern 7pt}
e \indJ^k &= \Big\{ - \d_0 E_k + \ep^{kl} \d_l B \, \Big\}
 - {e^2N^2 \over 2\pi m} \, \ep^{kl} \d_l B \cr
&\hskip 2.cm  + {e^2 N\over 2\pi} \, \ep^{kl} E_l
 - {N\over m} \, \ep^{kl} \, \d_l \,(div\, {\bf E}) + \cdots \cr}
   \eqn\LondonTwo $$
where $rot\, {\bf E}= \d_1 E_2 - \d_2 E_1$.  Notice that the dominant terms in
Eq. (\Londonone) represent an integer quantum Hall effect in the
system.\myref{\HosoChak}

Inserting (\LondonTwo) into the Maxwell equations
$$\eqalign{
div \, {\bf E} &= e \indJ^0 \cr
-\d_0 E_k + \ep^{kl} \, \d_l B &= e \indJ^k ~~~,  \cr}
  \eqn\MaxwellOne   $$
one recognizes that the terms in parenthesis $\big\{ ~ \big\}$ in
(\LondonTwo) exactly cancell the left sides of the Maxwell equations.
Crucial in this cancellation is the fact that the coefficient of  the induced
Chern-Simons term for $a^{(1)}_\mu+eA_\mu$ in  (\ourGLthree) is exactly the
negative of the coefficient of the bare Chern-Simons term for $a_\mu$.
We thus arrive at equations
$$\eqalign{
\Big( 1 - {m\over e^2 n_e} \, \nabla^2 \Big) \, B
 - {Nm\over 2\pi n_e} \, div\,{\bf E}
 - {m\over e^2n_e} \, \d_0\,(rot\,{\bf E}) &=0  \cr
E_k - {2\pi\over e^2m} \, \d_k\, (div \,{\bf E}) - {N\over m}\, \d_kB &=0 \cr}
   \eqn\MediumEq   $$
which describe electromagnetic fields in anyon fluids.

\sectionskip
\secno=16  \meqno=1

\line{\bf 16.  Meissner effect at $T=0$ \hfil}
\vglue 5pt
We examine a response of a charged anyon fluid to an external
magnetic field.  If the anyon fluid is   a superconductor,
a sufficiently small magneic field must be expelled from the system.
It must have a Meissner effect.

We shall show that it is indeed the case at least at $T=0$ within RPA
and SCF.  There are two ways to demonstrate it, one in the real configuration
space with an external magnetic field applied outside the body, and the other
by
introducing a test current of a $\delta$-function type in the middle of the
body.  The former approach corresponds to solving the Ginzburg-Landau equation
in the  BCS superconductors, whereas the latter to a linear response
theory.\myref{\SCReview}  We discuss both.

First we suppose that an anyon fluid occupies a half plane ($x_1 \ge 0,
 -\infty < x_2 < +\infty$) and an external magnetic field is applied such
that $B_3=B(\x)=B_\ext$ for $x_1<0$.  The problem is to determine
the magnetic field configuration $B(\x)=B(x_1)$ inside the anyon fluid.
One expects damping behaviour $B(x_1) \propto \exp (- x_1/\lambda)$ if the
anyon fluid is a superconductor.

In this section we consider a ``sufficiently small'' $B_\ext$.
We expect that deviations of both Chern-Simons and Maxwell fields from
the ground state values are small so that Eq. (\MediumEq) may be
employed.  Together with the boundary condition  $B(0)=B_\ext$,
the magnetic field  $B(x_1)$ ($x_1>0$) inside the system is determined.

For configurations under consideration, Eq.$\,$(\MediumEq) becomes
$$\eqalign{
\Big( 1- {m\over e^2 n_e} \d_1^2 \Big) \, B - {Nm\over 2\pi n_e}\, \d_1E_1
 &=0   \cr
E_1 - {2\pi\over e^2 m} \, \d^2_1 E_1 - {N\over m}\, \d_1 B &=0 \cr
E_2 &=0 ~~. \cr}
    \eqn\MeissnerEqOne  $$
At this point one has to examine numerical values of various parameter.
We give a summary of numerical values in Sections 18 and 20.   It will be seen
that to very good accuracy (\MeissnerEqOne) is approximated by
$$\eqalign{
\Big( 1- {m\over e^2 n_e} \d_1^2 \Big) \, B &\sim 0  \cr
E_1  &\sim 0 \cr
E_2  &=0 ~~~. \cr}
    \eqn\MeissnerEqTwo  $$
Hence the solution is
$$\eqalign{
B(x_1) &= B_\ext \, e^{-x_1/\blam} \hskip 1cm {\rm for}~ x_1 >0 \cr
\noalign{\kern 6pt}
\lambar &= \sqrt{ {m \over e^2 n_e} }  ~~~. \cr}
  \eqn\MeissnerOne  $$
The magnetic field is exponentially damped from the surface.  The charged
anyon fluid exhibits a Meissner effect at $T=0$.  The penetration depth
coincides with the London penetration depth in BCS superconductors.
The persistent current flows along the boundary.

Next we shall examine the same problem in the linear response theory.  We
imagine that a charged anyon fluid occupies the entire space in the $x_1$-$x_2$
plane.  We introduce an external current of a $\delta$-function type
at $x_1=0$:
$$\eqalign{
e J_\ext^2(x) &= - 2B_0 \, \delta(x_1)   \cr
J_\ext^0(x) &= J_\ext^1(x) =0   \cr}
   \eqn\extCurrent  $$
which generate an external magnetic field
$$ B_\ext(x) = B_0 \, \ep(x_1) ~~~. \eqn\extMagnetic $$
In the momentum space
$$\eqalign{
e J_\ext^2(\omega,\q)
     &= - 2B_0 \cdot (2\pi)^2 \, \delta(\omega) \delta(q_2)   \cr
B_\ext(\omega,\q)
   &= {i\over q_1} \, e J_\ext^2(\omega,\q)  \cr
A^2_\ext(\omega, \q) &= {1\over q_1^2} \, e J_\ext^2(\omega, \q)
   ~~~,~~~ A^0_\ext = A^1_\ext =0 ~~~.  \cr}
  \eqn\extField  $$

The response of the system to an external perturbation is described by
the response function $Q_c$ determined in the preceeding sections.
The relation to the induced current is given by (\linearCharged):
$${J_\ind^\mu}^{\rm linear} = - \Gamma^{\mu\nu} ( a^{(1)}_\nu + e A_\nu)
   = Q_c^{\mu\nu} ~ e A_\nu^\ext ~~.
    \eqn\responseOne $$
For the configuration (\extField)
$$\eqalign{
\indJ^0 &= - \, Q_c^{02} ~e A^2_\ext
  = - \, {e^2\over q^2_1} ~ Q_c^{02} \, J^2_\ext  \cr
\noalign{\kern 2pt}
\indJ^1 &= ~ 0  \cr
\noalign{\kern 2pt}
\indJ^2 &= - \, Q_c^{22} ~e  A^2_\ext
  = - \, {e^2\over q_1^2} ~ Q_c^{22} \, J^2_\ext \cr}   \eqn\responseTwo $$

At this point we need to evaluate $Q_c^{\mu\nu}$ for $\omega=0$, $q_2=0$,
and small $q_1=q$.  We notice that
$$\eqalign{
\Pi_0, \Pi_1, \Pi_2 &= {\rm O}(\,1\,) \cr
\Pibar_1, \Pibar_2 &= {\rm O}(q^2) \cr
\Pi_1^2 - \Pi_0 \Pibar_2 &= {\rm O}(\,1\,) \cr
\squarePibar - \Pi_0 \Pibar_2 &= {\rm O}(q^2)  ~~~.  \cr}  $$
Hence in $\Delta_c$ in (\defineDeltaC), only the second and fourth terms are
relevant.  Explicitly
$$\eqalign{
\Delta_c & \sim {e^4\over q^2} \, \Big( {N\over 2\pi} \Big)^2 \,
 \bigg\{ 1 + \sqlamb q^2 - {1\over 2} \, |N| (ql)^2 + {2\pi\over me^2}\, q^2
 + {\rm O}(q^4) \, \bigg\} \cr
\noalign{\kern 5pt}
&\sim {e^4\over q^2} \, \Big( {N\over 2\pi} \Big)^2 \, (1 + \sqlamb q^2)
    ~~~. \cr}    \eqn\DelCsmallq $$
In the second line we have suppressed numerically negligible terms.
$\lambar$ is  given in (\MeissnerOne).

{}From (\finalQc)
$$\eqalign{
Q_c^{22} &= {1\over \Delta_c} \, \big\{ \, -\Pibar_2
  + e^2 (\Pi_1^2 - \Pi_0  \Pibar_2) \, \big\}  \cr
Q_c^{02} &= {iq\over \Delta_c} \, \Big\{ - \Pi_1 + {2\pi\over N} \,
   (\Pi_1^2 - \Pi_0  \Pibar_2) \, \Big\} ~~.  \cr}   $$
It is straightforward to see that in our approximation
$$\eqalign{
Q_c^{22} &\sim {q^2\over e^2} \, {1\over \strut 1+ \sqlamb q^2} \cr
Q_c^{02} &\sim {iN\over 4e^4 n_e} \, {q^5\over \strut 1+\sqlamb q^2}  \cr}
            \eqn\staticQcOne  $$

The total current in the presence of the perturbation (\extCurrent) becomes
$$\eqalign{
J^2_\tot &= \indJ^2 + J^2_\ext \cr
&=\Big( 1 - {e^2\over q_1^2} \, Q_c^{22} \Big) ~ J^2_\ext  \cr
&= {\sqlamb q_1^2 \over \strut 1 + \sqlamb q_1^2} ~ J^2_\ext  ~~~.  \cr}
  \eqn\MeissnerCurrent $$
Notice that $J^2_\tot$ vanishes at $q_1=0$, i.e.~the external current is
completely shielded.  The total magnetic field is given by
$$\eqalign{
B_\tot(\omega,\q) &= {i\over q_1} \, eJ^2_\tot \cr
&= -2iB_0 \, {\sqlamb q_1\over \strut 1+ \sqlamb q_1^2} \cdot
  (2\pi)^2 \, \delta(\omega) \delta(q_2)  \cr}
  \eqn\MeissnerFieldOne $$
so that in the configuration space
$$B(x) = B_0 \, \ep(x_1) \, e^{-|x_1|/\blam} ~~~.
         \eqn\MeissnerFieldTwo  $$
We have reproduced the same result as in (\MeissnerOne).
The Meissner effect is complete at $T=0$ for a sufficiently small
external field.

For completeness we look at a response to an external static charge:
$J^0_\ext \not= 0$, $J^k_\ext=0$.  One finds that, for $\omega=0$,
$$
Q_c^{00} = - {1\over \Delta_c}\, \big\{ q^2 \Pi_0
  + e^2 (\Pi_1^2 - \Pi_0 \Pibar_2) \big\}
\sim - {q^2\over e^2}   ~~~.   \eqn\staticQcTwo  $$
Noticing
$$A^0_\ext = {1\over q^2} \, eJ^0_\ext ~~~, $$
we find that
$$
J^0_\ind = Q_c^{00} \, eA^0_\ext  = {e^2\over q^2} \, Q_c^{00} \, J^0_\ext
\sim - J^0_\ext ~~~.     \eqn\induceCharge $$
An external charge is completely shielded, as it should.

\sectionskip


\secno=17  \meqno=1

\line{\bf 17.  $T\not= 0$ -- homogeneous fields \hfil}
\vglue 5pt
The behaviour of anyon fluids at finite temperature is particularly
interesting to know.  In a sense the behaviour of anyon fluids at zero
temperature is very similar to that of conventional superfluids or
superconductors.  For instance, a charged anyon fluid exhibits a
Meissner effect for sufficiently small magnetic fields with the same
penetration depth as in BCS superconductors.  It is very difficult to see
an effect of the unique structure of the ground state, namely the complete
filling of Landau levels in the Hartree-Fock approximation.  (There is a
tiny effect of $P$- and $T$-violation, which we shall briefly touch on in
Section 23.)

What would happen at finite temperature?  Does a charged anyon fluid
behave quite differently from BCS superconductors?  Is a charged anyon
fluid really a ``high'' $T_c$ superconductor?  We shall show in this and
following sections that there is an indication that a charged anyon fluid
has $T_c$ around 100 K, but not around 5 K or 1000 K.  After all the most
important feature of observed high $T_c$ superconductors
(cuprate superconductors) is that they have $T_c \sim 50 - 100$ K.

How is the order of a charged anyon fluid destroyed?  In this respect
it is suitable to consider effects of both finite magnetic fields and
temperature.\myref{\HHL}  Recall that at zero temperature with no external
fields  Landau levels with respect to the Chern-Simons magnetic field are
completely filled in the Hartree-Fock approximation.  Particles, or holons,
feel
only the  sum of Chern-Simons and Maxwell gauge fields.  They interact with the
gauge fields in the  combination of  $a_\mu + eA_\mu$.

If an external uniform magnetic field $B_\ext$ is applied in the direction
perpendicular to the two-dimensional plane, to the first approximation,
particles feel the total magnetic field $b_\tot = b^{(0)} +e B_\ext$, with
which the Landau levels are not completely filled any more.  If the relative
sign between $b^{(0)}$ and $eB_\ext$ is negative, less states are available
per Landau level so that some perticles must be put in the higher energy level.
On the other hand, if the relative sign is positive, more states are
available so that there appear vacant states in the top filled level.  Hence,
if the complete filling were essential for superconductivity, an external
magnetic field would destroy it.

Similarly, at finite temperature, the levels are not completely filled due to
thermal excitations.  Superconductivity should be destroyed at sufficiently
high temperature.

The anaysis at finite fields and temperature, however, is complicated by the
plausible breakdown of the approximation in which  homogeneous
configurations are supposed.    It is likely
that a finite uniform external magnetic field creates vortices in anyon
fluids, giving rise to inhomogeneous, though patterned, configurations.
At finite temperature vortex-antivortex pair creation would become
important.

Effects of vortex formation  has been examined by Kitazawa and
Murayama\myref{\KitaMura} in the case of neutral anyon fluids.  They have
argued
that vortices bring a stability to the superfluidity of neutral anyon fluids.
At the moment the existence of real (electromagnetic) vortices in charged
anyon fluids is yet to be established.

In this and following sections we shall examine effects of finite fields and
temperature, ignoring contributions of vortices.  This is a drastic
approximation, which has to be improved in future.   However,  we shall find
that even in this approximation anyon fluids exhibit interesting behaviour
which is quite different from that of conventional (type I) superconductors.
Some of the behaviour will be modified by the incorporation of vortices,
but all this is certainly essential for full understanding of anyon fluids.

The first evaluation of finite temperature effects in the model under
consideration was given by Randjbar-Daemi, Salam, and Strathdee.\myref{\RDSS}
Hetrick, Hosotani, and Lee\myref{\HHL} subsequently confirmed
their result, discussing more physical implications with additional effects of
finite  fields.  Later  many authors, particularly, Fetter and
Hanna,\myref{\Fetter} recovered the  same result by different methods.

The evaluation consists of two parts.  First thermodynamic quantities are
evaluated for uniform field configurations at finite temperature, from
which self-consistent uniform fields are determined.  Secondly, inhomogeneous
deviations of the fields from the self-consistent uniform configuration are
incorporated in perturbation theory at finite temperature.

With a uniform  magnetic fields $b^{(0)}$ and $B^{(0)}$, Landau levels are
formed with  respect to $b^{(0)}+ eB^{(0)}$.  In this section we suppress
the superscript ${}^{(0)}$ to simplify the notation. It will be recovered
when inhomogeneous  configurations are examined  in Section 19.
The number of states per area per Landau level is
$$n_L =   {\st  b+ eB \st \over 2\pi}  \eqn\newLevelDensity  $$
which defines the magnetic length
$$l(B)^2 = {1\over\st b+ eB \st }  ~~. \eqn\newLength  $$
Energy levels are given by
$$\ep_n(B) = \Big( n+{1\over 2} \Big) \, \ep \next
\ep \equiv {1\over m \,l(B)^2}
  \hskip .8cm (n=0,1,2,\cdots).
    \eqn\newEnergyLevel  $$
One of Chern-Simons equations implies that
$$b = {2\pi n_e \over N}     \eqn\CSfield  $$
is still valid.  Therefore the filling factor as a whole is
$$\eqalign{
\nu &= {n_e\over n_L} = 2\pi n_e \cdot l^2
   = \bigg| ~{Nb \over \, b+eB \, }~ \bigg| \cr
\noalign{\kern 6pt}
&= |N| \cdot \bigg| ~ 1 + {eB\over b } ~ \bigg|^{-1}
{}~~.      \cr}   \eqn\newFilling $$
For $|eB/b| \ll 1$,
$$\nu = |N| \cdot \bigg( 1 - {NeB\over 2\pi n_e} + \cdots \bigg) ~~~.
   \eqn\newFillingTwo $$
We keep the above definition of $\nu$ at $T\not= 0$.

The distribution function $\rho_{np}$ at level $n$ with the second index $p$
(defined in Section 6)
is given by
$$\rho_{np} = {1\over e^{(\ep_n -\mu)/T} + 1}
  \equiv \rho_n   \eqn\distributionF  $$
where $\mu$ and $T$ are the chemical potential and temperature, respectively.
The chemical potential is fixed by the condition
$$n_e = {1\over V} \sum_{n,p}  ~\rho_{np}  ~~.  \eqn\fixMu  $$
The summation over $p$ gives $V/2\pi l^2$, as is most easily seen in the
Landau gauge.  Hence
$$\nu = 2\pi l^2 \cdot n_e =
\sum_{n=0}^\infty  ~\rho_n  ~~~.  \eqn\nurhoRelation  $$

The energy density $\E$ and entropy density $\S$ are given by
$$\eqalign{
\E &= {1\over V} \sum_{n,p} ~ \ep_n \rho_n \cr
&={2\pi n_e^2\over \nu^2 m} \sum_n ~\Big( n+{1\over 2} \Big) \, \rho_n  \cr
\S &= - {1\over V} \sum_{n,p} ~ \big\{ \, \rho_n \ln \rho_n +
 (1-\rho_n) \ln (1-\rho_n) \, \big\}   \cr
&= - {n_e\over \nu} \sum_n ~ \big\{ \, \rho_n \ln \rho_n +
    (1-\rho_n) \ln (1-\rho_n) \, \big\}   \cr}   \eqn\entropy  $$
The free energy density $\F[B]$ is
$$\F[B] = \E - T \, \S ~~~.  \eqn\FEnergyDensity  $$

A good approximation to $\rho_n$ is obtained by examining numerical values
of various parameters.  As we shall see in the next section, with typical
values of $m \sim 2 m_e$ and $n_e \sim 2 \times 10^{14} \,{\rm cm}^{-2}$,
$$\eqalign{
\ep &= {\,2\pi n_e\,\over |N|m}
  ~\sim~ {2\over |N|} \cdot 2800 ~{\rm K}  \cr
{1\over e} \, b &= {\,2\pi n_e\,\over |N|} ~\sim~ \, {2\over |N|} \cdot
  1200 ~ {\rm T} ~~. \cr}  \eqn\valueOne  $$
With this choice, $T_c$ will turn out about 100 K.  In other words,
with $N=\pm 2$ and for
$T < 200 $ K and $B < 30$ T, the filling factor $\nu$ is very close to $|N|$
and  thermal excitations are appreciable only to the $|N|$-th Landau level:
$$
\rho_n \sim \cases{
  1 &~~for $n \le |N|-2$\cr
  \noalign{\kern 5pt}
  0 &~~for $n \ge |N|+1$~~.\cr}    \eqn\approximateFilling  $$
Hence Eq. (\nurhoRelation) becomes
$$x \equiv \nu - |N| = (\rho_{|N|-1} -1 ) + \rho_{|N|} ~~.
  \eqn\nurhoApproximate  $$
It is called the two-level approximation.  The condition for its validity is
$$ e^{-\ep/T} \ll 1 ~~.  \eqn\TwoLevelCondition  $$

Since
$$
\rho_{|N|-1} = {1\over e^{-\ep/2T} e^{(|N|\ep-\mu)/T} +1} \next
\rho_{|N|} = {1\over e^{+\ep/2T} e^{(|N|\ep-\mu)/T} +1} ~~,  $$
Eq. (\nurhoApproximate) is solved by
$$\eqalign{
z(x,T)^{\pm 1} &\equiv e^{\pm  (|N|\ep- \mu)/T}  \cr
&= {1\over  1 \pm x} ~ \bigg( ~
 \sqrt{ 1 + x^2 \sinh^2 {\ep\over 2T} } \mp x \cosh {\ep\over 2T}  ~\bigg) ~~,
  \cr}    \eqn\determineMu  $$
which determines the chemical potential $\mu$ with given $T$ and $B$.  (Note
that $\ep$ is a function of $B$.)

In particular,
$$\mu(T) = |N| \, \ep  \hskip 1cm {\rm at~} B=0 ~~.  \eqn\chemicalOne $$
There is no $T$-dependence in the two-level approximation.  The correction has
been evaluated by Randjbar-Daemi et al.~to be\myref{\RDSS}
$$\mu (B=0) = |N| \, \ep - {T\over 2} \,
   e^{- |N| \ep/T} + \cdots ~~, \eqn\chemicalTwo   $$
which is exponentially small.

We also remark that $T \go 0$ limit is singular, as can be seen from the
presence of $\sinh (\ep/2T)$ or $\cosh (\ep/2T)$ in (\determineMu).
At zero temperature for small $|x|$
$$\mu(T=0) = \cases{ (|N|-{1\over 2}) ~\ep &for $x<0$\cr
                     \noalign{\kern 5pt}
                     (|N|+{1\over 2}) ~\ep &for $x>0$\cr}
   \eqn\chemicalThree  $$
which also follows from the consideration of the Fermi level.
Thermodynamic quantities at $T=0$ have a singularity at $x=0$.  We shall come
back to this point in the next section.

Let us define
$$\rhobar = \rho_{|N|} = {1\over z\, e^{\ep/2T}  +1 } ~~, \eqn\rhobarDef  $$
in terms of which the energy and entropy density in the two-level approximation
are given by
$$\eqalign{
\E &= {2\pi n_e^2\over \nu^2 m} ~ \Big\{ ~{1\over 2} |N|^2 + \Big( |N|- {1\over
2} \Big) x + \rhobar ~ \Big\} \cr
\noalign{\kern 5pt}
\S &= - {n_e\over \nu} ~ \Big\{ ~\rhobar \ln \rhobar + (1-\rhobar) \ln
(1-\rhobar) \cr
&\hskip 1.2cm +(\rhobar-x) \ln (\rhobar-x)
 + (1-\rhobar +x) \ln (1-\rhobar +x) ~ \Big\} \cr}
   \eqn\ESapprox  $$

There are two parameters, $x$ and $T$.  For experimentally available
magnetic fields we always have $|x| \ll 1$.  However,
$z(x,T)$ in (\determineMu) depends on $T$ sensitively with
small, finite $x$.  It is easy to see that
$$\vcenter{
\halign{  $\big #$ \hfil & $\big #$ \hfil
  &\hskip .5cm $\big #$ \hfil   &\hskip .5cm $\big #$ \hfil
  &\hskip .5cm $\big #$ \hfil  \cr
1) ~|x| \ll 1 ~, &|x|~ e^{\ep/2T} \ll 1 ~:  \cr
\noalign{\kern 5pt}
  &~~z= 1   &\rho_{|N|-1} = {1\over e^{-\ep/2T} +1}
           &\rho_{|N|} = {1\over e^{\ep/2T} +1 }  \cr
\noalign{\kern 10pt}
2) ~|x| \ll 1 ~, &|x|~ e^{\ep/2T} \gg 1 ~: \cr
\noalign{\kern 5pt}
{}~~ x>0 ~: &~~z= {1\over x} \, e^{-\ep/2T}
  &\rho_{|N|-1} =1  &\rho_{|N|} = x \cr
\noalign{\kern 5pt}
{}~~ x<0 ~:  &~~z= -x\, e^{\ep/2T}
  &\rho_{|N|-1}= 1+x  &\rho_{|N|}=0 ~~~. \cr}
}  \eqn\xTdependence  $$
Note that with the numerical values (\valueOne) the boader line
defined by $|x|~ e^{\ep/2T} =1$ is given by, for $N=\pm 2$,
$$\vcenter{
\halign{ $x=~$\hfil $#$ & $\times 10^{-#}$
           &\hskip .5cm $B=~ #$\hfil &#~G
           &\hskip 1.cm $T=~$\hfil $#$~K \cr
\noalign{\kern 4pt}
-6 &11  &10^{-3} &&60  \cr
-6 &9   &10^{-1} &&74  \cr
-8.3 &7 &12.5    &&100 \cr
-6 &6   &10^{+3} &&146 \cr
-6 &3   &10^{+5} &&279 \cr
\noalign{\kern 4pt}
}   }  \eqn\crossover $$

It is interesting that the crossover takes place around 70 -- 150 K for
moderate magnetic fields.  However, it should be borne in mind that the
above result is in the approximation which ignores contributions of vortices.

For completeness we evaluate the specific heat (per volume) and pressure
in the two-level approximation.  From (\ESapprox) one finds
$$C_v = \Big( {\d\E\over \d T} \Big)_V
  = {2\pi n_e^2\over \nu^2 m} ~  \Big( {\d\rhobar \over \d T} \Big)_V ~~.
   \eqn\SpecificHeat  $$
It follows from (\determineMu) and (\rhobarDef) that
$$\eqalign{
\Big( {\d\rhobar \over \d T} \Big)_V &=
{\ep\over 2T^2} {1\over (z\,e^{\ep/2T}  +1) \,(z^{-1} \, e^{-\ep/2T} +1) } \cr
\noalign{\kern 12pt}
&\hskip 3cm \times \bigg\{ 1 - { x \sinh (\ep/ 2T) \over
  [ 1+ x^2 \sinh^2 (\ep /  2T) ]^{1/2} } ~\bigg\}  ~. \cr}
  \eqn\SpecificHeatTwo  $$

The expression for the pressure at finite $x$ is lengthy.  The  result for
$x=0$ is simple.$$\eqalign{
P &= - \Big( {\d F\over \d V} \Big)_T \cr
\noalign{\kern 6pt}
&= {2\pi n_e^2 \over |N|^2 m} ~ \Big( {1\over 2} \, |N|^2 + \rhobar \Big)
- \Big( {\d\rhobar\over \d V} \Big)_T \, V \,
\Big( {2\pi n_e^2\over |N|^2 m} +  {2n_eT\over |N|} \, \ln {\rhobar\over
1-\rhobar} \Big)  ~.\cr}   \eqn\TwoLevelPressure  $$
Insertion of (\rhobarDef) with $z=1$ shows that the two terms in the
last parenthesis cancell each other.   Hence
$$P= {\pi n_e^2\over m} ~ \Big( 1 + {2\over |N|^2} \, \rhobar \Big)
   \hskip 1cm {\rm at ~~} x=0 ~.  \eqn\TwoLevelPressureTwo $$
There results only a tiny correction.  However, this may be an artifact of
the uniform field approximation.  We shall see in Section 21
a sign of instability in a neutral anyon fluid.

\sectionskip

\secno=18  \meqno=1

\line{\bf 18.  de Haas -- van Alphen effect in SCF \hfil}
\vglue 5pt
Charged anyon fluids have the structure of Landau levels, and therefore
should exhibit a de Haas -- van Alphen effect\myref{\Alphen} when external
magnetic fields are applied.  Of course, an implicit assumption is that the
system remains uniform in the presence of uniform fields, which is probably not
true even with a modest external field.  Observed high $T_c$
superconductors are of type II, i.e.~vortices are formed.  Nevertheless,
it is worthwhile to examine how the system respond to external fields
in the uniform field approximation.  We shall see that the Meissner effect
at $T=0$ for sufficiently small fields can be understood as a part of a
de Haas -- van Alphen effect, and that an important departure from the
Meissner effect of the BCS type results both at modest external fields and
at finite temperature.\myref{\HHL}

In the previous section we have evaluated in the two-level approximation
the energy and entropy densities as functions of temperature $T$ and magnetic
field $B$.  In terms of the free energy dnesity $\F=\E- T \S$, the
magnetization is given by
$$M(T,B) = - {\d \F(T,B) \over \d B}  ~~~.  \eqn\magnetization $$
$B$ is the total magnetic field (magnetic induction).  The relation to an
external field (thermodynamic field) $H$ is given by
$$B = H + M(T,B) \eqn\induction $$
which defines a relation between $H$ and $B$.

Let us see first what happens in the $T\go 0$ limit in the uniform field
approximation.  The mean field energy density is easily computed to be
$$\eqalign{
\noalign{\kern 5pt}
\E(0,B) &= {\pi n_e^2\over m} \, \bigg\{ 1+
 {(|N|-\nu)(\nu-|N| \pm 1)\over \nu^2} \, \bigg\}  \cr
\noalign{\kern 7pt}
&= {\pi n_e^2\over m} \, \bigg\{ 1+
{|eB|\over 2\pi n_e} - |N|(|N| \mp 1) \Big( {eB\over 2\pi n_e} \Big)^2 \,
\bigg\}\cr
\noalign{\kern 10pt}
&\hskip 2cm {\rm for}~~
   \cases{ |N|-1\le \nu\le |N| &$(NeB >0)$~, \cr\cr
           |N|\le \nu\le |N|+1 &$(NeB <0)$~. \cr}   \cr
\noalign{\kern 5pt}
}                           \eqn\EnergyB  $$
Notice the appearance of $|eB|$ in the expression.  The energy density
has a cusp at $B=0$, as was first noticed by Chen et al.\myref{\CWWH}
The magnetization is found to be
$$\eqalign{
eB>0 \qquad
M(0,B)&= - {en_e\over 2m} \bigg\{ 1 - 2 |N| (|N|\mp 1) \,
    {eB\over 2\pi n_e} \bigg\}  \cr
\noalign{\kern 10pt}
eB<0 \qquad
M(0,B)&= + {en_e\over 2m} \bigg\{ 1 + 2 |N| (|N|\mp 1) \,
    {eB\over 2\pi n_e} \bigg\} ~~. \cr}
   \eqn\zeroMagnetization  $$
It has a discontinuity at $B=0$.  The magnitude decreases from the value
$M(0,0^\pm)$ as $|B|$ increases.  For magnetic fields available in
laboratories, we always have $|eB/2\pi n_e| \ll 1$.  Therefore, to good
accuracy
$$M(0,B) = \mp {en_e\over 2m}  \hskip 1cm {\rm for~~}
   \cases{ eB>0&\cr \cr eB<0&\cr} ~~.  \eqn\zeroMagnetizationTwo   $$

This is nothing but a de Haas -- van Alphen effect.
(See Fig.~10.)  The only difference
from the standard one is that even in the absence of magnetic field ($B=0$),
we have a Chern-Simons magnetic field ($b\not=0$) such that we are at
the integer filling $\nu=|N|$.  The magnetization reaches its maximum
$|e|n_e/2m$ at discontinuous points.

{\figure
\vglue 17cm
\centerline{Fig. 10 ~~ \vtop{ \hsize=9cm \noindent
{\eightit B} vs {\eightit M} as {\eightit T} varies.
(a)  At {\eightit T}=0.
(b)  For moderate {\eightit B} as {\eightit T} varies.  }}
}
\vskip 12pt

What does this mean?  When an external magnetic field $H$ is
applied in the direction perpendicular to the plane, the solution to
Eq. (\induction) is
\vglue 3pt
$$
\left. \eqalign{
&H \le - H_c' \cr -H_c'\le &H \le ~~H_c' \cr  H_c' \le &H \cr} \right.
\hskip 1cm
\left. \eqalign{
&B= H+H_c'~, \cr &B=~ 0 ~, \cr &B=H-H_c'~, \cr}  \right.
\hskip 1cm
\left. \eqalign{
&M= +H_c' ~, \cr &M=-H~, \cr &M=-H_c' ~, \cr} \right.
   \eqn\BHMrelation  $$
\vglue 3pt
where the critical field is given by
$$H_c'={|e|n_e\over 2m} ~~.   \eqn\criticalH  $$
So long as $|H| < H_c'$, there is no magnetic field ($B=0$) in bulk.
It is a Meissner effect.  However, if $|H|$ exceeds $H_c'$, a part of the
external magnetic field penetrates inside the anyon fluid. (See Fig.~11.)
As a consequence of the de Haas -- van Alphen effect, there is the maximum
for $|M|$.

{\figure
\vglue 8.5cm
\centerline{Fig. 11 ~~ {\eightit H} vs {\eightit B} as {\eightit T} varies.}
}
\vskip 12pt

Of course, all of these results have been  obtained in the uniform
approximation.    In reality the formation of vortices would invalidate
the above picture.  Then arises a question which one is smaller, $H_{c1}$
or $H_c'$.

So far all quantities are defined in the effective two-dimensional space.
We have been supposing that three-dimensional material has a layered
structure with interplanar spacing $d \sim 5 \Ang$.  Two-dimensional
quantities (denoted by $(~~)_{d=2}$) are related to three-dimensional
quantities (denoted by $(~~)_{d=3}$) by
$$\eqalign{
\Big( {e^2\over 4\pi} \Big)_{d=2}
  &= {1\over d} \, \alpha_{d=3}
   = {1\over 137} \cdot {1\over d} \cr
B_{d=2} &= \sqrt{d} \, B_{d=3} \cr
b_{d=2} &= ~ b_{d=3} \cr
n_e^{d=2} &= d \, n_e^{d=3}  ~~~. \cr}
   \eqn\relateTwoThree $$
Hence the Chern-Simons magnetic field $b$ in (\CSfield) and the critical
field $H_c'$ in (\criticalH) are given by
$$\eqalign{
\Big( {1\over e } \,b \Big)_{d=3}
  &= {1\over e_{d=3} } \,b_{d=2}
= {1\over N}  \sqrt{ {\pi\over \alpha}} \, n_e^{d=2}  \cr
(H_c')_{d=3} &= \delta^{-1/2} (H_c')_{d=2}
 = {\sqrt{4\pi\alpha} \, n_e^{d=3} \over 2m}  ~~~. \cr}
  \eqn\bHcformula  $$

The hole density $n_e^{d=2}$ and $n_e^{d=3}$, and the spacing $d$ are
measured directly.  Typical values are
$$\eqalign{
n_e^{d=2} &= (1 \sim 5) \times 10^{14} \, {\rm cm}^{-2} \cr
n_e^{d=3} &= (2 \sim 10) \times 10^{21} \, {\rm cm}^{-3} \cr
d~ &= 5 \Ang = 5 \times 10^{-8} {\rm cm} ~~. \cr}
    \eqn\typicalDensity  $$
We also note the conversion formulas:
$$\eqalign{
m_e &= 5.1 \times 10^5 \, {\rm eV} = 2.6 \times 10^{10} \, {\rm cm}^{-1} \cr
1 \, {\rm G} &= 1.779 \times 10^8 \, {\rm cm}^{-2}
 = 6.903 \times 10^{-2} \, {\rm eV}^2  \cr
1 ~{\rm K} &= 8.617 \times 10^{-5} ~ {\rm eV}
  = 4.367  ~ {\rm cm}^{-1} ~~. \cr}
    \eqn\conversionFormula $$

The effective mass $m$ is not directly measurable.  It can be fixed from
observed values for the penetration depth at $T=0$ or $T_c$.  ($T_c$ is
discussed in Section 22.)  Recalling (\MeissnerOne),
one finds
$$\lambda(T=0)= \lambar = \sqrt{ {\strut m\over e^2_{d=2} n_e^{d=2}}}
= \sqrt{ {\strut m\over e^2_{d=3} n_e^{d=3} } } ~~.
     \eqn\penetrationThreeD $$
Therefore
$$ m= 4\pi\alpha \, n_e^{d=3} \, \lambda(0)^2 ~~~.
  \eqn\effectiveMassOne   $$
If one substitutes the values (\typicalDensity) and
$$\lambda(0)=1400 \Ang \quad \Longrightarrow \quad
  m = (1.4 \sim 6.9) ~ m_e ~~~.
   \eqn\effectiveMassTwo  $$
We shall see in Section 22 that
$$T_c \sim 100\, {\rm K} \quad \Longrightarrow \quad
  m = (1 \sim 5) ~ m_e~~~.
   \eqn\effectiveMassThree  $$
(Indeed, the ratio $n_e^{d=3}/m$ is related to $\lambda(0)$ or $T_c$.)

Upon making use of (\effectiveMassOne), $H_c'$ can be written as
$$\eqalign{
H_c' &= {1\over 4 \sqrt{\pi\alpha}} \, {1\over \lambda(0)^2} \cr
&=2 \cdot {2\pi \hbar c\over 2e} \cdot {1\over 4\pi\lambda(0)^2}  ~~~.  \cr}
    \eqn\HcformulaTwo $$
In the Ginzburg-Landau theory of conventional superconductors of type II
$H_{c1}$ is approximately given by\myref{\SCReview}
$$H_{c1} \sim {2\pi \hbar c\over 2e} \cdot {1\over 4\pi\lambda^2}
  \cdot ln {\lambda\over \xi} ~~~,  \eqn\HconeGL $$
where $\xi$ is a coherence length.  If one uses the GL parametrization to
high $T_c$ superconductors, one typically finds\myref{\Burns} $\lambda/\xi \sim
100$ so that $\ln (\lambda/\xi) \sim 4.6$.  In other words,
$${\rm roughly~~~} H_c' \sim H_{c1} ~~~.  \eqn\HcEquality $$

The values for $b$ and $H_c'$ in three dimensions are
$$\eqalign{
{1\over e } \,b
  &=  {2\over N}  \, {n_e^{d=2} \over 2 \times 10^{14} \,{\rm cm}^{-2}}
  ~\cdot ~1.2 \times 10^7  ~ {\rm G}  \cr
\noalign{\kern 5pt}
H_c' &=  {n_e^{d=3}\over 4 \times 10^{21}~{\rm cm}^{-3} }
 \, {2m_e \over m} ~\cdot~ 66 ~{\rm G}  ~~~. \cr}
   \eqn\bHcvalueOne $$
Or, with the aid of (\HcformulaTwo) one has
$$H_c' = \bigg( {1400 \Ang \over \lambda(0)} \bigg)^2 ~\cdot~ 47 ~{\rm G}~~~.
    \eqn\HcvalueTwo  $$

The Chern-Simons magnetic field is huge ($\sim 1000$ T), but $H_c'$ turns
out to be in a modedst range ($\sim $ 50 G).  The rough equality
(\HcEquality) suggests that vortices are formed in anyon fluids, and
that the uniform magnetic field inside the fluid, $B$ in (\BHMrelation),
may represent the average field over the vortex lattice.

The huge magnitude of $b$ might be related to $H_{c2}$ of high $T_c$
superconductors at $T=0$, which is known to be much larger than 100 T.
Related to the huge Chern-Simons magnetic field is the energy spacing in
the Landau level.
$$\eqalign{
{1\over ml^2} &= {~2\pi n_e^{d=2}\over |N|m}  \cr
&= {2\over |N|}\, {2m_e\over m}
  \, {n_e^{d=2}\over 2 \times 10^{14} ~{\rm cm}^{-2}}
 ~\cdot~ 2800 ~{\rm K} ~~(0.24 ~ {\rm eV}) ~.  \cr}
   \eqn\GapValue  $$

Generalization to finite temperature ($T\not=0$) is straightforward.
Magnetization $M(T,B)$ is determined by (\magnetization) with (\ESapprox).
Then, the $B$ vs $H$ relation is found from (\induction).  We have given the
result in Figs. 10 and 11.

As is seen from the figures, magnetization alomost vanishes around 100 K.
This is presumably related to the phenomenon observed in (\crossover), and
therefore need more elaboration by incorpolating vortices.

\sectionskip


\secno=19  \meqno=1

\line{\bf 19.  $T\not= 0$ -- inhomogeneous fields \hfil}
\vglue 5pt
The effective action or free energy at finite temperature can be evaluated
in perturbation theory.  One way is to write down
the partition function in the path integral representation as we did in
Section 9 for $T=0$.  This amounts to rotating a time $t$ to an imaginary
time $\tau=it$ ($0\le \tau \le \beta=1/T$) with appropriate boundary
conditions on the fields imposed.  In other words, we have Matsubara's
finite temperature Green's functions in place of time-ordered Green's
functions at $T=0$.

Most of the arguments in Sections 9 -- 11 remain intact, provided that
the frequency $\omega$ is replaced by Matsubara frequency $i\omega_n$.
[ $\omega_n =2n\pi T$ or $(2n+1)\pi T$ for bosons or fermions, respectively,
where $n$ is an integer.]  In particular, the decomposition
(\GammaDecomposition) of the kernel $\Gamma$ in terms of $\Pi$'s, and the
relation (\finalQn) or (\finalQc) between $\Pi$'s and the response function
$Q_n$ or $Q_c$ are valid after the Wick rotation.  However, one should note
that
all these are for  $\tau$-ordered Matsubara's Green's functions so that to
relate them to, for instance, a response function in real time, one need to
make
necessary  transformations.  There are also  studies in the real-time
formalism of finite temperature Green's functions.\myref{\Eliashvili}

Again we integrate the fermion field $\psi$ first to obtain the effective
theory for Chern-Simons and Maxwell fields.   We expand the gauge fields around
constant magnetic fields.    In the Landau gauge
$$\eqalign{
a^k(\x) &= - \,b^{(0)}\, x_2 + a^{(1)k}(\x) \cr
A^k(\x) &= - B^{(0)} \, x_2  + A^{(1)k}(\x)  \cr}
  \eqn\ExpandCSEM  $$
where
$$b^{(0)} = {2\pi n_e\over N}  \eqn\bZeroth  $$
and $B^{(0)}$ is a constant background magnetic field.  According to
(\newLevelDensity) -- (\newFilling)
$$ \ep(N) \, (b^{(0)} + e B^{(0)}) =  {1\over l^2} = {2\pi n_e\over \nu}  ~~.
     \eqn\newMagLength $$

The fermion part of the Hamiltonian is
$$
H_e[a+eA] =  \int d\x ~ \Big\{ \, {1\over 2m} (D_k\psi)^\dagger (D_k \psi)
  + (a_0+ eA_0) \psi^\dagger \psi \, \Big\}    \eqn\eHamiltonian $$
where $D_k = \d_k - i (a^k + eA^k)$.
We decompose it into the free and interaction parts:
$$\eqalign{
\noalign{\kern 3pt}
H_e[a & +eA] = H_0 + H_\inte = H_0 + V_1 + V_2 \cr
\noalign{\kern 11pt}
H_0 &= H_e[a^{(0)}+eA^{(0)}]
= \int d\x ~  {1\over 2m} (\Dbar_k\psi)^\dagger (\Dbar_k \psi) \cr
\noalign{\kern 5pt}
V_1 &= \int d\x~ \Big( {i\over 2m} \, a^{(1)k}_\tot \, \big\{
\psi^\dagger  \cdot \Dbar_k \psi -(\Dbar_k \psi)^\dagger \cdot \psi \big\}
 + a^0_\tot \, \psi^\dagger \psi \, \Big) \cr
\noalign{\kern 5pt}
V_2 &= \int d\x ~ {1\over 2m} \, \big( a^{(1)k}_\tot \big)^2 \,
  \psi^\dagger \psi   \cr
\noalign{\kern 3pt}
}
   \eqn\decomposeEHam $$
where
$$\left.
\Dbar_k =\d_k -  \delta_{k1} \,i \ep(N) {x_2\over l^2} ~~,
\right. \hskip .5cm
\left. \eqalign{
a^{(1)0}_\tot &= a_0 + e A_0 ~~,\cr \noalign{\kern 3pt}
a^{(1)k}_\tot &= a^{(1)k} + e A^{(1)k} ~~. \cr} \right.   $$

The matter part of the free energy with given static ($t$-independent)
gauge field configurations is defined by
$$ e^{-\beta F_e[a+eA]} = {\rm Tr}_{\, \rm canonical}~ e^{-\beta H_e[a+eA]} ~~.
     \eqn\FreeEnergy  $$
In the previous section we have evaluated the zeroth order free energy:
$$ e^{-\beta F_0} = {\rm Tr}_{\, \rm canonical}~ e^{-\beta H_0} ~~.
     \eqn\FreeEnergyZero  $$
In these formulas  the trace is
taken over states with a fixed number of particles, i.e.~over a canonical
distribution.  One can consider, instead, a grand canonical distribution
to define the thermodynamic potential $\Omega_e$:
$$\eqalign{
 e^{-\beta \Omega_e[a+eA]} &= \tr e^{-\beta (H_e[a+eA]-\mu \N)} \cr
\Omega_e &= F_e - \mu \,  N_e         \cr}
             \eqn\ThermoP  $$
where
$$\eqalign{
\N &= \int d\x ~ \psi^\dagger \psi  \cr
N_e &= n_e \, V = \la \N \ra \cr
\la {\cal Q} \ra &= \tr {\cal Q} \, e^{\beta (\Omega_e-H_e-\mu \N)} ~~.
      \cr}   \eqn\NumberOperator  $$
We have adopted notation $\N$ for the number operator to distinguish it from
the coefficient of the Chern-Simons term, $N$.

The original computation of ref. (\RDSS) was performed for a grand canonical
distribution.  In view of the $n_e$-dependence of $H_0$ through
$b^{(0)}$,  a perturbation theory for a canonical distribution was
employed in ref. (\HHL).  So long as macroscopic physical quantities are
concerned,  there arises no difference between the two.   Even at the diagram
level there is not much difference except for a minor change in the fermion
propagator.

In this article we adopt a perturbation theory for a grand canonical
distribution, which is summarized in the book of Abrikosov et
al..\myref{\Abrikosov}
We outline the  argument in the operator formalism, supplementing
expressions in the path integral formalism.

We define finite temperature Heisenberg field operators by
$$\eqalign{
&\left. \eqalign{
- {\d\over \d\tau}\, \psi(\tau,\x) &= {\cal M} \, \psi(\tau,\x) \cr
+ {\d\over \d\tau}\, \psibar(\tau,\x)
   &= {\cal M}^*  \psibar(\tau,\x) \cr}   \right. \hskip 1cm
\left. \eqalign{
\psi(0,\x) &= \psi(\x)  \cr
\psibar(0,\x) &= \psi^\dagger (\x)  \cr}  \right.      \cr
&\hskip 2cm {\cal M}= - {1\over 2m} \, D_k^2 + a_0+eA_0 -\mu \cr}
 \eqn\THeisenbergEq $$
where $\psi(\x)$ and $\psi^\dagger(\x)$ are the operators in the Schr\"odinger
representation.

If the Hamiltonian $H_e$ is $\tau$-independent,  then the equations can be
integrated as
$$\left. \eqalign{
\psi(\tau,\x) &= e^{\tau(H_e-\mu \N)} \, \psi(\x) \, e^{-\tau(H_e-\mu \N)} \cr
\psibar(\tau,\x)   &= e^{\tau(H_e-\mu \N)} \,
           \psi^\dagger(\x) \, e^{-\tau(H_e-\mu \N)} \cr}  \right.
\hskip .5cm
 {\rm if ~~} {\d\over \d \tau} \, H_e =0  ~~.    \eqn\psitau  $$

At this point we make one technical generalization.  We allow that gauge fields
$a^{(1)}_\mu$ and $A^{(1)}_\mu$ may depend on $\tau$, provided that they are
periodic with a period $\beta$:
$$ a^{(1)}_\mu(\beta,\x) +eA^{(1)}_\mu (\beta,\x)
   = a^{(1)}_\mu (0,\x) +eA^{(1)}_\mu (0,\x) ~~.
   \eqn\periodicCond $$
(We assume that the zeroth order parts are $\tau$-independent.)
This is a technical device which enables us to probe dynamical properties,
namely time-dependent phenomena, of the system at finite temperature, through
appropriate analytic continuation.

The transformation matrix for a general $\tau$-dependent $H_e$ is defined by
$$U(\tau_2,\tau_1) =  ~ {\rm T}_\tau \,
 \exp \bigg\{ - \int_{\tau_1}^{\tau_2} d\tau ~
  \big( H_e(\tau) - \mu\N \big)  \bigg\}  ~,    ~~(\tau_2 > \tau_1)
    \eqn\transformationM  $$
where ${\rm T}_\tau$ indicates the $\tau$-ordering.  The specification of the
ordering is necessary, since $[H_e(\tau), H_e(u)] \not= 0$ for $\tau$-dependent
gauge fields.   By definition
$$U(\tau_3, \tau_2) \, U(\tau_2,\tau_1) = U(\tau_3,\tau_1)
\qquad (\tau_3 > \tau_2 >\tau_1) ~~. \eqn\Udecomposition  $$
Further  $U(\tau) \equiv U(\tau,0)$  satisfies
$$\eqalign{
 {\d\over \d \tau} \, U(\tau) &= - \big( H_e(\tau) - \mu\N \big) ~
U(\tau) \cr
 {\d\over \d \tau} \, U(\tau) ^{-1}
&= + U(\tau)^{-1} \, \big( H_e(\tau) - \mu\N \big) ~~~. \cr} \eqn\Uidentity $$

In terms of $U(\tau)$, Eq. (\THeisenbergEq) is integrated to yield
$$\eqalign{
\psi(\tau,\x) &= U(\tau)^{-1} \, \psi(\x) \, U(\tau) \cr
\psibar(\tau,\x) &= U(\tau)^{-1} \, \psi^\dagger(\x) \, U(\tau) ~~. \cr}
            \eqn\ExpressPsitau  $$
Indeed,
$$\eqalign{
\Big( - {\d\over \d \tau} \Big) \, U(\tau)^{-1} \, \psi(\x) \, U(\tau)
&= U(\tau)^{-1} \, [\psi(\x), H_e(\tau) -\mu\N] ~ U(\tau) \cr
&= U(\tau)^{-1} \, {\cal M} \,\psi(\x) ~ U(\tau) \cr
&= {\cal M} \,\psi(\tau,\x) ~~~.  \cr}  $$
In the last equality we have made use of the fact that the differential
operator
${\cal M}$ commutes with $U(\tau)$.

The next step is to introduce the interaction representation.  With the free
Hamiltonian $H_0$ in (\decomposeEHam), field operators in the interaction
representation are defined by
$$\eqalign{
\psi_\inte(\tau,\x)
  &= e^{\tau(H_0-\mu \N)} \, \psi(\x) \, e^{-\tau(H_0-\mu \N)}  \cr
\psibar_\inte(\tau,\x)
  &= e^{\tau(H_0-\mu \N)} \, \psi^\dagger(\x) \, e^{-\tau(H_0-\mu \N)} ~~. \cr}
   \eqn\IntPsi  $$
The transformation matrix in the interaction representation is given by
$$\eqalign{
S(\tau) = & ~{\rm T}_\tau ~ \exp \bigg\{ - \int_0^\tau d\tau' ~
 \hat H_\inte(\tau')  \bigg\} \cr
\noalign{\kern 8pt}
\hat H_\inte(\tau)
&=e^{\tau(H_0-\mu \N)} \, H_\inte \, e^{-\tau(H_0-\mu \N)}  \cr
&= H_\inte[\psi_\inte(\tau,\x),\psibar_\inte(\tau,\x); (a+eA)(\tau)]   \cr}
   \eqn\TransitionMatrix  $$
The fundamental operator identity is
$$U(\tau) = e^{-\tau(H_0- \mu\N)} \, S(\tau) ~~.
   \eqn\FundamentalIdentity  $$

The diagram method is developed on the basis of (\FundamentalIdentity).
 Let us denote
$$\eqalign{
e^{-\beta \Omega_0 }&= \tr e^{-\beta (H_0 -\mu\N)}   \cr
\la {\cal Q} \ra_0 &= \tr {\cal Q} \, e^{\beta (\Omega_0-H_0 +\mu \N)} ~~. \cr}
    \eqn\OmegaZero $$
We define the Euclidean effective action $I_E[a+eA]$ by
$$e^{-I_E[a+eA]} = \tr U(\beta) ~~.  \eqn\EuclidAction  $$
Taking a trace of (\FundamentalIdentity), one finds
$$I_E[a+eA] = \beta \, \Omega_0 -  \ln \la S(\beta) \ra_0  ~~. $$
The Bloch-De Dominics theorem is applied to $\la S(\beta) \ra_0$, leading to
$$I_E[a+eA] = \beta \, \Omega_0 -
      \Big\{ \la S(\beta) \ra_c - 1 \Big\}      \eqn\EActionOne  $$
where the subscript $c$ indicates that only connected diagrams
be taken into account. For static gauge field configurations, one has
$$
\Omega_e = {1\over \beta} \, I_E
= \Omega_0 - {1\over \beta} \Big\{ \la S(\beta) \ra_c - 1 \Big\}
\hskip .5cm  \hbox{for static $a+eA$}~.
            \eqn\OmegaIdentityOne  $$

The path integral representation is obtained for $\tr U(\beta)$ in
(\EuclidAction) by the standard technique:
$$\eqalign{
e^{-I_E[a+eA]}
&= \int_{\rm B.C.} \D \psibar \D \psi
{}~ \exp \bigg\{ - \int_0^\beta d\tau  \int d\x ~  \Lbar_e[a+eA]  ~ \bigg\}
\cr
\noalign{\kern 6pt}
\Lbar_e &=  \psibar \dot \psi +  {1\over 2m} \,D_k^* \psibar D_k \psi
          + (a_0+ eA_0 - \mu ) \psibar \psi   \cr
{\rm B.C.} &: ~~~ \psi(\beta,\x) = - \psi(0,\x) ~~,~~
     \psibar(\beta,\x) = - \psibar(0,\x)  \cr}
    \eqn\pathintegralEAction  $$
As it stands, this expression is formally obtained from (\defineSf) by
Wick-rotating the time axis through 90 degrees and imposing the anti-periodic
boundary condition on the fermion fields.  One should remember however that
the rigorous derivation follows from (\EuclidAction), and that the
expression (\pathintegralEAction) is for grand canonical distributions.

The free propagator for the fermion field $\psi$ is defined by
$$G(\x,\y;\tau_1 - \tau_2) = - \la {\rm T}_\tau [ \psi_\inte(\x,\tau_1)
  \,\psibar_\inte(\y,\tau_2) ] \ra_0
 \hskip .5cm (0<\tau_1,\tau_2 < \beta) ~.  \eqn\Tpropagator  $$
It is easy to see
$$G(\x,\y;\tau) = - G(\x,\y;\tau+ \beta) \hskip .5cm {\rm for} ~~
                 -\beta<\tau<0  ~~. \eqn\TpropagatorTwo $$
In the Landau gauge, analogously to (\GreenTwo), we find
$$\eqalign{
G(\x,\y; \tau)
&={1\over l L_1} \sum_{n,p}  e^{-2\pi ip(x_1-y_1)/L_1}~
v_n\Big({x_2-\bar x_2 \over l}\Big) v_n \Big({y_2-\bar y_2 \over l}\Big) \cr
\noalign{\kern 2pt}
&\hskip 1.8cm \times e^{-(\ep_n -\mu)\tau} \,
 \cases{ \rho_n(\beta) -1  &for $~~ 0<\tau<\beta$     \cr
         \rho_n(\beta) &for $-\beta< \tau <0$  \cr}   \cr
\noalign{\kern 9pt}
&=e^{i \phi(x,y) }  \,\cdot \, G_0(\x-\y,\tau) ~~~. \cr
\noalign{\kern 4pt}   }
   \eqn\TGreenTwo $$
Here
$$\eqalign{
\phi(x,y) &= - \ep(N) \, {1\over 2l^2} \,(x_1-y_1)(x_2+y_2)  \cr
\noalign{\kern 8pt}
G_0(\x,\tau) &= \sum_n {1\over 2\pi l^2}
\int_{-\infty}^\infty dz ~ e^{-izx_1/l} ~
  v_n[z-\bar z(x_2) ] \, v_n [ z+\bar z(x_2) ] \cr
&\hskip 1.8cm \times e^{-(\ep_n -\mu)\tau} \,
 \cases{ \rho_n(\beta) -1  &for $~ 0<\tau<\beta$     \cr
         \rho_n(\beta) &for $-\beta< \tau <0$  \cr}   \cr
\noalign{\kern 7pt}~~
 &\hskip .5cm \bar z(x_2) =  \ep(N) ~{x_2\over 2l} ~~.  \cr}
    \eqn\TGreenThree   $$
As in the $T=0$ case, the Green's function $G(\x,\y;\tau)$ is not manifestly
translation invariant, but is invariant  up to a gauge transformation.

In the Fourier space
$$\eqalign{
&G_0(\x,\tau) = {1\over \beta} \sum_r \int {d\p\over (2\pi)^2} ~
     G_0(\omega_r,\p) \, e^{-i\omega_r \tau + i\p\,\x}  \cr
\noalign{\kern 9pt}
&G_0(\omega_r,\p) =  \int_0^\beta d\tau \int d\x
 ~ G_0(\x,\tau) \, e^{i(\omega_r \tau  - \p \,\x)}  \cr
\noalign{\kern 6pt}
&\hskip .5cm =\sum_{n=0}^\infty ~ {1\over i \omega_r - \ep_n + \mu}
{}~ \int {dx_2\over l} ~ e^{-ip_2x_2} ~
 v_n[p_1l+\bar z(x_2)] \, v_n[p_1l - \bar z(x_2)]  \cr
\noalign{\kern 8pt}
&\hskip 6.cm {\rm where~~} \omega_r = {(2r+1)\pi\over \beta} ~~~.  \cr}
   \eqn\TFourierGreen  $$

We also note that
$$\eqalign{
\la \psibar(\x,\tau)&\psi(\x,\tau) \ra_0
= G(\x,\x;0^-)  \cr
&= \sum_n {1\over 2\pi l^2} \int_{-\infty}^\infty dz ~
    v_n(z)^2 ~ \rho_n(\beta)
= {1\over 2\pi l^2} \sum_n \rho_n \cr
&= n_e ~~~.   \cr}
   \eqn\TDensity  $$
Similarly
$$\la \psibar \cdot \Dbar_k \psi - \Dkstarbar\psibar \cdot \psi \ra =0  ~~~.
   \eqn\TCurrent  $$
(\TDensity) and (\TCurrent) appear in the first order perturbation.

As in (\SfExpansion), one can expand the Euclidean effective action $I_E$ in a
power series of $a^{(1)} +eA^{(1)}$:
$$\eqalign{
I_E&[a+eA]\cr
\noalign{\kern 5pt}
&= \beta \,\Omega_0
 +  \int dx ~ n_e (a_0 + eA_0)(x)  \cr
&\hskip .5 cm -  \int dxdy ~
 {1\over 2} (a+eA)^{(1)}_\mu(x) \,\Gamma_E^{\mu\nu}(x-y)
  (a+eA)^{(1)}_\nu(y)  + \cdots  \cr
\noalign{\kern 8pt}
&= \beta \,\Omega_0
 +  \int dx ~ n_e (a_0 + eA_0)(x)  \cr
&\hskip .5 cm - {1\over \beta} \sum_r \int {d\p\over (2\pi )^2} ~
 {1\over 2} (a+eA)^{(1)}_\mu(-p)
\,\Gamma_E^{\mu\nu}(p)
  (a+eA)^{(1)}_\nu(p)  + \cdots  \cr}
    \eqn\OmegaExpansion  $$
where
$$\eqalign{
&x = (\tau_1,\x) \next  y=(\tau_2, \y)  \next
 \int dx = \int_0^\beta d\tau \int d\x\cr
&p=(\omega_r, \p) \next \omega_r = {2\pi r\over \beta}  ~~~. \cr}   $$

A relationship between $\Gamma_E^{\mu\nu}$ and the response function
at finite temperature can be found easily.  Let us denote
$$\eqalign{
\jzbar(\tau,\x) &=  ~~\psibar(\tau,\x) \psi (\tau,\x)    \cr
\jkbar(\tau,\x) &= - {i\over 2m} \Big\{ \psibar(\tau,\x)\cdot D_k \psi(\tau,\x)
  -  D_k^* \psibar(\tau,\x) \cdot \psi(\tau,\x)  \Big\} ~. \cr}
     \eqn\jbarDef  $$
Then from  (\transformationM) and (\ExpressPsitau) it follows that
$$\eqalign{
{\delta \tr U(\beta) \over \delta a_0(\tau,\x)}
&= - \tr U(\beta,\tau) \, j^0(\x) \, U(\tau,0)  \cr
&= - \tr U(\beta,0) \, \jzbar(\tau,\x)  \cr
&= - \la \jzbar(\tau,\x) \ra \cdot \tr U(\beta)   \cr}  $$
so that
$$\eqalign{
{\delta I_E[a+eA] \over \delta a_0(\tau,\x)} &=+ \la \jzbar(\tau,\x) \ra  \cr
\noalign{\kern 6pt}
{\delta I_E[a+eA] \over \delta a^k(\tau,\x)} &=- \la \jkbar(\tau,\x) \ra
{}~.\cr}
    \eqn\jbarIE  $$
Obviously
$$\la \jmubar (\tau,\x) \ra = \la j^\mu(\x) \ra
\hskip .6cm \hbox{for static $a+eA$} .  \eqn\staticJIE $$

We denote the thermal average of the induced current by
$$\Jmubar (\tau,\x) = \la \jmubar(\tau,\x) \ra - \delta^{\mu 0} n_e ~~~.
    \eqn\indJtau   $$
Making use of (\OmegaExpansion) and (\jbarIE), one finds
$$\Jmubar (x) = - \int dy ~ \Gamma_E^{\mu\nu}(x-y) \, (a+eA)_\nu^{(1)} (y) +
  \cdots  ~~~, $$
or in the Fourier space
$$\Jmubar (\omega_r,\p) = -  \Gamma_E^{\mu\nu}(\omega_r,\p) \,
   (a+eA)_\nu^{(1)} (\omega_r,\p) + \cdots  ~~~.
   \eqn\JbarGamma $$
As in the zero temperature case, the self-consistent field approximation (SCF)
is defined by Eq. (\JbarGamma) and the field equations for the gauge fields
with the source replaced by the thermal average $\la \jmubar(\tau,\x) \ra$.

Of course, for  general $\tau$-dependent field configurations, appropriate
analytic continuation of the field equations is necessary.  In this article
we mostly restrict ourselves to physics for static configurations
($\omega_r=0$), for which field equations are the same as those at $T=0$.  In
particular, the response functions $Q_n$ and $Q_c$ are defined in the same way
as at $T=0$, by introducing static external fields.  In the linear
approximation,
in which higher order terms in (\JbarGamma) are neglected, one has
$$\Jmubar^{\rm linear} = Q_n^{\mu\nu} \, a_\nu^\ext ~~~ {\rm or} ~~~
   Q_c^{\mu\nu} \, a_\nu^\ext ~~~ .    \eqn\JbarQ  $$
The relation among $\Gamma$ , $Q_n$, and $Q_c$  remains intact for
$\omega_r=0$.   (The relation is valid even for $\omega_r\not= 0$ upon
the substitution $\omega \go i \omega_r$.)

We examine the current conservation:
$$\eqalign{
i\, {\d\over \d \tau} \, \jzbar(\tau,\x) &= U(\tau)^{-1} i \,
[H_e(\tau)-\mu\N, j^0(\x) ] ~ U(\tau) \cr
&=  U(\tau)^{-1}  \Big( - \nabla_k j^k(\x) \Big|_{(a+eA)(\tau)} \Big) \,
   U(\tau)  \cr
&= - \nabla_k \jkbar(\tau,\x)  ~~.   \cr}   $$
Hence
$$  i\,{\d\over \d \tau} \, \jzbar + \nabla_k \jkbar =0 ~~.
   \eqn\TcurrentConservation $$
For the kernel $\Gamma^{\mu\nu}$ it implies that
$$i \omega_r \Gamma^{0\nu}_E - p_k \, \Gamma_E^{k\nu} =0
=i \omega_r \Gamma_E^{\mu 0} - p_k \, \Gamma_E^{\mu k}~~~.
   \eqn\TGammaConservation  $$

Other relations such as $\Gamma_E^{\mu\nu}(p)= \Gamma_E^{\nu\mu}(-p)$
remain intact.
The decomposition of $\Gamma_E^{\mu\nu}$ is given by
$$\eqalign{
\noalign{\kern 3pt}
\Gamma_E^{00}(\omega_r,\q) &= q^2 \Pi^E_0 \cr
\noalign{\kern 3pt}
\Gamma_E^{0j}(\omega_r,\q)
   &= +i\omega_r q_j \Pi^E_0 - i \ep_{jk} q_k \Pi^E_1 \cr
\noalign{\kern 3pt}
\Gamma_E^{j0}(\omega_r,\q)
   &= +i\omega_r q_j \Pi^E_0 + i \ep_{jk} q_k \Pi^E_1  \cr
\noalign{\kern 3pt}
\Gamma_E^{jk}(\omega_r,\q) &= - \delta_{jk} \,\omega_r^2 \Pi^E_0
  - \ep_{jk}\omega_r \Pi^E_1  - (q^2 \delta_{jk} -q_jq_k)\, \Pi^E_2 \cr
\noalign{\kern 3pt}
}     \eqn\TGammaDecomposition $$
where all $\Pi_j$'s are functions of $\omega_r^2$ and $\q^2$ only.
In a  frame $\q= (q, 0)$
$$\eqalign{
\noalign{\kern 3pt}
&\Gamma_E^{\mu\nu} = \left( \matrix{
q^2 \Pi^E_0 & i\omega_r q \Pi^E_0 & +iq \Pi^E_1 \cr
\noalign{\kern 4pt}
i\omega_r q \Pi^E_0 & -\omega_r^2 \Pi^E_0 & -\omega_r \Pi^E_1 \cr
\noalign{\kern 4pt}
- iq \Pi^E_1 & + \omega_r \Pi^E_1 & -\omega_r^2 \Pi_0 - q^2 \Pi^E_2 \cr}
     \right) ~~. \cr
\noalign{\kern 3pt} }
  \eqn\TGammaMatrix $$
Mathematically
$$\Gamma^{\mu\nu}_E = \Gamma^{\mu\nu}(i\omega_r,\q)  \next
\Pi^E_k = \Pi_k(-\omega_r^2,q^2) ~~~.  \eqn\relateGandGE $$

As in the $T=0$ case, we need to evaluate one loop diagrams in Fig. 8 in
Section 10 to find $\Gamma_E^{\mu\nu}$.   Computations are completely
parallel to those in Section 12.  The diagram (a) yields the linear term
in (\OmegaExpansion).  The diagram (b) yields
$$\Gamma_E^{{\rm (b)}jk} = - \delta^{jk} \, {1\over m} \,
   \la \psibar \psi \ra_0 = - \delta^{jk} \, {n_e \over m} ~~~.
   \eqn\TGb   $$

For the diagrams (c), (d), and (e), the phase factor $\phi(x,y)$ in the
propagator $G(x,y)$, (\TGreenTwo), completely cancells. We have
$$\eqalign{
\Gamma_E^{{\rm (c)}00}
&= - {1\over \beta} \sum_s \int{d\p\over (2\pi)^2} ~
   G_0(p) \, G_0(p-q)  \cr
\Gamma_E^{{\rm (d)}0j}
&= - {i\over 2m\beta} \sum_s \int{d\p\over (2\pi)^2} ~
  \Big\{  G_0(p) \cdot D_j^- G_0(p-q)  + D_j^+ G_0(p) \cdot G_0(p-q) \Big\} \cr
\Gamma_E^{(e)jk}
&= {1\over 4m^2\beta} \sum_s \int{d\p\over (2\pi)^2} ~
\Big\{
D^-_k G_0(p) \, D^-_j G_0(p-q) + D^+_j G_0(p) \, D^+_k G_0(p-q) \cr
&\hskip 2.cm +D^+_jD^-_k G_0(p) \cdot  G_0(p-q) + G_0(p)
      \,  D^-_jD^+_k G_0(p-q) \Big\}  \cr}
    \eqn\TGcde $$
where
$$\eqalign{
&p=(\omega_s,\p) \next \omega_s = {2\pi (s+{1\over 2})\over \beta} \cr
&q=(\omega_r, \q) \next \omega_r = {2\pi r\over \beta}~~~.  \cr}  $$
$G_0(\omega_s,\p)$ is defined in (\TFourierGreen).  Without confusion we have
adopted the same notation for the propagator as in the $T=0$ case.

The only technical change to be made in comparison with the computations in
Section 12 is the infinite sum over frequencies.  Employing the formula
$$\sum_{s=-\infty}^\infty g(s) = - \sum_{{\rm poles}\,a_j\,{\rm of}\,g(z) }
{\rm ~Res~} \big( \pi \cot \pi z \cdot  g(z) \, , \, a_j \,\big)  ~~~, $$
one easily finds that
$$\eqalign{
f(\omega_r; n,m) &\equiv - {1\over \beta} \sum_s
 {1\over \strut [ \, i\omega_s - \ep_n +\mu \, ] ~
[\, i(\omega_s-\omega_r) - \ep_m +\mu \, ]}    \cr
\noalign{\kern 10pt}
&=\cases{
\big ~~ \beta \, \rho_n \, (1-\rho_n) &for $\omega_r=0$ and $n=m$,\cr\cr
\big -{\rho_n - \rho_m\over \ep_n-\ep_m - i\omega_r} &otherwise.  \cr} \cr}
   \eqn\omegaSum  $$
$\rho_n(\beta)$ is the distribution function for the $n$-th Landau level.
This is the only place where $\rho_n$ shows up in the computation of
$\Gamma_E^{\mu\nu}$.  In other words, finite temperature effects
in the linear approximation are
contained solely in the discrete sum above.  We shall see that the diagonal
component at zero frequency, $n=m$ and $\omega_r=0$, leads to unique behavior
of anyon fluids at $T\not= 0$.

Working in the frame $\q=(q,0)$, one finds  $\Pi^E_k$'s  to be
$$\eqalign{
q^2 \, \Pi^E_0 = \Gamma^{00}_E
&= {1\over 2\pi l^2} \sum_{n=0}^\infty  \sum_{m=0}^\infty
  f(\omega_r; n,m) \, C^{(0)}_{nm}(ql)^2 ~~, \cr
\noalign{\kern 10pt}
q \, \Pi^E_1 = -i \, \Gamma^{02}_E
&={\ep(N)\over 2\pi ml^3}  \sum_{n=0}^\infty  \sum_{m=0}^\infty
  f(\omega_r; n,m) \, C^{(1)}_{nm}(ql)\,C^{(0)}_{nm}(ql) ~~,\cr
\noalign{\kern 10pt}
- \omega_r^2 \Pi^E_0 - q^2 \Pi^E_2 = \Gamma^{22}_E
&=- {n_e\over m} +
{1\over 2\pi m^2l^4}  \sum_{n=0}^\infty  \sum_{m=0}^\infty
  f(\omega_r; n,m) \, C^{(1)}_{nm}(ql)^2 ~~.\cr}
   \eqn\TPis  $$
We are going to examine implications of the above result, particularly
in the static case ($\omega_r=0$), in the following three sections.

\sectionskip


\secno=20  \meqno=1

\line{\bf 20.  Thermodynamic potential in inhomogeneous fields \hfil}
\vglue 5pt
In this section we first summarize the result in the previous section in the
form of free energy or thermodynamic potential
 for slowly varying static gauge fields
confugurations, from which it follows that the Meissner effect becomes partial
at finite temperature at least in the self-consistent field method (SCF).
Although it may be an artifact of the approximation which neglects
vortices,   the behavior found here is unique and seems essential for
understanding properties of anyon fluids.

The infinite sum in (\TPis) can be performed for  static,
slowly varying gauge field configurations.  Let us define
$$\eqalign{
\Delta_p &= \sum_{n=0}^\infty \big( n+ \hbox{${1\over 2}$} \big)^p ~ \rho_n \cr
\delta_p~ &= \sum_{n=0}^\infty \big( n+ \hbox{${1\over 2}$} \big)^p ~
       \rho_n \, (1-\rho_n) ~~. \cr}
         \eqn\DefDeltap  $$
It follows from (\newFilling) and (\nurhoRelation) that
$$\eqalign{
&\Delta_0(T,B) = \nu \cr
&\Delta_0(T,0) = |N| ~~~. \cr}   \eqn\DeltaZero  $$
At $T=0$ and $B=0$, $\rho_n=0$ or 1 so that
$$\eqalign{
\Delta_0 &= |N| \next \Delta_1 = {N^2\over 2} \next
 \Delta_2 = {|N| (4N^2 -1)\over 12} \next \cdots \cr
\noalign{\kern 6pt}
\delta_n &=0 ~~~,\cr
&\hskip 5cm {\rm at~~~} T=0 ~,~ B=0 ~~~.  \cr}  \eqn\zeroDelta  $$
In evaluating off-diagonal sums ($n\not=m$) in (\TPis), one also needs
$$\sum_{n=0}^\infty I_n = S[\,I\,] $$
where
$$\vcenter{
\halign{ $\big #$\hfil &\qquad\quad $\big #$ \hfil \cr
\noalign{\kern 5pt}
\qquad ~~I_n & ~~~ S[\,I\,] \cr
\noalign{\kern 6pt}
(\rho_n - \rho_{n-1}) \, n   & - \Delta_0 \cr
\noalign{\kern 4pt}
(\rho_n - \rho_{n-1}) \, n^2 & - 2\Delta_1 \cr
\noalign{\kern 4pt}
(\rho_n - \rho_{n-1}) \, n^3
        & - 3 \Delta_2 - \hbox{${1\over 4}$} \Delta_0 \cr
\noalign{\kern 4pt}
(\rho_n - \rho_{n-2}) \, n(n-1)
        & - 4\Delta_1 \cr
\noalign{\kern 4pt}
(\rho_n - \rho_{n-2}) \, n(n-1)(2n-1)
        & -12 \Delta_2 - 3  \Delta_0 \cr
\noalign{\kern 4pt}
(\rho_n - \rho_{n-3}) \, n(n-1)(n-2)
        & -9 \Delta_2 - \hbox{${15\over 4}$} \Delta_0 \cr
\noalign{\kern 5pt}
}   }  \eqn\sumIdentity $$

With these preparations, we start to evaluate $\Pi^E_0$ in (\TPis) at
$\omega_r=0$.   Employing (\omegaSum), one finds
$$
q^2 \Pi^E_0(0,q^2) =
 - {m\over 2\pi} \sum_{n\not= m} {\rho_n-\rho_m\over n-m}
  ~ C^{(0)}_{nm}(ql)^2
+{\beta\over 2\pi l^2} \sum_{n=0}^\infty
  \rho_n (1-\rho_n) ~ C^{(0)}_{nn} (ql)^2
  \eqn\PiEzeroOne $$
where we have made use of $\ep_n=(n+{1\over 2})/(ml^2)$.  For slowly varying
configurations ($ql \ll 1$) the expansions (\CformulaOne) -- (\CformulaTwo)
can be employed:
$$\eqalign{
\noalign{\kern 5pt}
q^2 \Pi^E_0 (0,q^2) =
 - {m\over \pi} \sum_{n>m} {\rho_n-\rho_m\over n-m}
  ~\bigg\{ {n\over 2}\, \delta_{n,m+1}\, (ql)^2 \hskip 4.5cm \cr
- \Big(  {n^2\over 4}\,
 \delta_{n,m+1} - {n(n-1)\over 16} \, \delta_{n,m+2} \Big) \, (ql)^4
  + \cdots \bigg\}&   \cr
 + {\beta\over 2\pi l^2} \sum_{n=0}^\infty
  \rho_n (1-\rho_n) ~\bigg\{ 1 - {2n+1\over 4} \, (ql)^2
 + {2n^2+2n+1\over 32}\, (ql)^4 + \cdots \bigg\} & \cr
\noalign{\kern 5pt}
}  $$
so that
$$\eqalign{
\noalign{\kern 5pt}
\Pi^E_0(0,q^2)
&= {\nu m\over 4\pi^2 n_e} \, \bigg\{ \nu  - {3\over 4} \,
\Delta_1 \, (ql)^2 + \cdots \bigg\}   \cr
&\hskip .5cm + {1\over q^2}\, {\beta n_e\over \nu}\, \delta_0
- {\beta \over 2\pi} ~ \bigg\{  \delta_1
 - \Big( {3\over 8}\, \delta_2 + {1\over 32}\,\delta_0 \Big) \, (ql)^2
 + \cdots \bigg\} ~~~.  \cr
\noalign{\kern 5pt}
}     \eqn\PiEzeroTwo  $$
$\Pi^E_1$ and $\Pi^E_2$ are similarly evaluated.  One finds that for $\Pi^E_1$
$$\eqalign{
\noalign{\kern 5pt}
\Pi^E_1(0,q^2) &=
 + {\ep(N)\over 2\pi} \, \bigg\{ \Delta_0
 - {3\over 2}\, \Delta_1 \, (ql)^2 + \cdots \bigg\} \cr
\noalign{\kern 3pt}
&\hskip 2cm - {\ep(N)\, \beta\over 2\pi ml^2 } \, \bigg\{ \delta_1 -
   \Big( {3\over 4} \, \delta_2 + {1\over 16}\, \delta_0 \Big) \, (ql)^2 +
   \cdots \bigg\} \cr
\noalign{\kern 6pt}
&= +\ep(N)\, {1\over 2\pi} \, \bigg\{ \nu
 - {3\over 2}\, \Delta_1 \, (ql)^2 + \cdots \bigg\}  \cr
\noalign{\kern 3pt}
&\hskip 1.5cm  - \ep(N)\,{\beta n_e\over m\nu } \, \bigg\{ \delta_1 -
   \Big( {3\over 4} \, \delta_2 + {1\over 16}\, \delta_0 \Big) \, (ql)^2 +
   \cdots \bigg\} ~~~.  \cr}
   \eqn\PiEone $$
For $\Pi^E_2$
$$\eqalign{
- q^2 \, \Pi^E_2 (0,q^2) = - {n_e\over m}
+ {1\over 2\pi ml^2} \bigg\{ \Delta_0 - 2 \Delta_1\, (ql)^2
  + \Big( {3\over 2}\,\Delta_2 + {1\over 8}\, \Delta_0 \Big) \, (ql)^4
  +\cdots \bigg\} & \cr
\noalign{\kern 5pt}
 + {\beta\over 2\pi m^2l^4} \, \bigg\{ \delta_2 \,(ql)^2
  - \Big( {1\over 2}\,\delta_3 + {1\over 8}\,\delta_1\Big) \, (ql)^4
 + \cdots \bigg\} & \cr
\noalign{\kern 3pt}
 } $$
so that
$$\eqalign{
\Pi^E_2 (0,q^2)
= +  {1\over 2\pi m} \bigg\{ 2 \Delta_1
  - \Big( {3\over 2}\,\Delta_2 + {1\over 8}\, \Delta_0 \Big) \, (ql)^2
  +\cdots \bigg\} & \cr
\noalign{\kern 5pt}
 -{\beta n_e\over m^2 \nu} \, \bigg\{ \delta_2
  - \Big( {1\over 2}\,\delta_3 + {1\over 8}\,\delta_1\Big) \, (ql)^2
 + \cdots \bigg\} &  ~~.\cr }
       \eqn\PiEtwo  $$

There are a few things to be recognized.   $\Pi^E_0$
develops a pole $(1/q^2)$ at $T\not=0$, which, as we shall see shortly,
leads to a partial Meissner effect.
We also argue in the next section that it determines the scale of the
phase transition temperature, $T_c$.  Secondly, the
diamagnetic term $n_e/m$ in $q^2\Pi^E_2$ is cancelled by the induced term
as at $T=0$.  No pole develops in $\Pi^E_2$.
Thirdly $\Pi^E_1(0,0)$ determines the
induced Chern-Simons term, which is not exactly $N/2\pi$ at $T\not=0$.
In other words, the cancellation between the bare and induced Chern-Simons
terms is not exact.   In some of the early literature in the anyon
superconductivity it was said that the exact cancellation is essential for
superconductivity, which, as we shall show, is rather misleading.

It follows from  (\OmegaExpansion) and (\TGammaDecomposition) that
$$\eqalign{
\Omega_e&[a,A] = \Omega_0[a,A] + \int d\x ~ n_e(a_0+eA_0) \cr
\noalign{\kern 4pt}
& - \int d\x ~ \bigg\{ {1\over 2} (a+eA)_0 ~q^2 \Pi^E_0~ (a+eA)_0
 - (a+eA)_0 ~\Pi^E_1~ (b^{(1)}+eB^{(1)}) \cr
&\hskip 4.3cm - {1\over 2} (b^{(1)}+eB^{(1)}) \Pi^E_2 (b^{(1)}+eB^{(1)})
\bigg\}   + \cdots~~,\cr}  $$
where $\Pi^E_j(0,q^2)=\Pi^E_j(0, -\nabla^2)$.
Insertion of (\PiEzeroTwo) -- (\PiEtwo) leads to
$$\eqalign{
&\Omega_e[a,A]
= \Omega_0[a,A] + \int d\x ~ \bigg\{ ~ n_e(a_0+eA_0) \cr
\noalign{\kern 5pt}
&\hskip .2cm -{\beta n_e\over 2\nu} \, \delta_0 \, (a_0+eA_0)^2
 - \bigg( {\nu^2 m\over 8\pi^2 n_e} - {\beta\over 4\pi} \, \delta_1
\bigg)  \big( \nabla (a_0+eA_0) \big)^2 \cr
&\hskip .4cm + \bigg( {3\nu^2 m\over 64 \pi^3 n_e^2}\, \Delta_1
-{\nu\beta\over 256 \pi^2 n_e} \,(\delta_0+12\delta_2) \bigg)
  \big( \nabla^2 (a_0+eA_0) \big)^2 \cr
\noalign{\kern 5pt}
&\hskip .2cm + \ep(N) \,\bigg( {\nu\over 2\pi} - {\beta n_e\over m\nu}
\,\delta_1 \bigg) \, (a_0+eA_0) \, (b^{(1)}+eB^{(1)}) \cr
&\hskip .4cm - \ep(N) \,\bigg( {3\nu\over 8\pi^2n_e} \,\Delta_1 -
{\beta\over 32\pi m} \,(\delta_0+12\delta_2) \bigg) \,
 \nabla (a_0+eA_0) \nabla (b^{(1)}+eB^{(1)})  \cr
\noalign{\kern 5pt}
&\hskip .2cm + \bigg( {1\over 2\pi m} \, \Delta_1 - {\beta n_e\over 2m^2\nu}
\,\delta_2 \bigg) \, (b^{(1)}+eB^{(1)})^2 \cr
&\hskip .4cm -\bigg( {\nu\over 64\pi^2m n_e}\,(\nu+12\Delta_2) -
{\beta n_e\over 32\pi m^2}\,(\delta_1+4\delta_3) \bigg) \,
 \big( \nabla (b^{(1)}+eB^{(1)}) \big)^2 \cr
&\hskip .2cm  + \cdots ~ \bigg\} ~.  \cr}
     \eqn\finiteOmegaOne $$

In  applications at long wave length, dominant terms are given by
$$\eqalign{
\Omega_\tot& [a,A] =  \Omega_e[a,A]
+ \int d\x ~\bigg\{~ {1\over 2} (F_{0k}^2 + B^2) - en_e A_0
- {N\over 2\pi}\, a_0 \, b ~ \bigg\} \cr
\noalign{\kern 5pt}
= &{\rm (const)} + \int d\x ~\bigg\{~ {1\over 2} (F_{0k}^2 + B^2)
- {N\over 2\pi}\, a_0 \, b^{(1)}  \cr
\noalign{\kern 5pt}
&\hskip .2cm - {1\over 2} \Big( {\nu\over 2\pi} \Big)^2 \,
   \, e^2 \, c_0\, (a_0+eA_0)^2
 - {\nu^2 m\over 8\pi^2 n_e}( 1- c_1) ( f_{0k}+eF_{0k} )^2 \cr
\noalign{\kern 5pt}
&\hskip .2cm + \ep(N) \, {\nu\over 2\pi}\,(1-c_1)
       \, (a_0+eA_0) \, (b^{(1)}+eB^{(1)}) \cr
\noalign{\kern 5pt}
&\hskip .2cm + {\Delta_1\over 2\pi m} \, (1-c_2) \, (b^{(1)}+eB^{(1)})^2
+ ~ \cdots ~~ \bigg\} ~~~~~, \cr}
     \eqn\finiteOmegaTwo $$
where dimensionless constants $c_j(T,B)$'s are defined by
$$c_0={4\pi^2 n_e \beta\over e^2 \nu^3}\,\delta_0 \next
c_1={2\pi n_e\beta\over m\nu^2}\, \delta_1 \next
c_2={\pi n_e\beta\over m\nu \Delta_1} \,\delta_2 ~~~.
   \eqn\cDefinition $$
Equations are given by
$$
{\delta \Omega_\tot \over \delta a_\mu(\x)}  = 0
={\delta \Omega_\tot \over \delta A_\mu(\x)}   ~~~.
  \eqn\effectiveEquation $$

It is appropriate to examine numerical values of various coefficients.
With the values $m=2m_e$, $n_e= 2 \times 10^{14} \, {\rm cm}^{-2}$, and
$d=5 \A$,
$${m e^2\over \pi^2 n_e} = 48 \next
{e^2\over \pi m} = 1.1 \times 10^{-5}
\next {\pi n_e\over m^2} =2.4 \times 10^{-7}  ~~~.
   \eqn\valueOne $$
$\delta_p$ in $c_k$ is approximately given by, for $N=\pm 2$,
$$\delta_p \sim {\rm Max} \, \Big( {2eB^{(0)}\over b^{(0)}} \, , \,
e^{-\ep/2T} \Big) \sim
  {\rm Max} \, \Big( {B^{(0)}\over 6 \times 10^6 \,{\rm G}} \, , \,
10^{-6 \cdot (100 {\rm K}/T) }  \Big) ~. \eqn\valueTwo $$
Other relevant coefficients are, for $T=$ 100 K,
$${\pi^2 n_e\over 2 e^2 T} = 1.2 \times 10^6 \next
{\pi n_e\over 2mT} = 14 ~~~. \eqn\valueThree  $$

All $c_k$'s at $B^{(0)}=0$ are suppressed exponentially in the $T\go 0$ limit.
$c_1$ and $c_2$ are negligible ($\ll 1$) for $T < 200$ K, whereas $c_0$
suddenly  becomes large around $T= 100$ K.  (See the discussion in
Section 22, around (22.2).)    The dominant finite temperature effect is
contained in the $c_0$ term in (\finiteOmegaTwo), which represents an effect
similar to the Debye screening in plasmas.

\sectionskip

\secno=21 \meqno=1

\line{\bf 21.  Partial Meissner effect in SCF \hfil}
\vglue 5pt
Anyon fluids have quite unusual behavior at $T\not= 0$.  In this section we
examine a response against static inhomogeneous external perturbations, both
solving in real configurations and looking at response functions.\myref{\HHL}
We write Eq. (\effectiveEquation) in the form
$$\left. \eqalign{
- {N\over 4\pi} \, \eps f_{\nu\rho}
     &=  J_\ind^\mu +   n_e \delta^{\mu 0}\cr
\d_\nu F^{\nu\mu} &= e J_\ind^\mu  \cr}  \right.  ~~~,
\hskip .6cm
J_\ind^\mu = {\delta \Omega_e \over \delta a_\mu(\x) } ~~~.
   \eqn\effEquationTwo  $$
(Note that $a_k= - a^k$.)
The induced current $J_\ind^\mu(\x)$ is given by
$$\eqalign{
J_\ind^0&(\x) = \ep(N) {\nu\over 2\pi} (1-c_1) (b^{(1)}+eB^{(1)}) \cr
&\hskip .3cm -  \Big( {\nu\over 2\pi} \Big)^2 \, \, e^2 \, c_0\, (a_0+eA_0)
  - {\nu^2 m\over 4\pi^2 n_e} \, (1-c_1) \, \d_k (f_{0k} + e F_{0k} )
 + \cdots ~~, \cr
J_\ind^k&(\x) = \ep(N) {\nu\over 2\pi} (1-c_1) \ep^{kl} (f_{0l}+eF_{0l}) \cr
&\hskip .3cm
- {\Delta_1 \over \pi m} (1-c_2) \ep^{kl} \d_l (b^{(1)}+eB^{(1)})
 + \cdots ~~. \cr}
  \eqn\indCurrentOne $$
We remark that at finite temperature not only field strengths but also
the time component of the vector potentials, $a_0+eA_0$, appears in the
expression for $J_\ind^0$ in (\indCurrentOne).

The identities
$$\eqalign{
{N\over 2\pi} \, eb^{(1)} &= div\, {\bf E} ~~~,  \cr
{N\over 2\pi} \, ef_{0k} &= \d_k B  ~~~, \cr}  \eqn\identityFf $$
may be employed to eliminate the Chern-Simons fields.   Note that the
integration of the latter leads to
$${N\over 2\pi} \, ea_0(\x)  = - B(\x) + {\rm const} ~ , \eqn\identityFfTwo $$
where the constant has to be determined with the aid of, for instance, the
neutrality condition at one point in a given configuration.  (To be precise,
only the constant part of $a_0+eA_0$ is relevant in (\indCurrentOne).)
Substituting (\identityFf) and (\identityFfTwo) into (\indCurrentOne),
 one finds
$$\eqalign{
eJ^0_\ind = &  ({\rm const})
- \Big( {\nu e^2\over 2\pi} \Big)^2 \, c_0 \, A_0 +
 \Big( {\nu\over |N|} - (1-c_1) {\nu^2 m e^2 \over 4 \pi^2 n_e} \Big)
 \, div\, {\bf E} \cr
 & + \ep(N) \,{\nu e^2\over 2\pi}
 \bigg\{ \Big( 1+ {\nu\over |N|} c_0 -c_1 \Big) B -
{\nu e^2\over 2\pi} (1-c_1) {m\over e^2 n_e} \, \nabla^2 B \bigg\}
 ~, \cr
eJ^k_\ind = &\bigg\{ {\nu\over |N|} (1-c_1) - {\Delta_1 e^2\over \pi m} (1-c_2)
  \bigg\} \, \ep^{kl} \d_l B \cr
& + \ep(N) {\nu e^2\over 2\pi}   \ep^{kl}
 \, \bigg\{ (1-c_1) E_l - (1-c_2) {4\pi \Delta_1\over me^2 \nu |N|}
  \, \d_l (div\, {\bf E}) \, \bigg\}  ~~. \cr}
   \eqn\ourLondonOne $$

The  equations in (\ourLondonOne) correspond to the London equations
in the conventional superconductors.  Combined with the Maxwell equations,
they determine electromagnetic fields inside anyon fluids.  In general,
however, one more equation, Eq.~(\induction), has to be supplemented to fix the
constant part of the magnetic field.

To illustrate the problem, we consider a configuration in which an anyon fluid
occupies a half plane, say, $x_1 >0$.  We apply an uniform external magnetic
field $B_\ext$ in the empty space ($x_1 <0$).  The problem is to find
$B(\x)=B(x_1)$ for $x_1 >0$ with the boundary condition $B(0)=B_\ext$.

To extract the essence, we suppose that $eB_\ext \ll b^{(0)}$.  To good
accuracy one can approximate $\nu=|N|$.  With the aid of the numerical
evaluation
for various parameters given in Sections 18 and 20,  one finds
that the Maxwell equations become
$$\eqalign{
& (1+ c_0 ) B - {m\over e^2 n_e} \, \nabla^2 B
- {N e^2\over 2\pi}  \, c_0 \, A_0 +
{N m  \over 2\pi n_e}  \, div\, {\bf E} + ({\rm const}) =0 \cr
& - {2 \Delta_1 \over N m}\,  \d_l B
 +  E_l - {4\pi \Delta_1\over me^2 N^2}
  \, \d_l (div\, {\bf E}) =0 \cr}
   \eqn\MaxOne $$

For the configuration under consideration one expects
$$B(\x) = B_{\rm in} + (B_\ext - B_{\rm in}) \, e^{- x_1/\lam}
 \hskip 1cm (x_1 >0)~.   \eqn\insideB $$
$B_{\rm in}= B(+\infty)$ does not vanish at $T\not= 0$.  It is determined by
Eq.~(\induction)
$$B_{\rm in} = B_\ext + M(T,B_{\rm in}) ~~~.
   \eqn\BinBext $$
It follows from (\MaxOne) that $E_2=0$.

It is checked posterior that $|E_1/(B-B_{\rm in})| \ll 1$ so that the first
equation of  (\MaxOne) yields
$$(1+ c_0 ) (B- B_{\rm in}) - {m\over e^2 n_e} \, \d_1^2 (B- B_{\rm in})
    =0 ~~.   $$
Hence the damping length $\lam$ is approximately given by
$$\lam(T)^2 = {1\over 1+c_0} \, {m\over e^2 n_e}
  = {\sqlamb \over 1+c_0}   \eqn\damping $$
and
$${E_1(x_1) \over B(x_1) - B_{\rm in}} \sim  - {N\over m \lam}
 \sim 3 \times 10^{-6} \, ( 1+c_0)^{1/2} ~~~. \eqn\smallE  $$

Due to the non-vanishing $B_{\rm in}$, the damping length $\lam(T)$ should not
be confused with the penetration depth, which measures how fast the magnetic
field decreases in the material.  One complication in the calculation is
 that Eq. (\BinBext) cannot be solved analytically at $T\not= 0$.  We present
the result of numerical evaluation in Fig.~12.

{\figure
\vglue 9cm
\centerline{Fig.~12 ~~Magnetic field inside the anyon fluid in SCF}
}
\vskip 10pt

As can be seen from the figure, $B_{\rm in}$ is vanishingly small at low
temperature, but starts to increase around 70 K and becomes almost equal
to $B_\ext$ around 100 K.  Here we have only a partial Meissner effect,  at
least in SCF.  The magnetic field configuration is not a simple exponential
decay.

To avoid solving Eq.~(\BinBext), one may apply a spacially alternating
external magnetic field.  In the linear response theory it is reduced to
examinig $B_\ext(x_1)=B_0\,  \ep(x_1)$ applied to a system occupying
the whole space as we did in Section 16 at $T=0$.  We are going to show
that the partial Meissner effect  is observed in the response function, too.

The equations to be solved are the same as in Section 16, with $\Pi_k$'s
being replaced by $\Pi_k^E$'s.   Employing the expansions (\PiEzeroTwo),
(\PiEone), and (\PiEtwo), and keeping dominant terms, one finds,
instead of (\DelCsmallq) and (\staticQcOne),
$$\eqalign{
&\Delta_c =  {e^4\over q^4} \, \Big( {N\over 2\pi} \Big)^2 \,
 (1+ c_0 + \bar \sqlamb q^2)  \cr
&Q_c^{22} = {q^2 \over e^2} \,
   {1\over \strut 1+c_0 + \sqlamb q^2}   ~~~. \cr}
   \eqn\finiteKernel $$
As $T$ changes, the response function for a charged anyon fluid smoothly
varies.  However, its behavior is different from that in conventional
superconductors.

Recalling $eJ^2_\ext(\omega,\q)= - 2B_0 \cdot (2\pi)^2 \delta(\omega)
  \delta(q_2)$, one finds
$$\eqalign{
J^2_\tot &= J^2_\ext + J^2_\ind
    = \Big( 1 - {e^2\over q_1^2} \, Q_c^{22} \Big) \, J^2_\ext \cr
&={c_0 + \sqlamb q_1^2 \over 1+ c_0 + \sqlamb q_1^2} ~ J^2_\ext
{}~~~. \cr}
   \eqn\PartialMeissOne $$
The external current is not completely cancelled by the induced current:
$$J^2_\tot(\q=0) = {c_0\over 1+c_0}\, J^2_\ext(\q=0) \not= 0 ~~~.
\eqn\PartialMeissTwo $$

The magnetic field is
$$\eqalign{
B(\q) &= {i\over q_1} \, J^2_\tot  \cr
&= iq_1 \, \bigg\{ {c_0\over 1+c_0} \, {1\over q_1^2}
 + {1\over 1+c_0} \, {1\over q_1^2+ \lam^{-2} } \bigg\}
 \, J^2_\ext \cr}  ~~~.  \eqn\PartialMeissThree $$
A new pole develops at $\q=0$.  In the configuration space
$$B(x) = B_0 \, \ep(x_1) \, \bigg\{ {c_0\over 1+c_0}
   + {1\over 1+c_0} \, e^{-|x_1|/\lam} \bigg\} ~~~. \eqn\PartialMeissFour $$

At $T=0$, $c_0=0$ so that the Meissner effect is complete.  At $T \not= 0$,
$c_0 \not=0$, resulting a partial Meissner effect.  As we have seen
in the previous section,  $c_0(T)$ suddenly becomes very large around
$T=100$ K.  Therefore the Meissner effect effectively terminates around
this temperature.  In the approximation (SCF) in use, however, there
does not result a phase transition.  We shall argue in the next section
that a phase transition should result if vortices are incorporated.

With the aid of (\PartialMeissFour) one can define an effective penetration
depth, $\lambda_\SCF$,  which measures the rate of the change of the magnetic
field.  $\lambda_\SCF(T;d)$ is related to the change of $B(x_1)$ over a
distance
$d$ by
$$e^{-d/\lambda_\SCF} = {B(d) \over B_0} ~~~.
  \eqn\SCFlambdaOne  $$
It depends on $d$.  As a typical value we take $d=\lambar$.  Then
$${\lambar\over \lambda_\SCF} =
- \ln \Big\{ {c_0\over 1+c_0} + {1\over 1+c_0} \, e^{-(1+c_0)^{1/2} } \Big\}
  ~~~.  \eqn\SCFlambdaTwo $$
Approximately
$$
{\lambda_\SCF \over \lambar} =
\cases{ 1+ (e-1.5) c_0 &for $c_0 \ll 1$~,\cr \cr
        c_0            &for $c_0 \gg 1$~.\cr}    \eqn\SCFlambdaThree $$
The behavior of $\lambda_\SCF(T)$ is depicted in Fig.~13.
\def\smalllambda{\hbox{\eightmit\char'25}}
\def\smallSCF{\hbox{\sevenrm SCF}}
{\figure
\vglue 7.5cm
\centerline{Fig.~13 ~~
  \vtop{ \hsize=10cm \noindent
The penetration depth in SCF, {\smalllambda}$_{\smallSCF}$
defind in (\SCFlambdaTwo).    The small tail in
($\bar{\smalllambda}$/{\smalllambda}$_{\smallSCF}$)$^{2}$
around {\eightit T}=100  K is regarded as an artifact
of  the approximation.  See the discussion in Section 22.}  }
}
\vskip 14pt

As we have demonstrated, the dominant finite temperature effect is contanied in
$c_0(T)$.   The cancellation of the bare Chern-Simons term by the induced one,
for instance, is not exact at $T\not= 0$, since $c_1(T) \not= 0$. (See
Eq.~(\finiteOmegaTwo).)  However, its effect is numerically negligible.  For
the Meissner effect, ``$c_0$'' is important.  It makes the Meissner effect
partial at $T\not= 0$.

Before closing the section, we briefly mention about the subtlety in  neutral
anyon fluids.  The response function for a neutral anyon fluid at finite $T$ is
given by
$$\eqalign{
Q_n^{22}(\q) &\sim
  {q^2\over \big \bar c_0 + ( m \, q^2 / n_e)  } ~~~, \cr
\noalign{\kern 5pt}
\bar c_0 &= e^2 c_0 = {4 \pi^2 n_e\over \nu^3 T } ~ \delta_0 ~~~.\cr}
     \eqn\neutralResponseT $$
It follows that the two limits, $q\go 0$ and $T\go 0$ do not commute with
each other.   It may reflect an instability in the system of neutral
anyon fluids.

\sectionskip

\secno=22  \meqno=1

\line{\bf 22.  $T_c$ \hfil}
\vglue 5pt
How large is $T_c$, if there is a phase transition?   Having analysed
properties
of charged anyon fluids, we are in an awkward position.  In the linearized
SCF, or equivalently in RPA, we have seen no evidence for a phase transition,
or more precisely, mathematical singuralities in physical quantities, at
finite temperature.  For instance, we have seen in the previous section
that the penetration depth $\lambda_\SCF(T)$ rapidly increases around
$T_c' \sim 100$ K, but never diverges.

We argue that this is an artifact of the approximation in use, and that in
a full theory a charged anyon fluid should exhibit a phase transition around
$T_c'$.

Crucially missing in the previous treatment is a vortex.  It is missing,
because the linearized version of the SCF equations (\ourGLeq) are linear
in fields, and therefore do not admit a quantized flux.  In terms of
the effective theory obtained by  integrating  the fermion fields
$\psi$, one needs to retain higer order terms, cubic, quartic $\cdots$ in
the field $(a+eA)^{(1)}$.  It is a challenging problem to show how a vortex
solution comes out from such an effective theory.

Previously the quantization of vorticity in a neutral anyon fluid was
examined by Hanna, Laughlin, and  Fetter in the Hartree-Fock
approximation.\myref{\HLF}  They showed that an elementary excitation has a
vorticity given by (fundamental unit) /$|N|$, although it has an infinitely
large energy.   Kitazawa and Murayama\myref{\KitaMura} have
examined effects of vortex-antivortex pair
formation in a neutral anyon fluid at $T\not=0$.  They have contended that
there
is a Kosterlitz-Thouless phase transition at $T= {1\over 8}\, \ep$
for $|N|=2$.   The underlying assumption is that there are vortex pair
excitations with logarithmic interactions.

Having vortex-antivortex pair excitatons is one promissing way
of obtaining a phase transition in anyon fluids.  Supposing abundant
pair excitations, one still has to elaborate Kitazawa and Murayama's
argument for charged anyon fluids.

First of all interactions among
vortices are not logarithmic at low temperature.  The Meissner effect
is operating so that interactions are exponentially suppressed at large
distances. An energy of a single vortex, which is not known yet, must be a
dominant factor at low $T$.

The situation becomes more complicated as temperature increases.  As we have
observed in the previous section,  the Meissner effect effectively terminates
around $T_c'$.  There would be no screaning of magnetic fields any more.  The
interaction among vortices  becomes long-ranged, and the entropy
factor becomes important.  Whether or not this leads to a phase transition is
a matter subject to future investigation.  Any way there is not any trace of
superconductivity well above $T_c'$.  It is quite likely that $T_c$, which
separates the superconducting and normal states, turns out around $T_c'$.

$T_c'$ signifies a temperature where $c_0(T)$ becomes large.  From
(\cDefinition) to (\valueThree) one finds
$$c_0(T) = {\pi n_e^{d=2} d \over \alpha |N|^3 T} \,
\exp \Big( -{\pi n_e^{d=2}\over |N| m T} \Big) ~~~. \eqn\cZeroAgain $$
Numerically, for $N= \pm 2$,
$$\eqalign{
c_0 (T)
&= {121 \,{\rm K}\over T} \,{n_e^{d=2} \over 2 \times 10^{14} \,{\rm cm}^{-2}}
   \,{d\over 5 \A} \cr
\noalign{\kern 7pt}
&\hskip 1.5cm \times \exp \bigg\{ - 13.83
  \Big( {100\,{\rm K}\over T} \,
  {n_e^{d=2} \over 2 \times 10^{14} \,{\rm cm}^{-2}}
 \, {2m_e\over m} -1 \Big) \bigg\}  ~~~. \cr} \eqn\cZeroT  $$
Typical values are
\def\sp{\hskip .6cm }
\def\ssp{\hskip .1cm }
\def\msp{\hskip .3cm }
$$\vcenter{
\halign{ \hfil$#$\hfil &#\hskip .6cm
  &\hfil $#$ \hfil &#\ssp &\hfil $#$ \hfil&#\msp
  &\hfil $#$ \hfil &#\sp  &\hfil $#$ \hfil&#\sp &\hfil $#$ \hfil &#\sp
  &\hfil $#$ &#\sp  &\hfil $#$  \cr
\noalign{\kern 5pt}
T && 50 && 75 && 100 && 125 && 150 && 175 && 200 \cr
\noalign{\kern 7pt}
c_0(T)&&2.4\times 10^{-6} && 1.6\times 10^{-2}  &&1.2 &&15
   &&81   && 260 && 610   \cr
\noalign{\kern 5pt}
}   } $$
Note that if the values of both $n_e^{d=2}$ and $m$ are doubled, the value
of $c_0$ is also doubled.

The temperature dependence of $c_0$ is controled by the exponential
factor.   The critically important value is the ratio $n_e^{d=2}/m$.
With the given value in (\cZeroT), $c_0$ suddenly becomes large around 120 K.

With the assumption $T_c \sim T_c'$ we conclude
$$T_c \sim {2\over |N|}  {n_e^{d=2} \over 2 \times 10^{14} \,{\rm cm}^{-2}}
  {2m_e\over m} \times 120 \, {\rm K} ~~~.
  \eqn\TcValue  $$
Of course, the effective mass $m$ is very difficult to determine experimentally
so that one more piece of information is necessary to predict $T_c$.
One way is to express $m$ in terms of the penetration depth at $T=0$,
$\lambda(0)=\lambar$, with the aid of (\penetrationThreeD) and
(\effectiveMassOne).  Then
$$T_c \sim {2\over |N|} \,  {d\over 5\, \A} \,
 \bigg( {1400 \, \A\over \lambda(0) } \bigg)^2 \times 90 \, {\rm K} ~~~.
  \eqn\TcValueTwo $$

We stress that the value $T_c \sim 100\,$K is very natural in anyon fluids.
The dependence $T_c \propto n_e$ or $\lambda(0)^{-2}$ has been observed in high
$T_c$ superconductors.\myref{\Uemura}  This behavior, however,  is not
necessarily special to anyon superconductors.\myref{\TDLee}

Although the discussion in this section is only plausible and further
investigation is necessary, one might take (\TcValueTwo) as a very encouraging
result.

\sectionskip

\secno=23  \meqno=1
\def\subsec#1{ \vskip 8pt  \noindent {\bfit #1} \vglue 4pt }

\line{\bf 23.  Other important issues \hfil}
\vglue 5pt
In this article we have analysed some of the basic problems in anyon fluids,
attempting to summarize the first two years of the theory of
anyon superconductivity.  We have seen that various approaches are
equivalent, leading to many interesting physics consequences.  Nevertheless,
it is appropriate to say that so far we have had only partial understandings of
the  full theory.

There are many important issues left over.
We list them for readers' convenience.  For details readers should
consult original papers and other review articles.

\subsec{23.1.  Beyond RPA and the linearized SCF}

Going beyond RPA and the linearized SCF is important in many respects.
RPA and the linearized SCF fail to predict a phase transition.
It is essential to incorporate vortices in the theory.

Hanna, Laughlin, and Fetter\myref{\HLF} examined various quantities in the
Hartree-Fock approximation in the first quantized theory, as was described in
Section 7. They have found that, for $|N|=2$,  the correction to the phonon
spectrum in neutral anyon fluids  in the long wave
length limit is relatively small ($\sim 10$ \%).  More recently Dai,
Levy, Fetter, Hanna, and Laughlin\myref{\Dai} have performed a diagram
analysis at $T=0$ equivalent to the Hartree-Fock approximation.  In addition to
recovering the old result, they have shown that there results an important
modification  in the behavior of the spectrum at short wave lengths.
Furthermore
$\sigma_{xy}$ vanishes for $N=\pm 1$ thanks to higher order radiative
corrections.  RPA and the linearized SCF predict a non-vanishing $\sigma_{xy}$
even for $N=\pm 1$, which is certainly wrong as fermions are
merely converted to bosons, but not to genuine anyons.

Dai et al.~have solved the Schwinger-Dyson
equations numerically which involve about 100 different diagrams.  They have
noticed the importance of gauge invariance, and have observed many
cancellations
among various diagrams.  Their Feynman rules are based on the Hamiltonian
obtained after eliminating Chern-Simons fields.  As we have recognized in
Sections 9 -- 11, keeping the Chern-Simons gauge fields as auxiliary fields
greatly simplifies computations.  Gauge invariance is easily
implemented, and the notion of self-consistent fields (beyond the linear
approximation) can be established.

\subsec{23.2.  Pair-correlation}

Is there ``Cooper'' pairing in anyon superconductors?  The Hartree-Fock
ground state employed in the literature (see Sections 6 and 7) does not look
like  the BCS ground state.    Is there a different kind of pairing, then?
Is there an off-diagonal long range order?  The answer has not been known
for sure.

There is an indication for pairing in the $N=\pm 2$ theory, which, however,
is quite different from the Cooper pairing.  The unique feature of the
Hartree-Fock ground state is that the complete filling of the Landau levels
is achieved independent of the density $n_e$, provided that $N$ is an integer
($\not= 0$).

To be precise, suppose that each Landau level has $N_L$ available states so
that the total particle number is $N_e= n_e\cdot (vol) = |N| \cdot N_L$.
Assume that $|N| \ge 2$.
If one tries to add or delete one particle to or from the system, one
necessarily
has to put the particle in the next level, or make a hole in the top filled
level, in order to preserve the Landau level picture.  In other words,
the picture of the complete filling breaks down.

However, if a set of $|N|$ particles are added or deleted, one can still
maintain the complete filling.  One of the Chern-Simons field equations,
$(N/2\pi) \, b = j^0$, implies that the increase (decrease) of the particle
number leads to the increase (decrease) of the Chern-Simons magnetic field such
that precicely one more (less) state is available in each Landau level.

The states with the particle number $N_e$ and $N_e \pm |N|$ are very much
alike.   In a macroscopic system $N_e \gg 1$,  thermal fluctuations
give $\Delta N_e \sim \sqrt{ N_e}$.  It is quite likely that the real
ground state is not an eigenstate of the particle number,
but is a coherent state:
$$\Psi_G(\theta) = \sum_{k \atop |kN| < \sqrt{N_e}}
  e^{ik\theta} ~ \Psi_G(N_e + k |N|) ~~~. \eqn\coherentState  $$
In particular, for $N=\pm 2$, the structure (\coherentState) is exactly
the same as in the BCS theroy.\myref{\Josephson}

It is not clear if (\coherentState) implies pairing in the $N=\pm 2$ theory.
We should remember, however, that the structure of the coherent state is
indispensable in understanding many phenomena in superconductivity.

\subsec{23.3.  Flux quantization}

A magnetic flux is trapped by a superconducting ring.  The flux takes
quantized values in the unit of $2\pi \hbar c / 2 e$.  Can the $N=\pm 2$
anyon theory explain this behavior?  No convincing argument has been
provided.  Leggett\myref{\Leggett} has argued that the flux quantization is not
achieved in anyon theory at least in the Hartree-Fock approximation.

One needs to show two things.  First it must be shown that an energy is
locally minimized when a flux takes a quantized value.  Secondly,
the energy barrier height between the states with no flux and
with one unit of flux is proportional to the volume, but not the boundary area,
of the superconducting ring.

\subsec{23.4.  Vortices}

We have often mentioned in the preceeding sections that the establishment of
vortices in anyon superconductors is one of the major problems to be solved.
The previous analysis by Fetter, Hanna, and Laughlin must be elaborated.
Inclusion of electromagnetic interactions is essential to take account of
the Meissner effect.    Nonlinear terms in SCF must play an important role.
At finite temperture vortices might lead to a phase transition, too.

\subsec{23.5.  The Josephson effect}

The nature of the coherent state (\coherentState) is most important for the
Josephson effect.\myref{\Josephson}   A Josephson junction consists of two
superconductors separated by a barrier.  Electron tunneling through the barrier
brings about phase coherence over the entire system.  The energy is minimized
by volume if the two phases, which characterize $\theta$'s in
(\coherentState) of the two superconductors, coincide.  The difference
between the two phases should generate a current.

Experiments show that in order for anyon superconductivity to describe
high $T_c$ superconductors, $N$ must be equal to $\pm 2$.   It seems that
a Josephson effect should exist even for a junction between a BCS
superconductor and an anyon superconductor of $N=\pm 2$.

\subsec{23.6.  Interlayer couplings}

High $T_c$ superconductors have the layered structure characterized by
CuO planes.  Our analysis has been performed  in the effective
two-dimensional theory obtained by the dimensional reduction.  The implicit
assumption was that the system is uniform in the direction perpendicular to the
two-dimensional CuO planes.

Anyon theory has a parameter $N$, or the generated statistics phase
$\theta_{\rm statistics}= \pi/ N$.  Physics depends on $\exp (i\theta_{\rm
statistics})$.   The theory with $N$ differs from that with $-N$ if $|N| \ge
2$. The two theories are related by $P$ (parity) and $T$ (time
reversal) transformations. The idea behind the degeneracy of the ground state
is that $P$ and $T$ symmetry is spontaneously broken.

$P$ and $T$ invariant quantities such as the penetration depth and resistance
do not depend on the sign of $N$.  Moreover we have seen that even though
the Chern-Simons magnetic field $b/e$ is very large ($\sim$ 1000 T), the
dependence of the $P$- and $T$-odd magnetization $M(B)$ on the Maxwell magnetic
field $B$ is symmetric to good accuracy, $M(-B)\sim M(B)$.  The asymmetry
arises
only to the  order  O($eB/b$). (See Section 18.)

Nevertheless it is important to know how the sign of $N$ is ordered among
adjacent layers. Is it ordered ferromagnetically with the same sign
(FM ordering), or antiferromagnetically with the alternate sign (AFM ordering)?
Or, is it randomly distributed?

This problem has been examined by  Rojo, Canright,  and
Leggett.\myref{\interlayer}  Interactions among electrons in different layers
fix a pattern of the ordering. There are two types of interactions.  One is of
a
potential type, and the other is the hopping of electrons from one layer to
adjacent ones.

The detailed examination in the case of potential interactions has been
provided
by the above authors both numerically and analytically.  They have shown that
$T$-invariant potential interactions always prefer the AFM ordering.
The hopping interaction is expected to induce a Josephson effect and
lead to the FM ordering.  It is not clear which one is dominant.

\subsec{23.7.  $P$ and $T$ violation}

Many experiments have been performed to check $P$ or $T$ violation in
high $T_c$ superconductors.  The result is confusing, but a fair statement
is that so far there has been no solid evidence for $P$ or $T$ violation
in high $T_c$ superconductors.

As explained above, $P$ and $T$ are ordered either ferromagnetically or
antiferromagnetically among layers.  Most of the experiments done so far
measure $P$ and $T$ violation in bulk.   Therefore, if material has the
AFM ($P$-, $T$-) ordering, the effect cancells in bulk.

Experiments in this category are the electromagnetic wave polarization and Hall
voltage. Originally proposed by Wen and Zee, the polarization experiment tries
to
measure the $P$, $T$ violation effect, determinig the transmission and
reflection coefficients of injected polarized electromagnetic
waves.\myref{\WenZeeTwo-\Halperin}   The polarization vector
is rotated in anyon superconductors.  There is inconsistency among various
experiments,however.\myref{\ExpPolarization}  We note that even though Wen and
Zee argue that there must be appreciable effects in FM-ordered anyon
superconductors, a really microscopic computation of the magnitude of the
effect
is still lacking.   Also some criticisms have been provided on the
interpretation of experimental results.

The Hall voltage experiment measures the temperature dependence of the
transverse voltage (Hall voltage) when a current flows in a thin
film.\myref{\CWWH,\HRW,\Hetrick}  There is
a  microscopic calculation of the effect.  Theory predicts a peak in the
Hall voltage around $T_c$.  The peak value predicted is, for a thin film
of thickness 1000 \AA with a current $10^{-4}$ Amp, $V_{\rm Hall}
\sim 2 \times 10^{-7}$ Volt.  It is inversely proportional to the thickness.
The effect is tiny.  Preliminary experiments have been performed.
Due to the inhomogeneity of samples and also tiny temperature variation in
the samples, no conclusion has been obtained concerning the existence or
non-existence of the Hall voltage.\myref{\ExpHall,\Goldman}

There is one experiment which measures $P$, $T$ violation in one layer,
and therefore is sensitive even for AFM-ordered anyon superconductors.
It is the muon spin relaxation experiment.\myref{\HRW}  Injected muons are
stopped in high $T_c$ superconductors.  Since muons are charged, the
distribution of  electrons is deformed.   In effect, the distribution of
holons,
or our $\psi$ particles, deviates from the uniform value.  $\delta J^0(\x)
\not=
0$ results.   In anyon superconductors it induces a current $J^k_\ind \not= 0$,
since $Q_c^{k0} \not= 0$.
$J^k_\ind \not= 0$ in turn generates Maxwell magnetic field $B^3_\ind$, which
is felt by muon spins.  Muon spins start to precess, which can be observed
experimentally.

Halperin, March-Russel, and Wilczek gave a plausible argument, predicting
$B^3_\ind \sim 10$ G.  The experiment observed no effect.\myref{\ExpMuon}  We
remark that the magnitude of the effect can be determined more microscopically
from the knowledge of the response function $Q_c$ at both $T=0$ and
$T\not= 0$ without making the  long wave length approximation.

\subsec{23.8.  Anyons in spin systems}

In this article we have not discussed how anyon excitations arise in
material, particularly in spin systems.  We have started with the picture
that there are excitations called ``holons'' which obey half-fermion
($N=\pm 2$) statistics.

The derivation of anyons, or fractional statistics, in realistic spin
models for high $T_c$ material has been attempted by many
authors.\myref{\holonTwo-\Wiegmann}
The issue has not been settled yet, since the arguments involve many
approximations for which justification is not clear.  Readers are advised
to read original papers.  A closely related  subject is the existence of
the flux phase or chiral spin liquid state.

We note that Laughlin has given a spin model which has  Laughlin's
wave function in the fractional quantum Hall effect as an exact ground
state wave function.\myref{\LaughlinLattice}   This model provides the
exsistence proof of anyons in spin models.

\subsec{23.9.  Variations of anyon models}

We have analysed one particular anyon model, namely non-relativistic
spinless fermions with the minimal Chern-Simons interaction.  It is the
simplest model of anyons, and is based on the holon picture of Anderson's.

There are many variations.  They are interesting in their owm right.
Historically Chern-Simons gauge theory was first analysed in relativistic
field theory.  Along this tradition Lykken, Sonnenschein, and Weiss have
examined a relativistic anyon model, Dirac fields with Chern-Simons
interactions.\myref{\Banks,\Lykken}  They have argued that the neutral model
retains a superfluidity to all orders at $T=0$.  In passing,  Imai et
al.\ have shown that the  non-renormalization theorem for the induced
Chern-Simons coefficient holds even in non-relativistic
theory.\myref{\Ishikawa}  At finite temperature the relativistic  model behaves
in a fashion similar to, but not same as, the nonrelativistic  model.  To
discuss a Meissner effect, superconductivity etc.~in condensed matter systems,
one has to analyse nonrelativistic models.  Based on non-vanishing finite
temperature corrections to the induced Chern-Simons coefficient, Lykken et
al.~have incorrectly concluded that a superconductivity is lost at $T\not=0$.
 As we have seen in Sections 20 and 21, the important
finite temperature correction is the $c_0$ term, but not the $c_1$ term
(Chern-Simons coefficient), in the non-relativistic theory.

We have supposed that particles (anyons) have a single component, i.e.~they
have the same coupling to the Chern-Simons fields.  There might be
two kinds of anyons, a half of them having the $+$ coupling and the other half
having the $-$ coupling.  Furthermore, in addition to the minimal gauge
coupling to Chern-Simons fields, particles might have magnetic moment
interactions.  Such a model has been investigated.\myref{\Joe}

So far we have started with fermion fields $\psi$.  It is also possible
to start with boson fields.    It is not exactly the same as the fermion model,
since a bose field can condensate by itself  and there is no complete
filling of Landau levels.   As was pointed out by Boyanovsky et al.,
 boson models have  rich structures many of which
need to be clarified further.\myref{\Boyan}  It is also known that boson models
are particularly useful to construct phenomenological theory of fractional
quantum Hall effect.\myref{\CSinFQHE-\FQHEmoreTwo}

\sectionskip

\leftline{\bf Acknowledgements}
\vglue 5pt
The author would like to thank Jim Hetrick for preparing some of the figures.
This work was supported in part by the U.S. Department of Energy under
Contract No. DE-AC02-83ER-40105.

\sectionskip

\leftline{\bf References}
\vglue 5pt \frenchspacing

{\ninerm
 \immediate\closeout\reffile
	\input refs.tmp\vfill\eject\nonfrenchspacing   }

\vfil
\bye